\documentclass[a4paper,11pt]{article}
\pdfoutput=1 
\usepackage{jinstpub} 
\usepackage{hhline}

\usepackage{lineno}

\title{A neural network classifier for electron identification on the DAMPE experiment}

\author[a,1]{D.Droz,\note{Corresponding author.}}
\author[a]{A.Tykhonov,}
\author[a]{X.Wu,}
\author[b,c]{F.Alemanno,}
\author[d]{G.Ambrosi,}
\author[d,e]{E.Catanzani,}
\author[b,c]{M.Di Santo,}
\author[b,c]{D.Kyratzis,}
\author{S.Zimmer}

\affiliation[a]{University of Geneva, CH-1205 Geneva, Switzerland}
\affiliation[b]{Gran Sasso Science Institute (GSSI), Via Iacobucci 2, I-67100 L’Aquila, Italy}
\affiliation[c]{Istituto Nazionale di Fisica Nucleare (INFN) - Laboratori Nazionali del Gran Sasso, I-67100 Assergi, L’Aquila, Italy}
\affiliation[d]{Istituto Nazionale di Fisica Nucleare (INFN) - Sezione di Perugia, I-06123 Perugia, Italy}
\affiliation[e]{Dipartimento di Fisica e Geologia, Universita` di Perugia, I-06123, Perugia, Italy}

\emailAdd{david.droz@unige.ch}
\emailAdd{andrii.tykhonov@unige.ch}

\abstract{The Dark Matter Particle Explorer (DAMPE) is a space-borne particle detector and cosmic ray observatory in operation since 2015, designed to probe electrons and gamma rays from a few GeV to 10 TeV in energy, as well as cosmic protons and nuclei up to 100 TeV. Among the main scientific objectives is the precise measurement of the cosmic electron+positron flux, which, due to the very large proton background in orbit, requires a powerful particle identification method. In the past decade, the field of machine learning has provided us the needed tools. This paper presents a neural network based approach to cosmic electron identification and proton rejection and showcases its performance based on simulated Monte Carlo data. The neural network reaches significantly lower background than the classical, cut-based method for the same detection efficiency, especially at at the highest energies probed by the detector. Good agreement between simulation and real data is demonstrated.}


\arxivnumber{2102.05534} 

\begin{document}
\maketitle
\flushbottom

\section{Introduction}
\label{sec:intro}

The DArk Matter Particle Explorer (DAMPE - also known as Wukong in China) is a satellite-based cosmic ray observatory and gamma ray telescope \cite{TheDAMPE:2017dtc}. It can measure cosmic electrons up to an energy of 10 TeV and protons and heavier nuclei up to a hundred TeV. It is constituted of four subdetectors, from top to bottom: a plastic scintillator (PSD) for absolute charge measurement; a silicon-tungsten tracker-converter (STK) for precise direction measurement and for enabling photon pair production; a Bismuth Germanium Oxide imaging calorimeter (BGO) of about 32 radiation lengths, made of 308 hodoscopically arranged bars in 14 layers and used for energy measurement, particle identification, and trigger \cite{zhang2015design}; and a neutron detector (NUD) for improving the identification of hadronic showers \cite{TheDAMPE:2017dtc}. The DAMPE satellite was launched into a 500-km Sun-synchronous orbit on December 2015, and has been in stable operations since then \cite{ambrosi2019orbit}.

\begin{figure}[h]
    \centering
    \includegraphics[width=.45\linewidth]{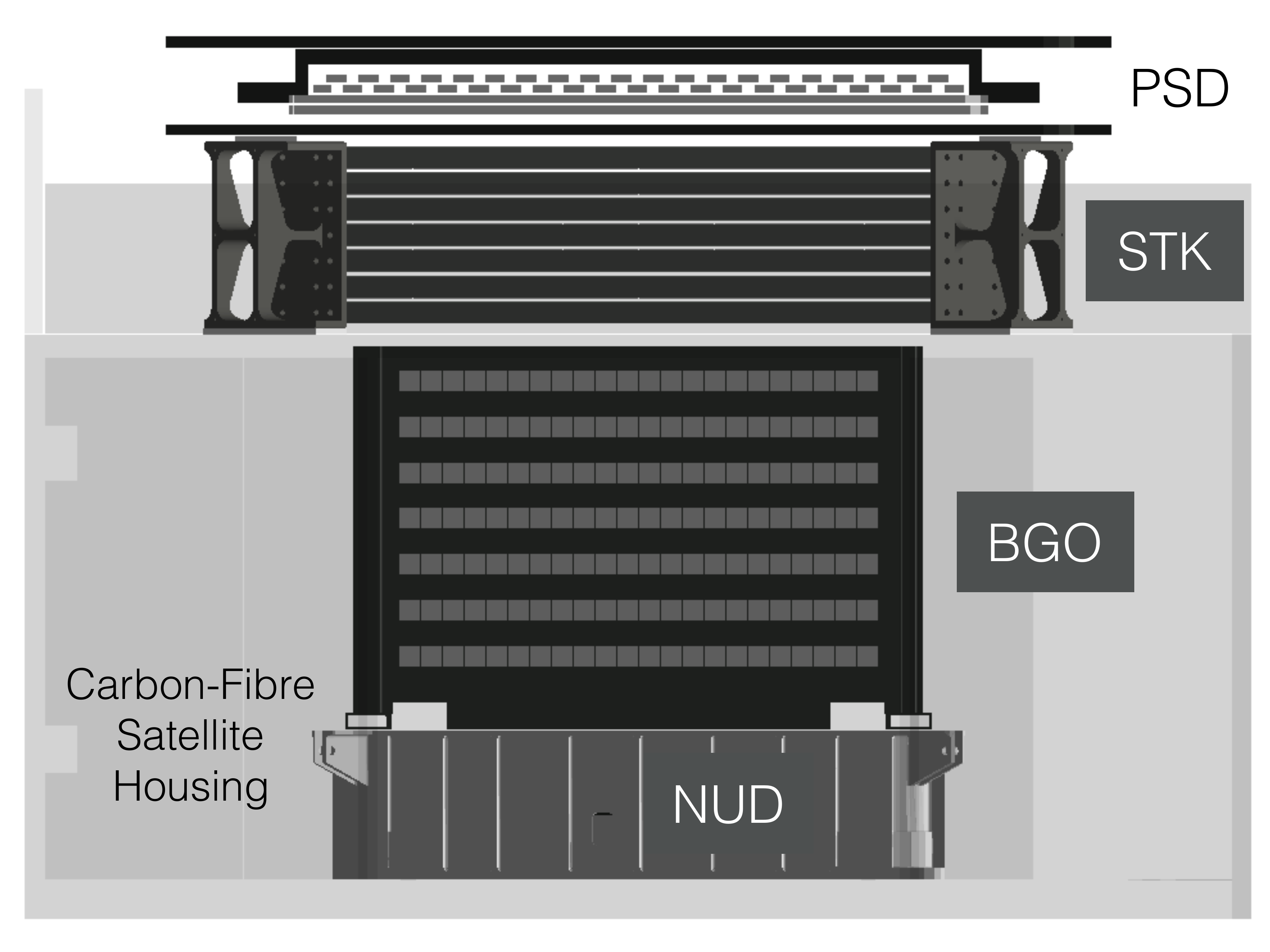} 
    \includegraphics[width=.45\linewidth]{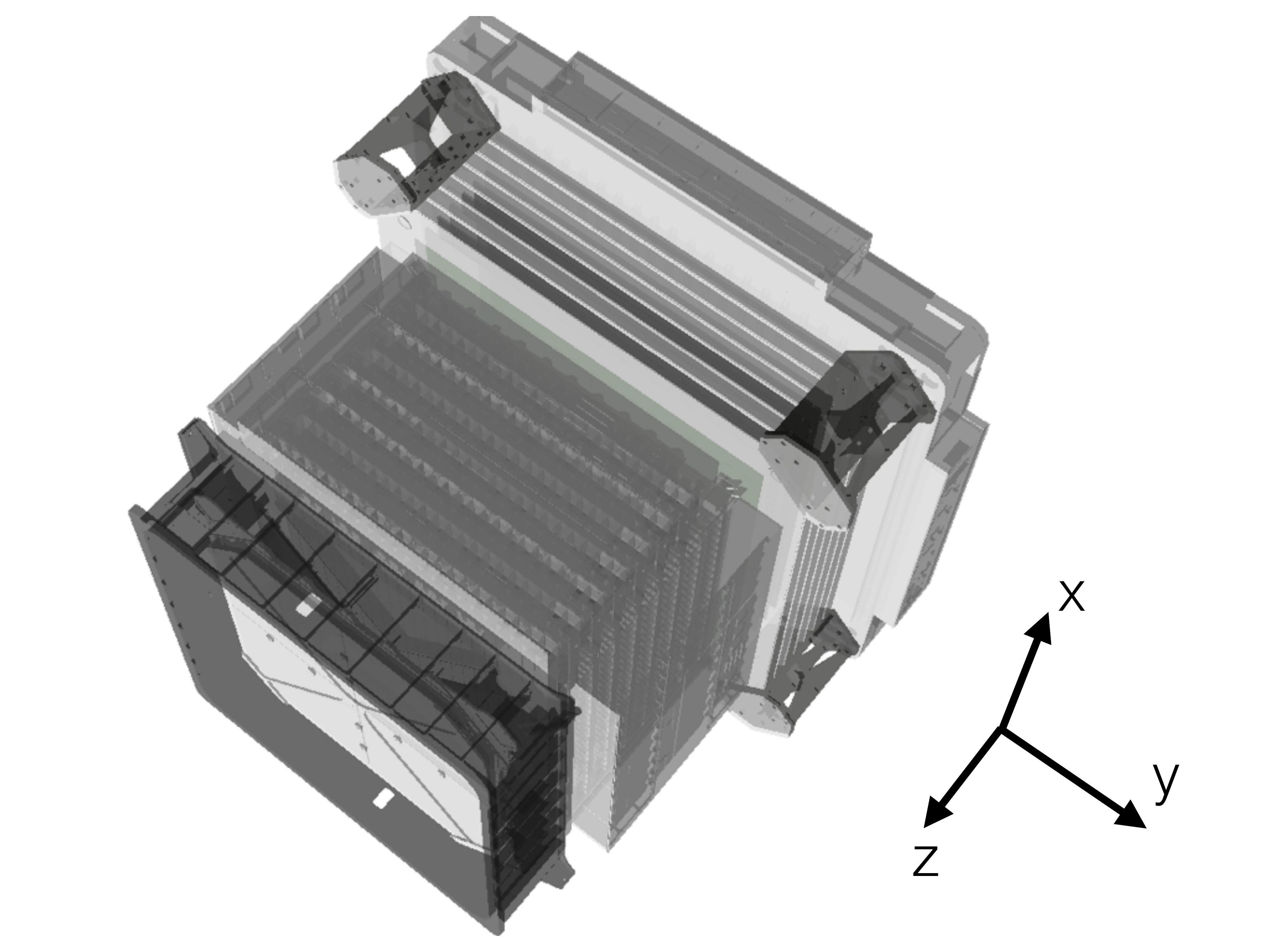}
    \caption{Layout of the DAMPE detector system. Figure from \cite{tykhonov2018internal}}
    \label{fig:DAMPE}
\end{figure}

Among the main scientific objectives of DAMPE is the study of cosmic ray electrons plus positrons (CREs) in the TeV range. Precise measurement of the spectrum of CRE is crucial for understanding the mechanisms of cosmic ray production and propagation \cite{meyer1969cosmic,moskalenko1998production,fan2010electron}. CREs represent a valuable probe of the most energetic process in the nearby Galaxy, such as supernovae explosions \cite{kobayashi2004most} and may enable the observation of phenomena such as dark matter decay or annihilation \cite{bertone2005particle,feng2010dark}.  The CRE spectrum was first measured by DAMPE in 2017 up to an energy of 4.6 TeV, with the direct detection of a spectral break in the TeV region \cite{Ambrosi:2017wek}. This result was obtained using a classical, cut-based technique for electron/proton discrimination. Said method, while powerful enough to reject proton background up to a few TeV with acceptable uncertainty, could not fully exploit the particle identification capability of the detector. 
After more than four years of operation and thus much higher accumulated data statistics, DAMPE is capable of measuring the CRE spectrum up to an unprecedented energy of 10 TeV with an instrument energy resolution better than 1.2\%~\cite{Ambrosi:2017wek}. However the standard identification technique does not have sufficient electron/proton discrimination power to be used at such extreme energies. A new method is therefore required, and we propose one coming from the ever-growing toolbox of data sciences and artificial intelligence.

In this paper we propose to use neural networks for the problem of electron identification on the DAMPE experiment. 
In section \ref{sec:electronidentification} we develop the electron identification technique based on a neural network classifier. In section \ref{sec:results:perf} we demonstrate the performances of the developed classifier on Monte-Carlo simulations and compare it with the standard cut-based technique. In section \ref{sec:results:val} we perform a validation of the developed classifier with the 4-years DAMPE orbit data and demonstrate that the new method allows a substantial enhancement of the electron identification, enabling the first CREs spectrum measurement with DAMPE in the 10 TeV energy domain. Finally, the work and the results are summarised in section \ref{sec:conclusions}.

\section{Electron identification}
\label{sec:electronidentification}

Measuring cosmic electrons in orbit requires first to identify them among the many sources of background \cite{grupen2005astroparticle}. Helium and heavier nuclei can be rejected by measuring the absolute electric charge, which is done in DAMPE using the signal deposited inside the plastic scintillator (PSD). The PSD cut is complemented by a charge measurement in the silicon tracker (STK) allowing to reject heavy nuclei outside of the PSD fiducial region. Gamma rays are another source of background, however their flux is orders of magnitude weaker than the CRE flux especially at energies higher than several hundred GeV. They are rejected thanks to their electric charge of zero: a gamma ray penetrating through the PSD will not leave any signal as opposed to other particle species.

\begin{figure}
    \centering
    \includegraphics[width=.9\linewidth]{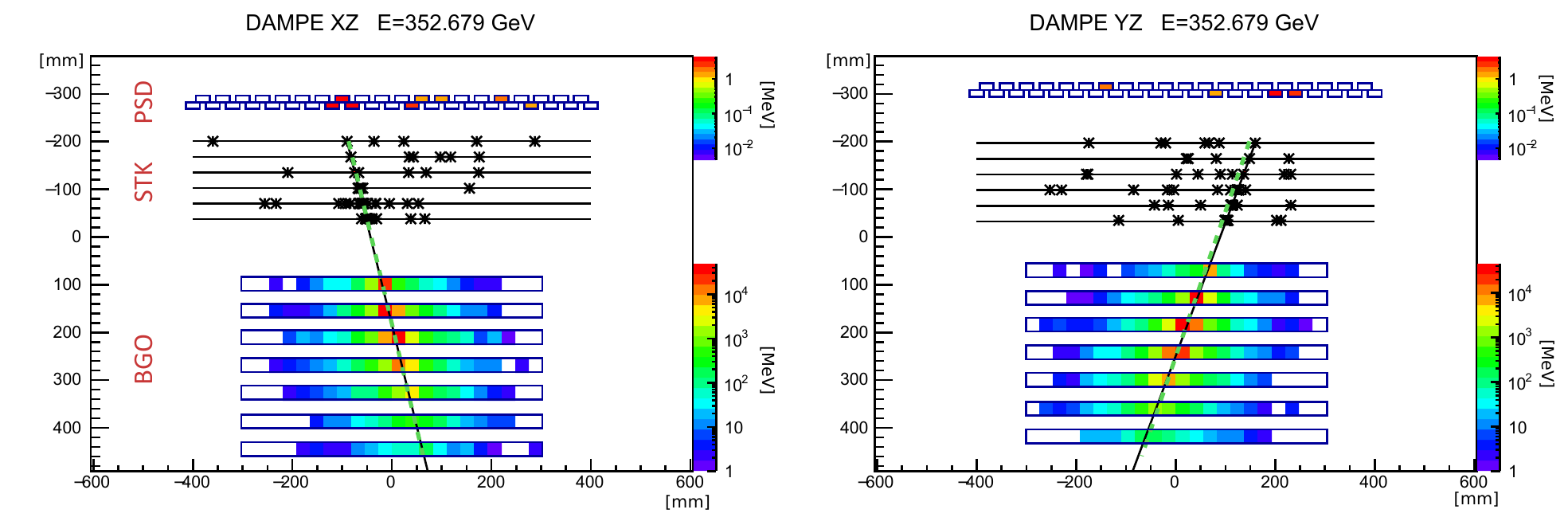}
    \includegraphics[width=.9\linewidth]{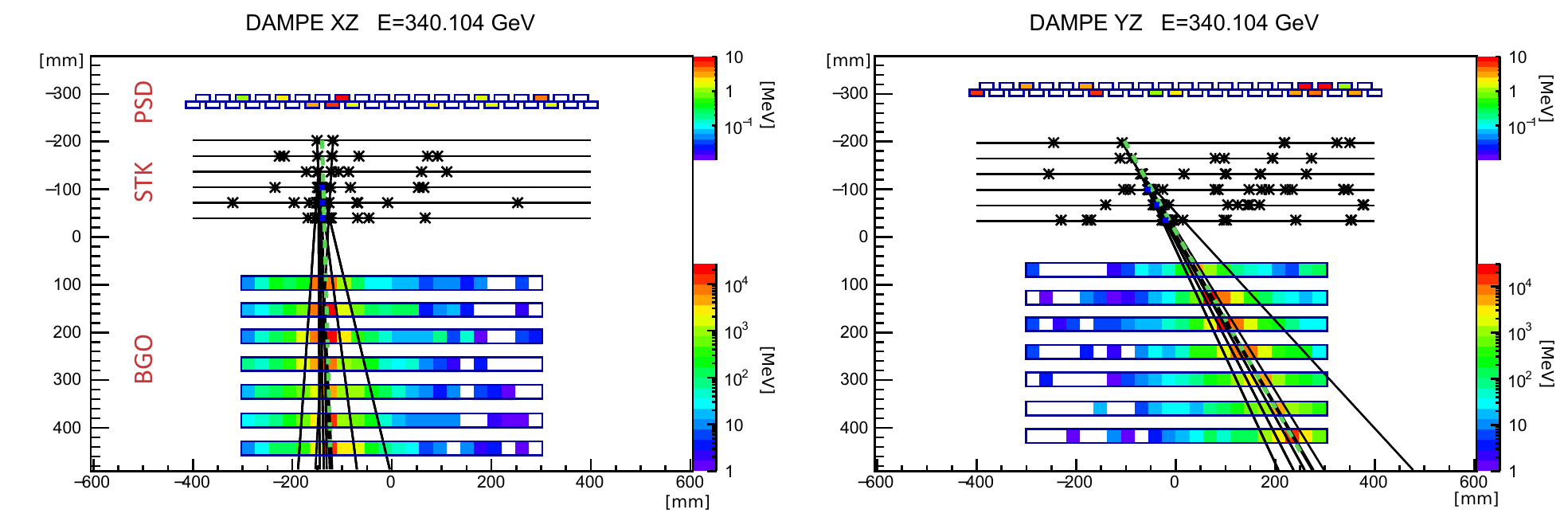}
    \caption{Display of simulated cosmic ray events inside DAMPE. The particle comes from the top and crosses in succession the PSD, STK and BGO detectors. Left and right panels show the X-oriented and Y-oriented views, respectively. Black solid lines are STK tracks, green dashed line is the BGO reconstructed trajectory. The stars in STK represent clusters of hits. Colour bars show the energy deposited in PSD and BGO.  \emph{Top:} Monte Carlo electron with a measured energy of 353 GeV. \emph{Bottom:} Monte Carlo proton with a reconstructed energy of 340 GeV.}
    \label{fig:evtDisplay}
\end{figure}

Protons have the same absolute electric charge as electrons, meaning they cannot be rejected with these methods. Moreover, protons are the most abundant comic ray species, with a flux orders of magnitude higher than that of electrons, drowning the signal in a high rate of background events. The exploitable difference between these two particles lies in the physical processes when interacting with matter: protons produce a wide and deep hadronic shower, while electrons produce narrower electromagnetic showers, resulting in different signal topologies in the BGO calorimeter. Figure \ref{fig:evtDisplay} shows a representation of electron and proton interactions within DAMPE: the simulated electron deposits most of its energy in the first half of the calorimeter whereas the proton shower develops both deeper and broader.

\begin{figure}
    \centering
    \includegraphics[width=.7\linewidth]{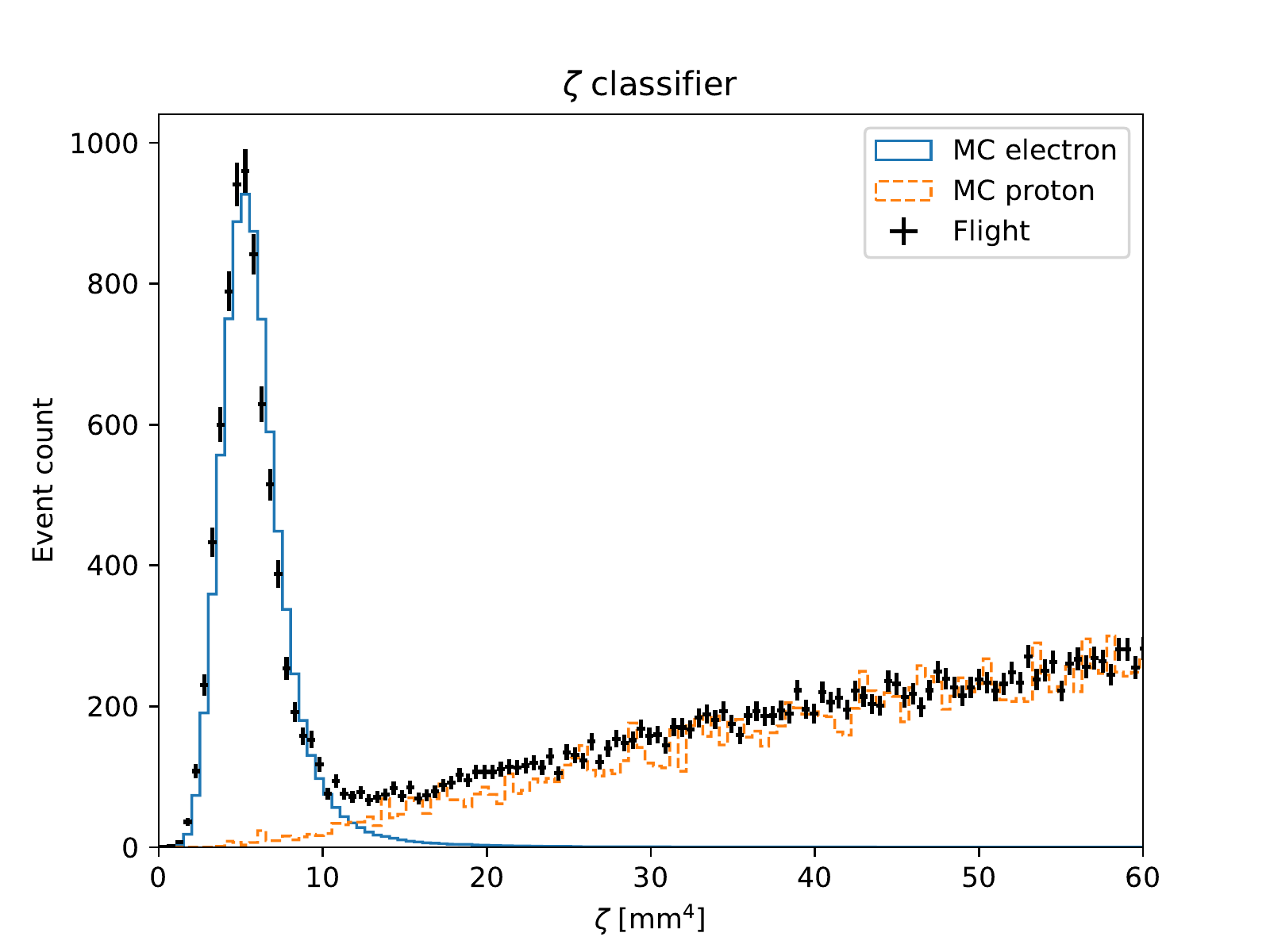}
    \caption{Distribution and Monte Carlo/data comparison of the classical $\zeta$ variable in the energy range 500 GeV to 1 TeV, out of a cleaned-up data to remove ions and require good shower containment (section \ref{sec:modeloptimisation}). }
    \label{fig:XTRL}
\end{figure}

Classically, one can build observables that quantify the shower shape inside the calorimeter, and use them to reject protons. The first DAMPE cosmic electron spectrum measurement  \cite{Ambrosi:2017wek} is based on a single observable named $\zeta$, quantifying the correlation between shower length and shower width~\cite{Chang:2008zzh}. It is defined as:
\begin{equation}
    \label{eq:Xtrl}
    \zeta = \frac{\left(\sum_i^\text{layers} (\texttt{RMS})_i \right)^4 \cdot \mathcal{F}_\texttt{Last}}{8 \cdot 10^6}
\end{equation}
where $\texttt{RMS}$ is the energy-weighted root-mean-square of hit positions in the calorimeter (shower width) and $\mathcal{F}_\texttt{Last}$ is the fraction of energy deposited in the last calorimeter layer with non-zero deposition (shower depth). This definition can be compared to the event displays in figure \ref{fig:evtDisplay}: the narrow electron shower naturally results in a $\zeta$ value much lower than for protons, as exhibited in their respective distributions in figure \ref{fig:XTRL}. Similar variables are commonly used for electron identification in calorimetric experiments.

While $\zeta$ proved powerful, it does not use the entirety of information available in the detector, including the possibility to exploit the strong correlations between topological variables used to describe the shower development in the imaging calorimeter. Furthermore its rejection power is limited above the TeV range where the topological development of hadronic and electromagnetic showers in the detector is less pronounced. A more powerful method is therefore required, and is presented in this paper.

\subsection{A neural network classifier}
\label{sec:neuralNets}

Deep learning and deep neural networks~\cite{DBLP:journals/nature/LeCunBH15,Goodfellow-et-al-2016} are at the forefront of current day developments in data sciences and artificial intelligence, often representing the state-of-the-art solution to a wide variety of problems. They have in particular found their way into both high energy physics~\cite{Baldi:2014kfa,guest2018deep}, where neural networks can exploit the large size and high dimensionality of the data, and into astrophysics \cite{schaefer2018deep,george2018deep} where their pattern recognition power allows for extraction of additional information from telescope images. However, neural networks have only been seldom used in cosmic ray physics, and never exploited for CRE direct detection experiments at multi-TeV energies. 

The classifier we propose to use is based on a regular neural network composed of a stack of densely connected layers (see below). Such algorithms are sometimes named as \emph{multilayer perceptron}, \emph{feedforward neural network}, or simply \emph{artificial neural network}. Other techniques were also studied in the DAMPE collaboration prior to this work: convolutional neural networks (CNN), suited for pattern recognition and image identification, showed promising performances.
While CNN demonstrated potential for further improvement with respect to the feedforward network, we opted for latter technique due to the better understood systematics of this type of networks, which is reflected, in particular, in data to Monte-Carlo agreement of a classifier score distribution  \cite{droz2019neural}.
Another technique studied by the collaboration are boosted decision trees, commonly used in high energy physics. They showed some improvement over the classical method, though optimised for the lower energy domain, around 10 to 100 GeV  \cite{zhao2018machine}.


An artificial neural network is a stack of densely connected layers of so-called neurons. A neuron is a mathematical unit that applies a non-linear function to the linear combination of its inputs. The function output is then used as input by all the neurons in the next layer, in the case of fully connected networks. Mathematically, if a neuron receives as input a set $\left\lbrace X_i \right\rbrace$, then its output $y$ is:
\begin{equation}
\left\lbrace X_i \right\rbrace ~ ~ \longrightarrow ~ ~ y = f\left( \sum_i w_i X_i \right) + b
\end{equation}
where $f$ is the non-linear activation function, $w_i$ the weights and $b$ the bias. The activation function $f$, as well as the number of neurons and layers, are characteristics of the model which are decided by the human programmer (section \ref{sec:modeloptimisation}). The values of $w_i$ and $b$ are set by the machine during the training procedure: the network is exposed to a set of labelled data (training data) where each observation/event is associated to a class (e.g. signal/background), and tries to minimise a given error metric on the set. The minimisation usually follows a so-called gradient descent or one of its flavours. In the case of a classification problem, such as discrimination between protons and electrons, the canonical error metric is the cross-entropy \cite{Goodfellow-et-al-2016}.

An artificial neural network takes as input a one-dimensional set of variables $\left\lbrace X_i \right\rbrace$. In a binary classification, it outputs a value between 0 and 1 which can be interpreted as the probability for an electron (signal) to produce $\left\lbrace X_i \right\rbrace$. This value is obtained as the output of a sigmoid or sigmoid-like function as the activation of the very last layer.

In-depth reviews of neural networks can be found in references \cite{DBLP:journals/nature/LeCunBH15,Goodfellow-et-al-2016}.

\subsection{Data and model optimisation}
\label{sec:modeloptimisation}


The data available for training and testing the model are Monte Carlo (MC) simulated events, created using the Geant4 package \cite{agostinelli2003geant4} interfaced to the DAMPE software \cite{wang2017offline} to emulate the detector response to the various cosmic rays species in orbit. The training data was prepared with a set of cleaning cuts to replicate the analysis chain applied on real data: this includes shower containment and fiducial volume selection criteria, as well as cuts to remove gamma rays, Helium ions, and heavier nuclei~\cite{Ambrosi:2017wek}. In this way, the only remaining background for electron identification are protons. The cut flow is completed by selecting only events with $\zeta < 100$, where $\zeta$ is the classical classifier (equation \ref{eq:Xtrl}). This conservative criterion eliminates events with obvious proton-like topology while retaining 99.95\% of electrons. The motivation behind this final cut is to get rid of easily identified events, such that the neural network can focus on more complex ones. This preselection yields a neural network with higher discrimination power.

Out of this cleaned-up data, we built a training set of 140'000 events, balanced 50/50 between electrons and protons. A further 120'000 events are kept for model validation.
The data is normalised by dividing each input variable by its maximum value through the dataset, therefore scaling them to the $[0;1]$ range.

\begin{figure}
    \centering
    \includegraphics[width=.32\linewidth]{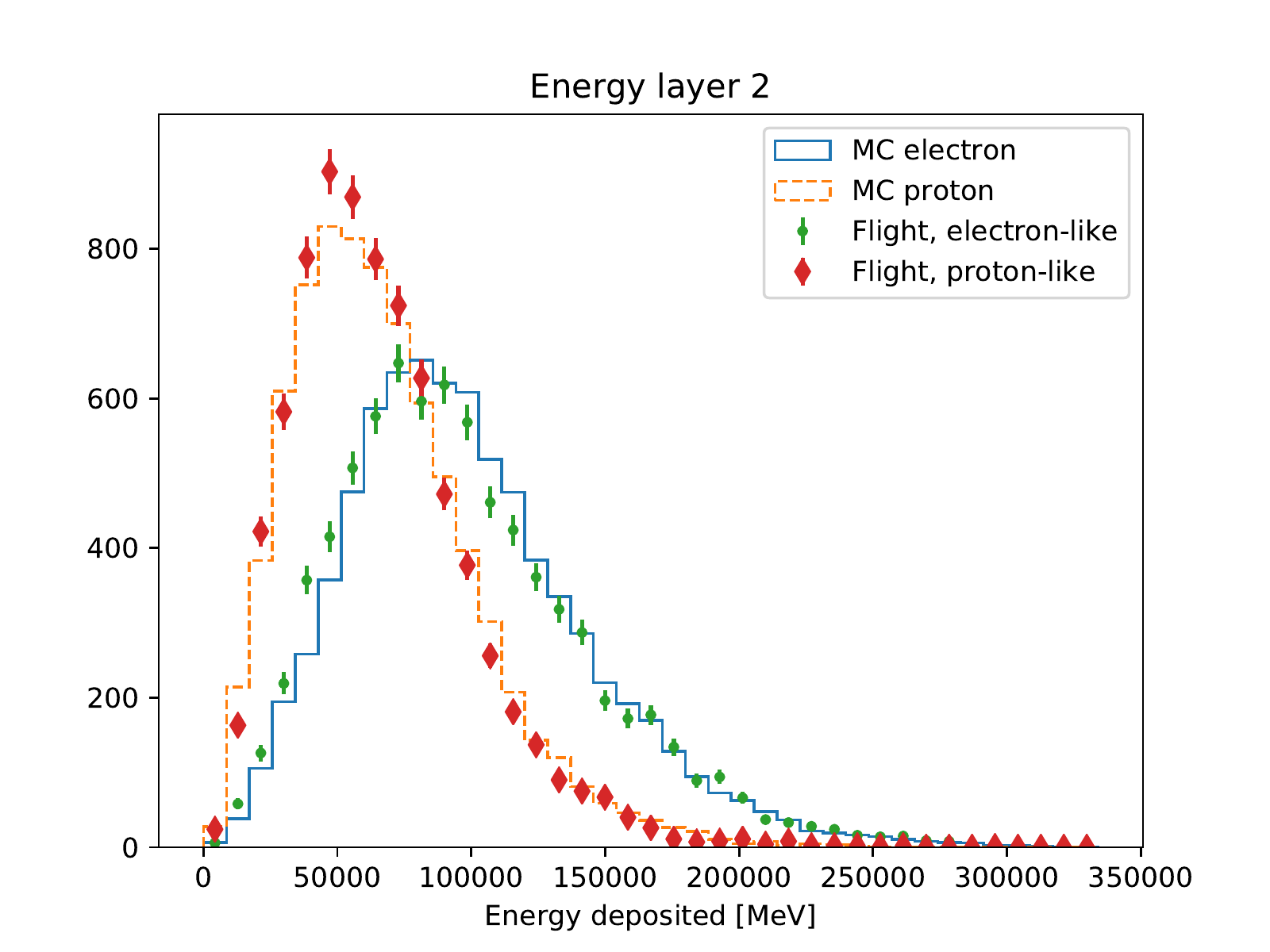}
    \includegraphics[width=.32\linewidth]{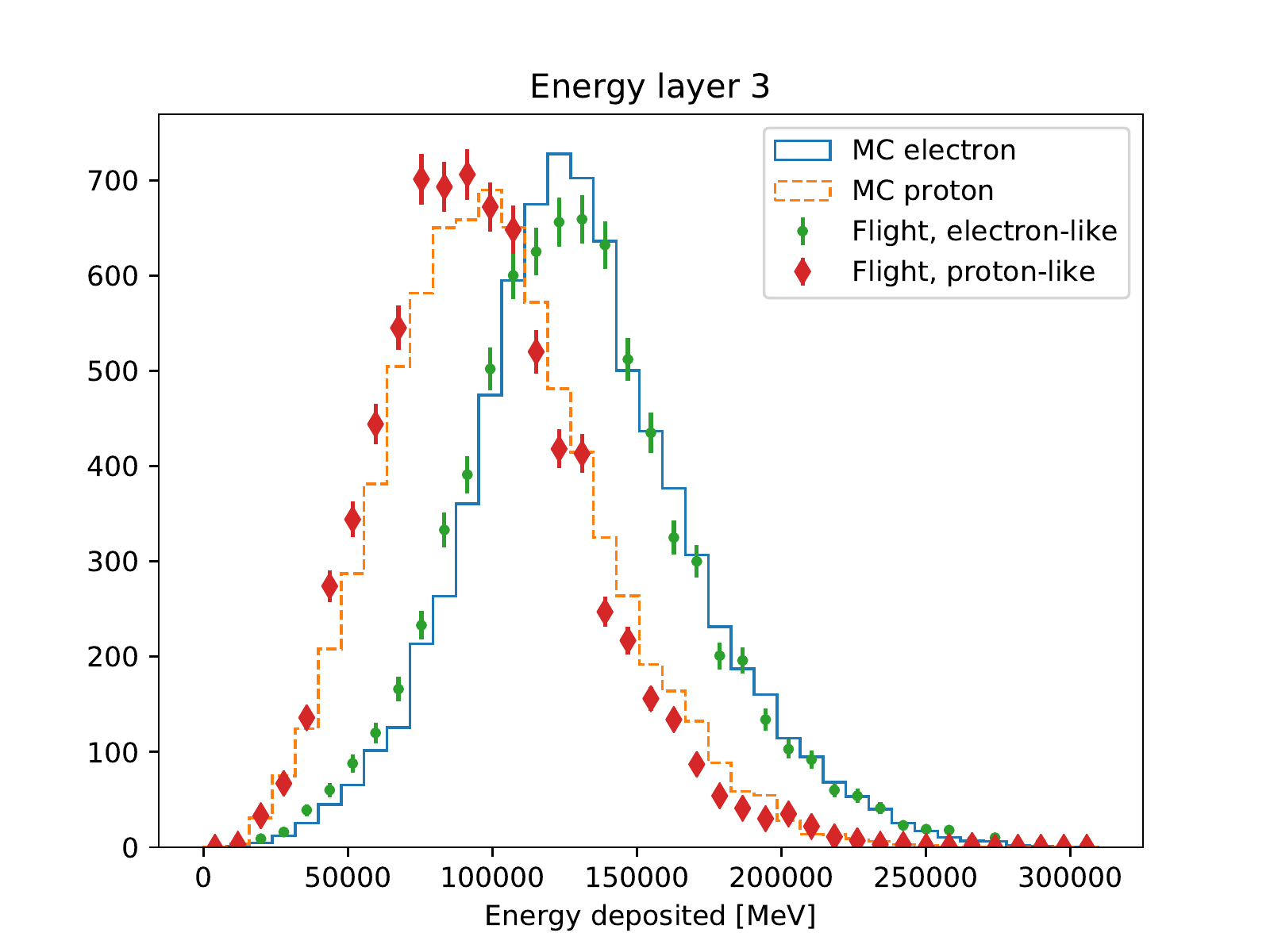}
    \includegraphics[width=.32\linewidth]{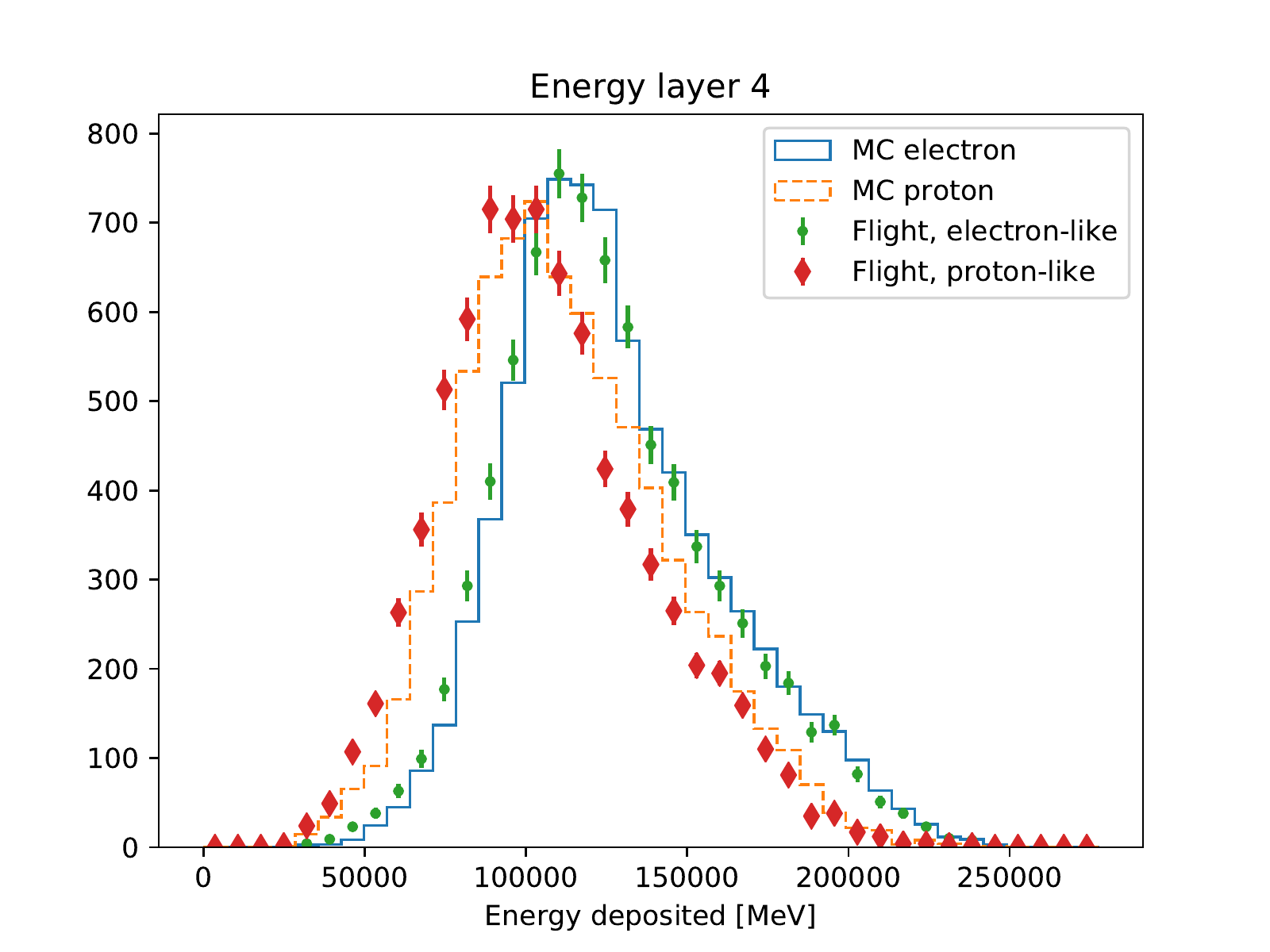}
    \includegraphics[width=.32\linewidth]{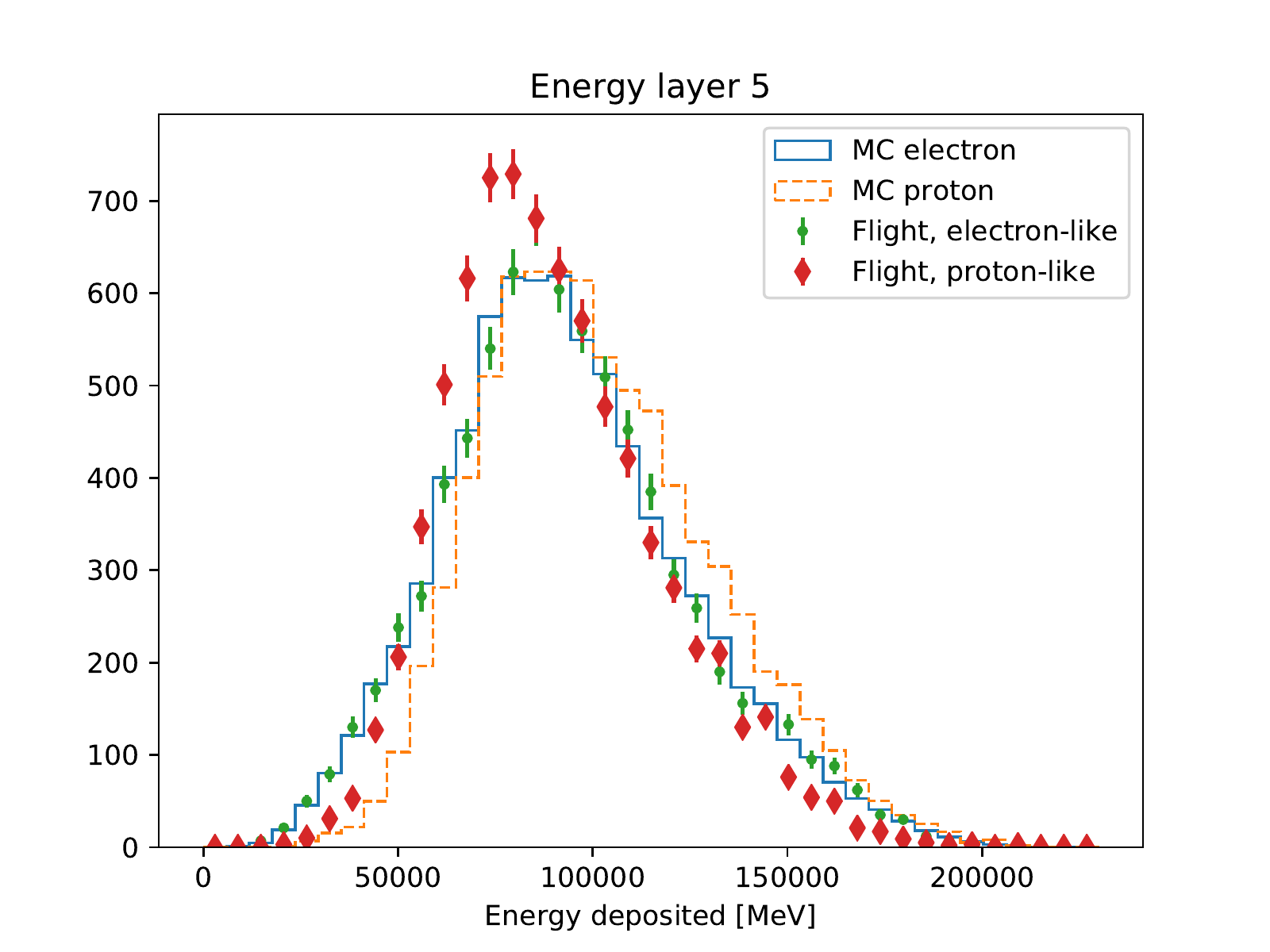}
    \includegraphics[width=.32\linewidth]{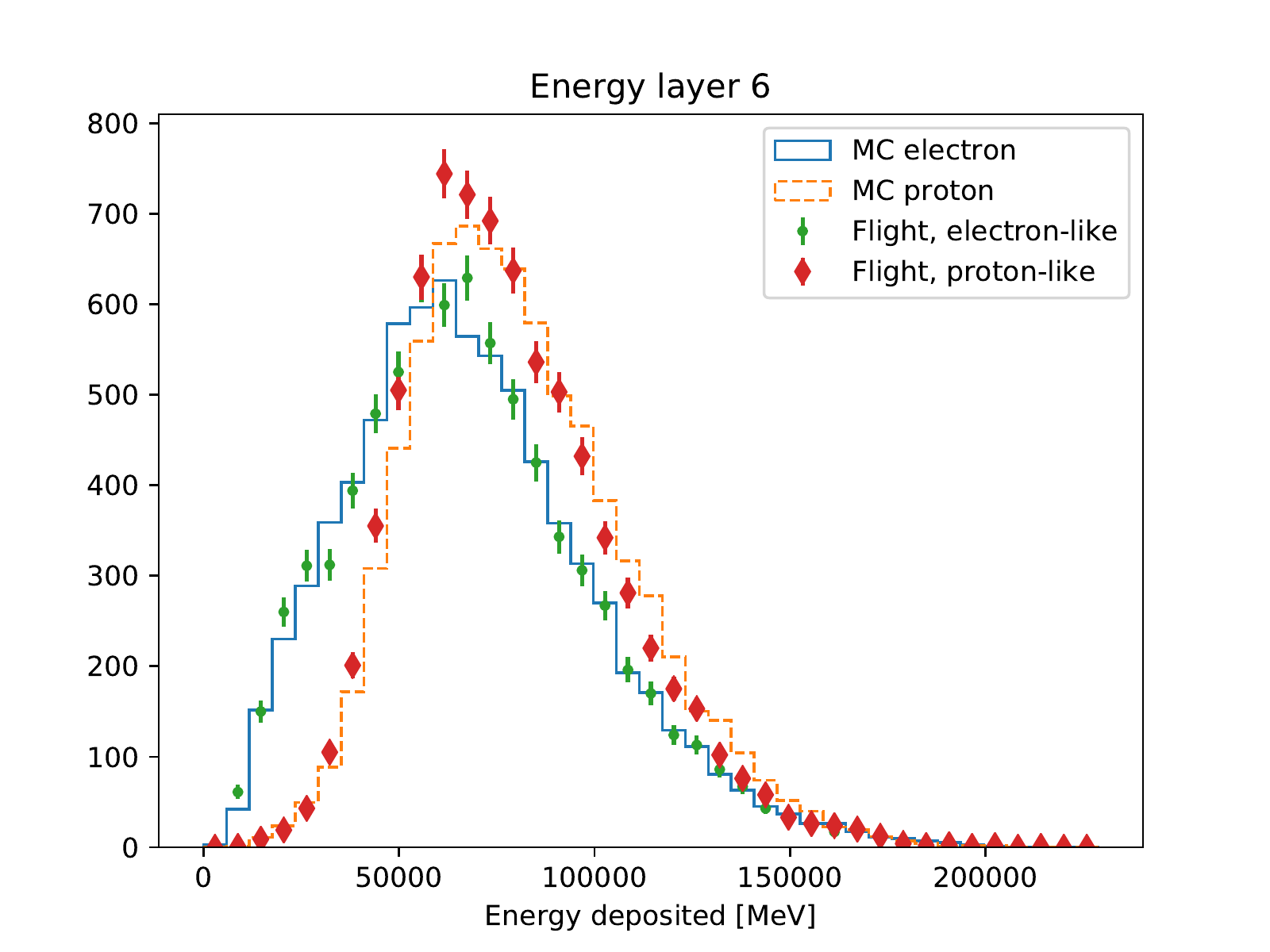}
    \includegraphics[width=.32\linewidth]{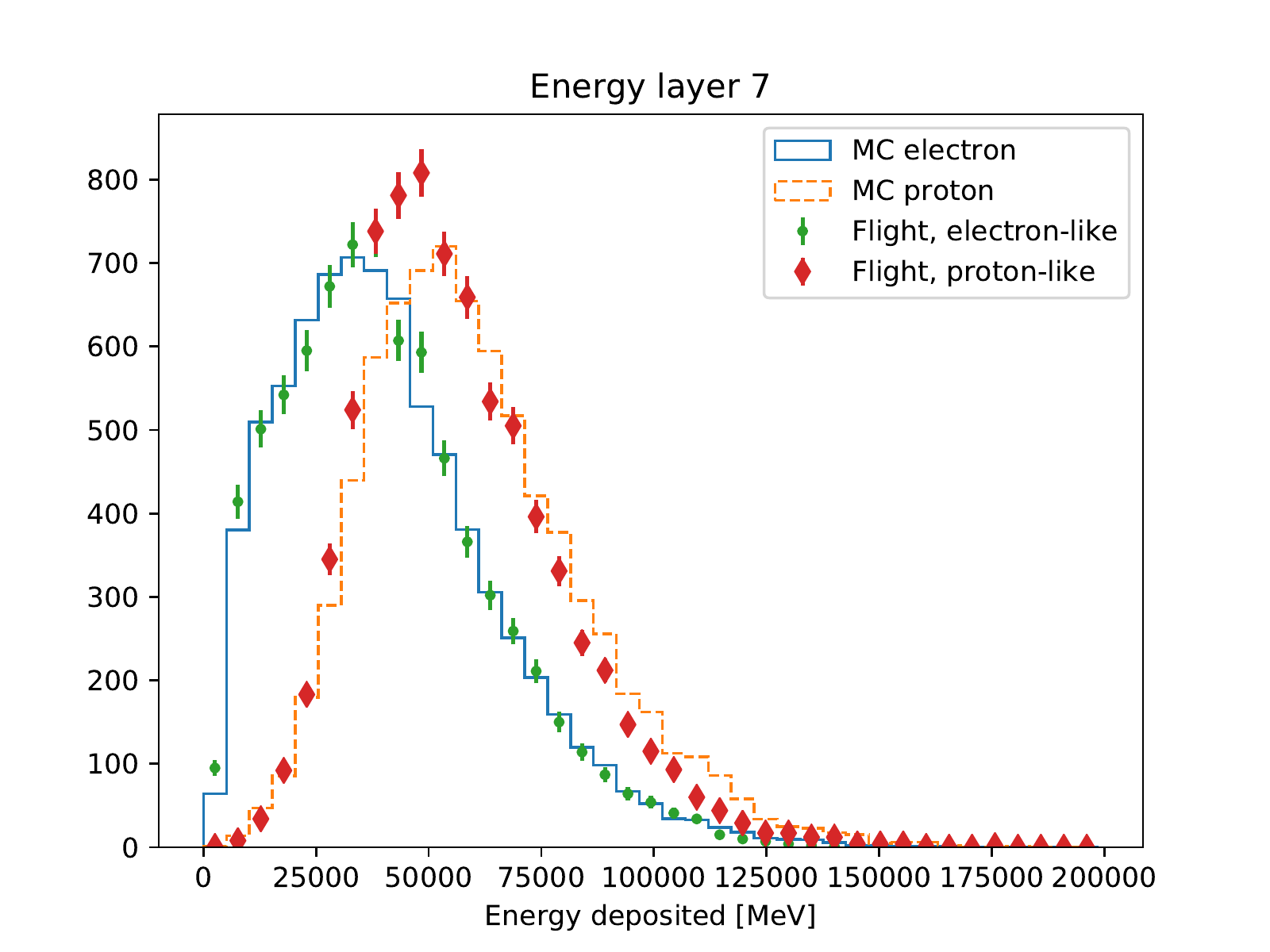}
    \includegraphics[width=.32\linewidth]{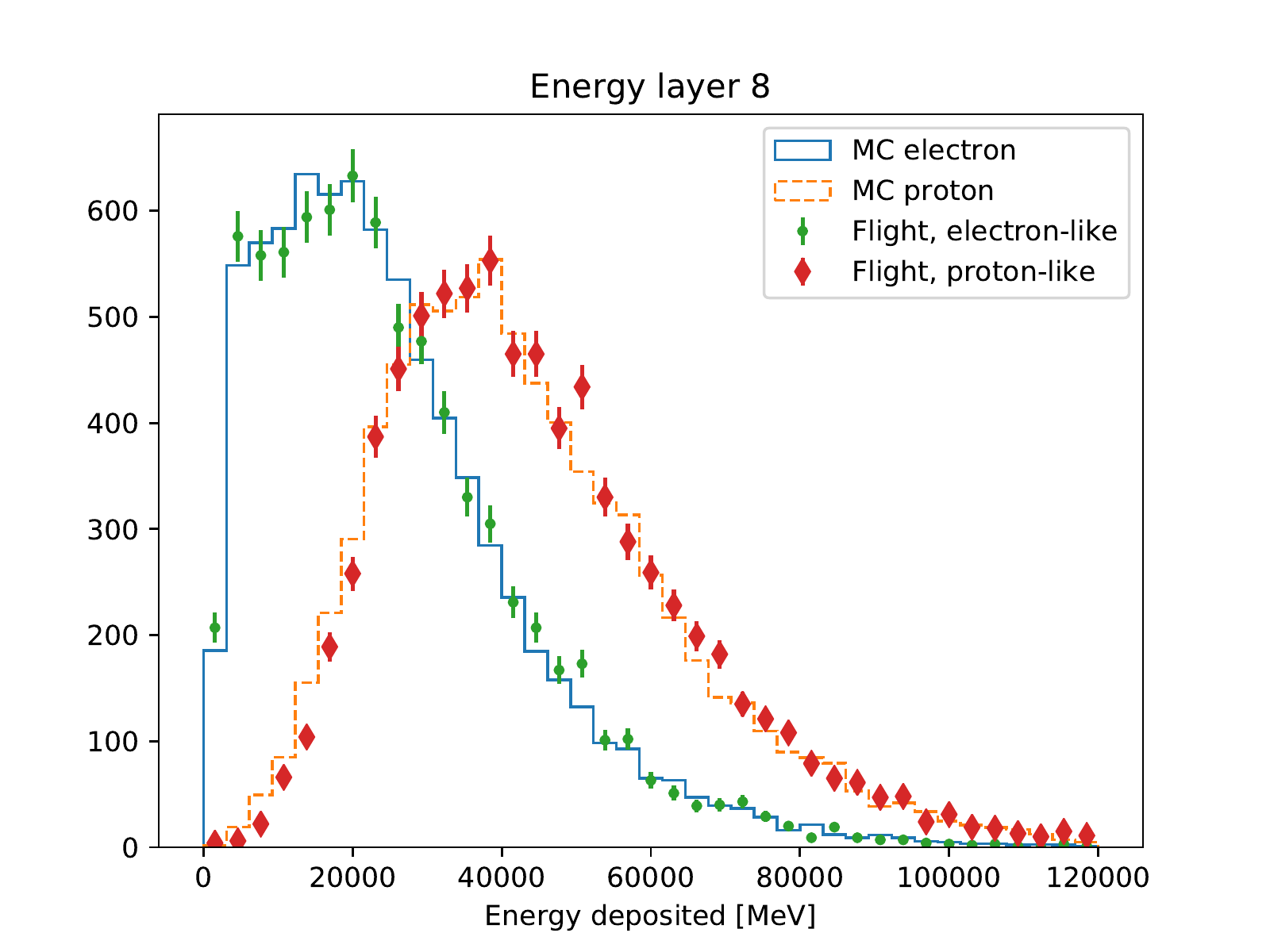}
    \includegraphics[width=.32\linewidth]{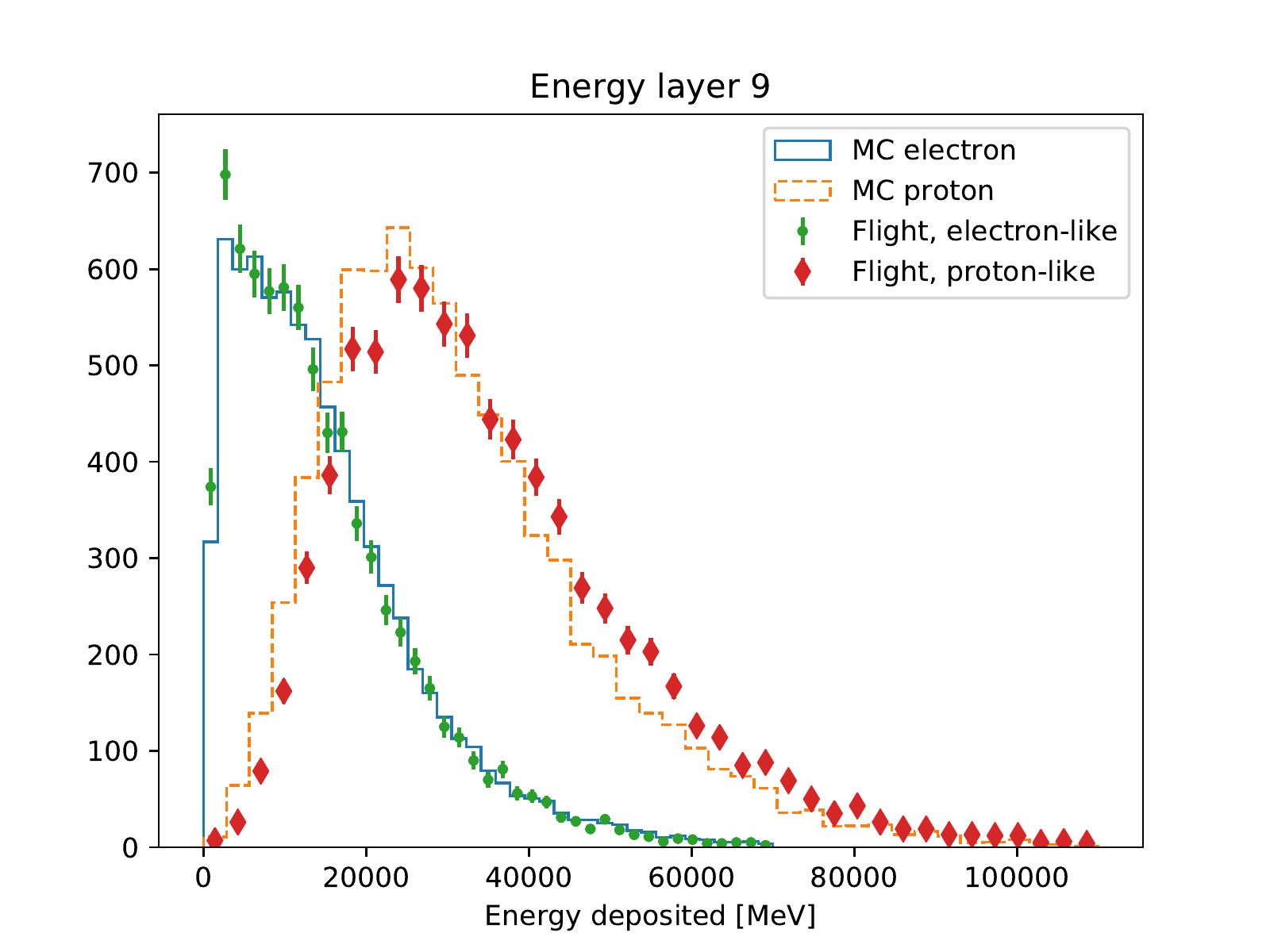}
    \includegraphics[width=.32\linewidth]{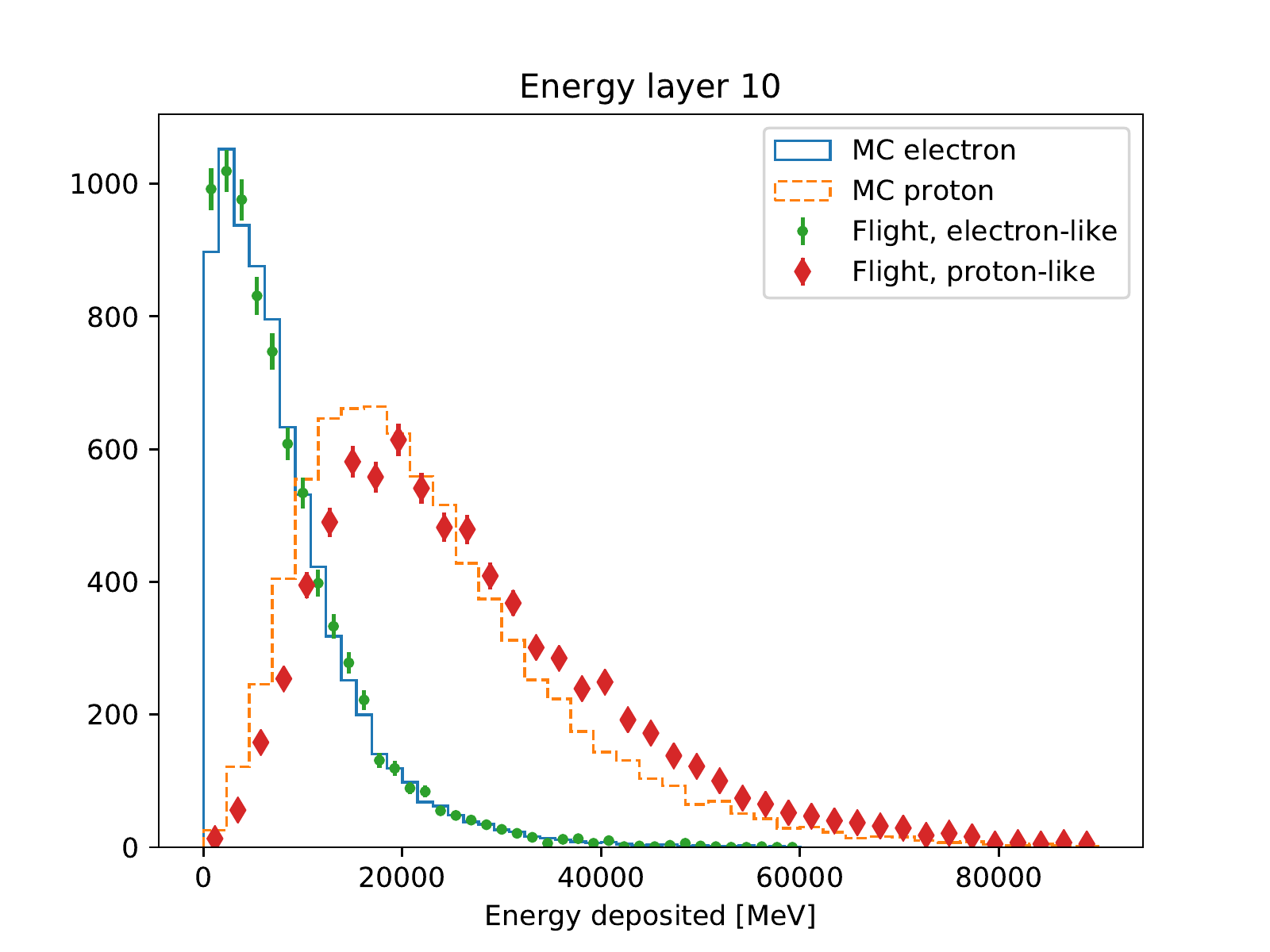}
    \includegraphics[width=.32\linewidth]{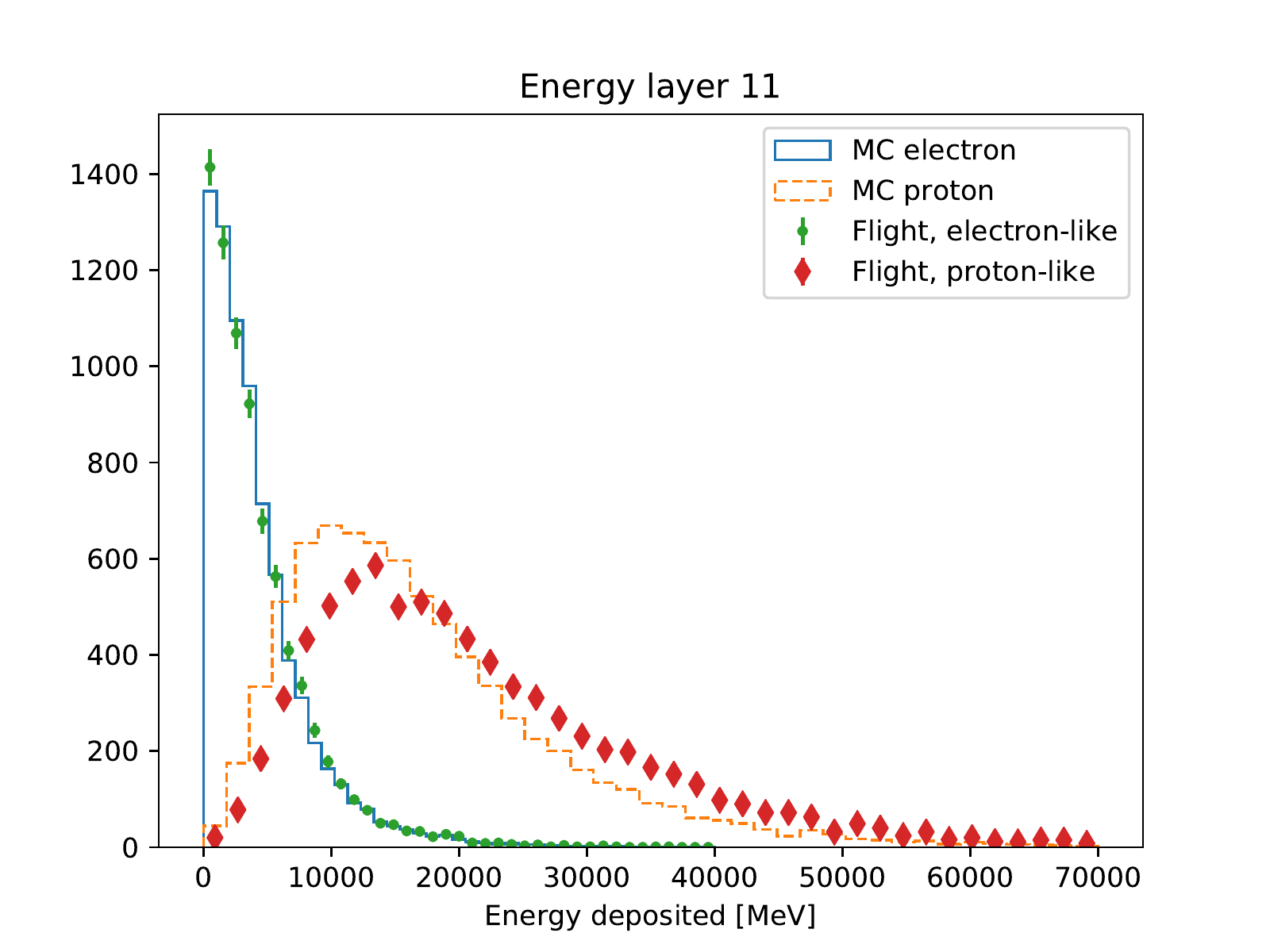}
    \includegraphics[width=.32\linewidth]{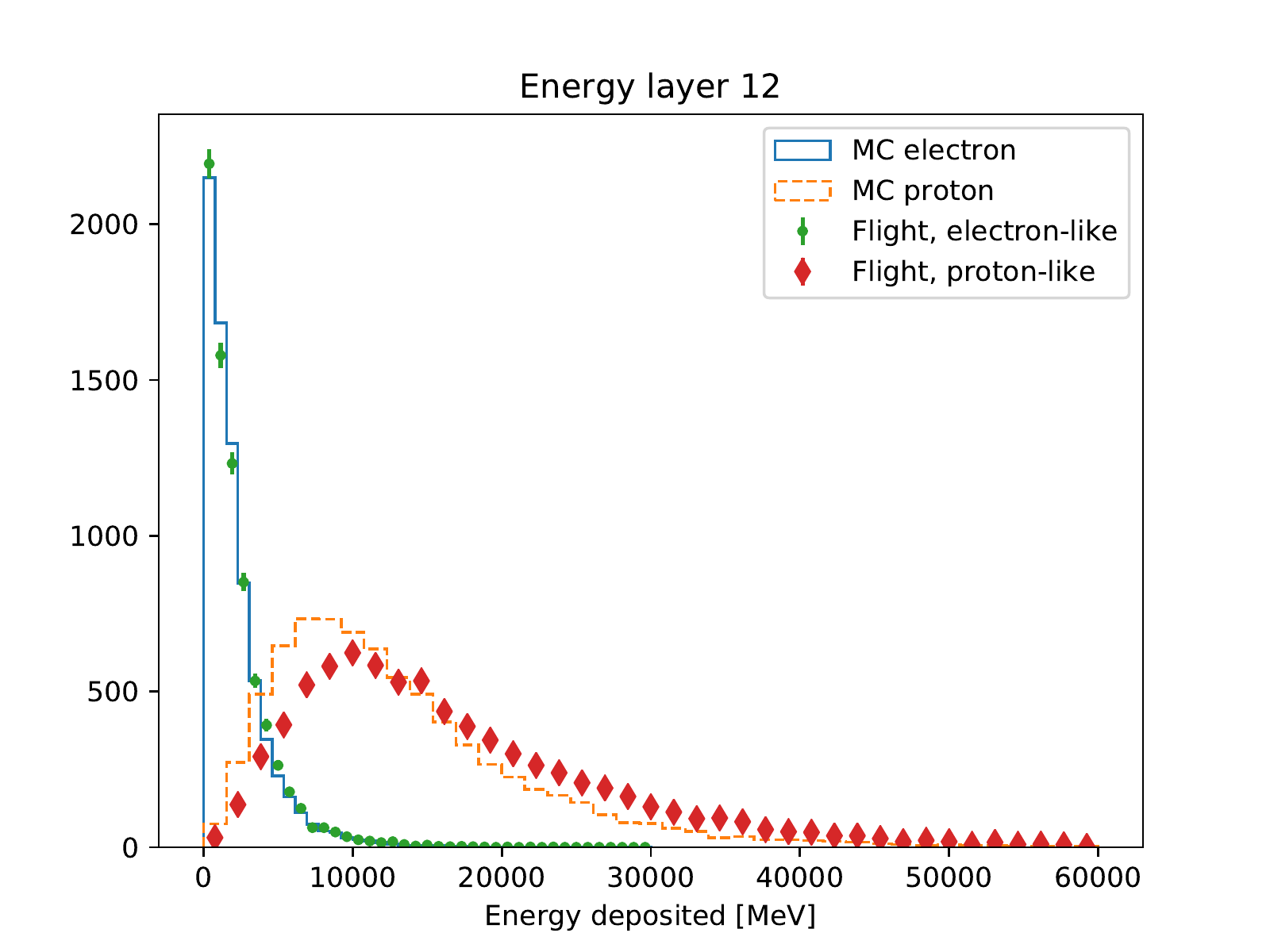}
    \includegraphics[width=.32\linewidth]{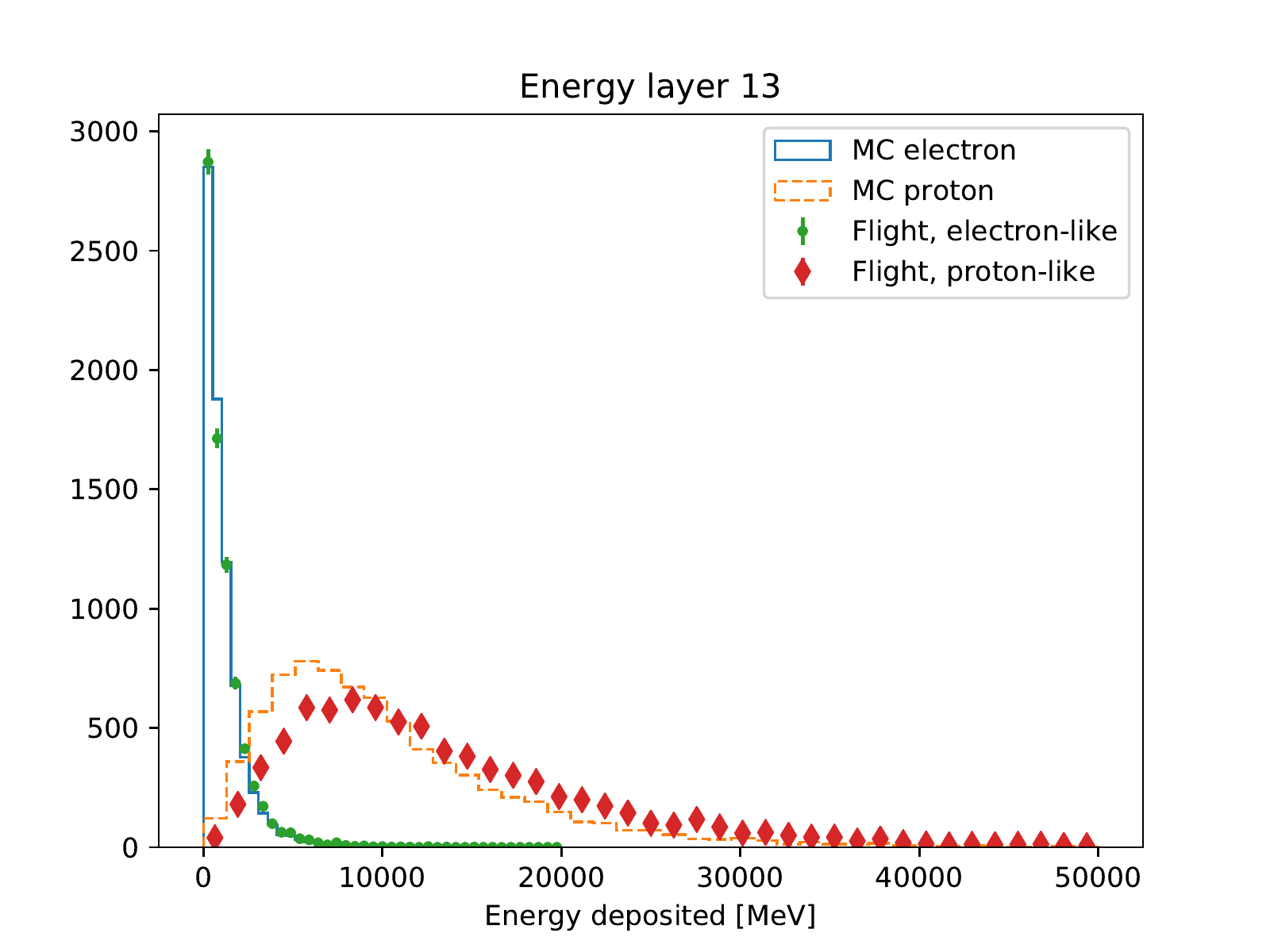}
    \caption{Distributions of selected DAMPE observables used as input of the neural network, comparing MC with real data (flight). Events are selected in the energy range 500 GeV to 1 TeV. Electron-like flight events are selected with $\zeta < 12$ and proton-like with $\zeta > 20$, where $\zeta$ is the classical discriminator (section \ref{sec:electronidentification}). The same cuts are applied on both MC and data. Error bars are purely statistical.}
    \label{fig:inputVariables1}
\end{figure}

\begin{figure}
    \centering
    \includegraphics[width=.32\linewidth]{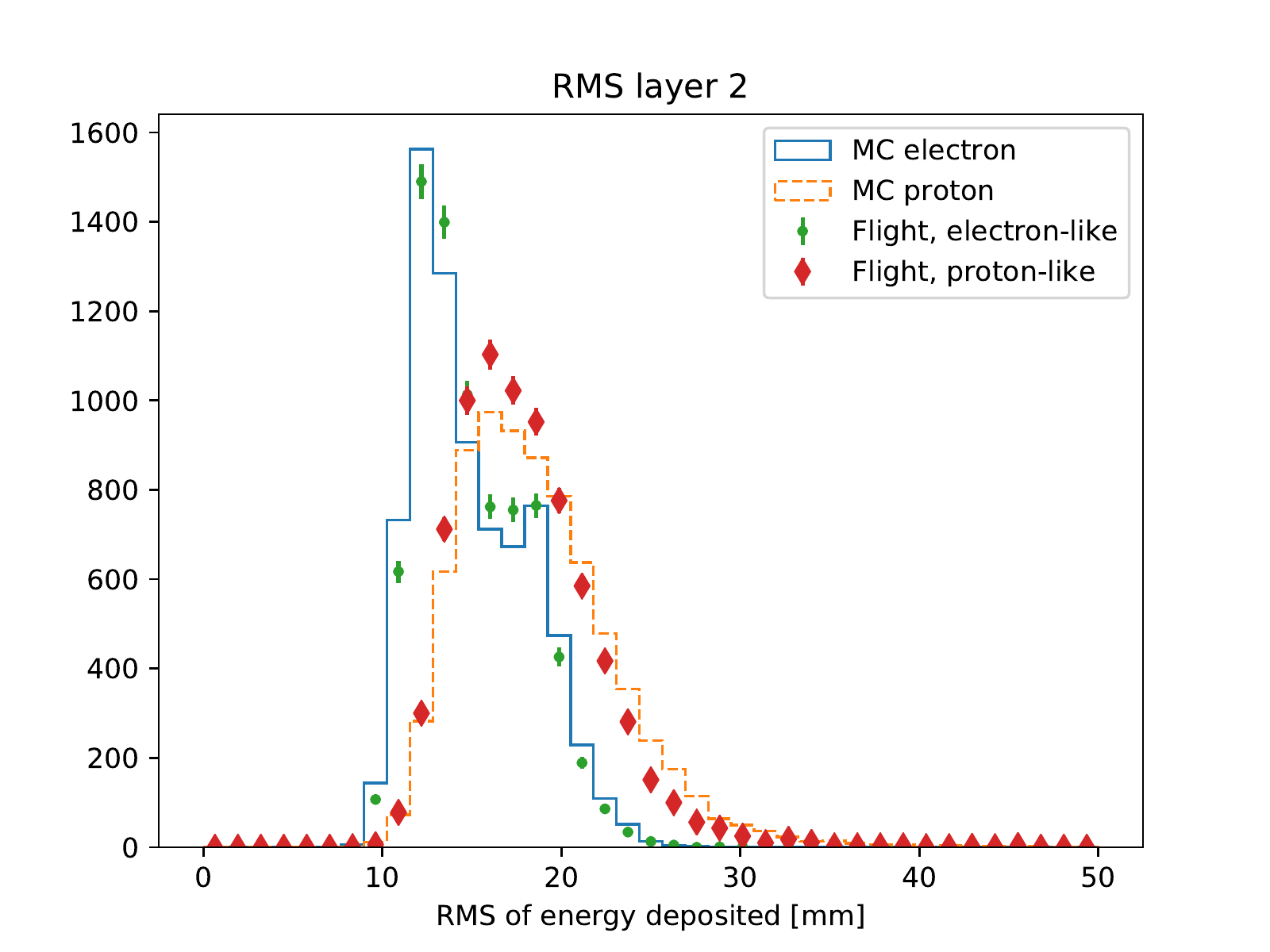}
    \includegraphics[width=.32\linewidth]{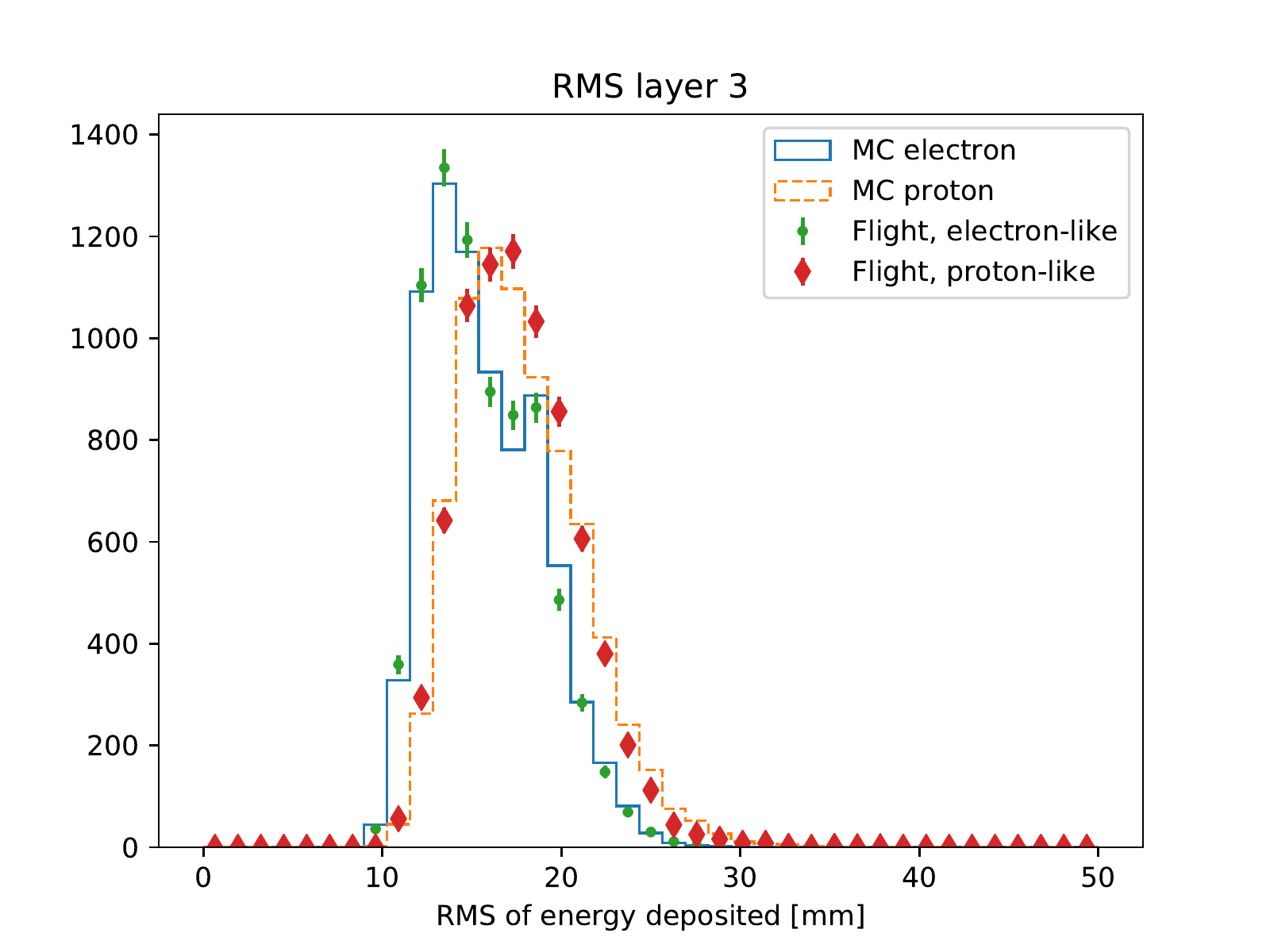}
    \includegraphics[width=.32\linewidth]{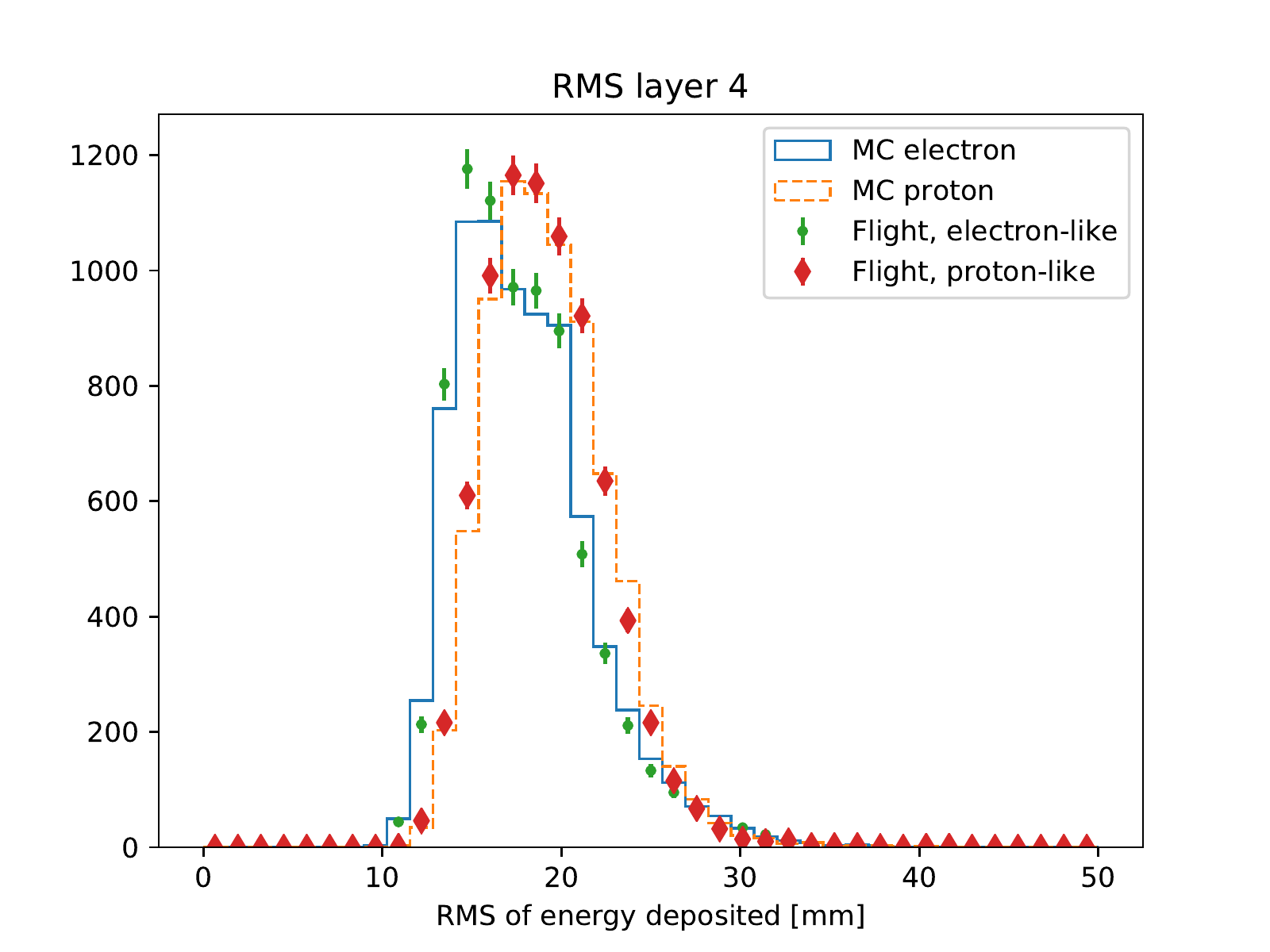}
    \includegraphics[width=.32\linewidth]{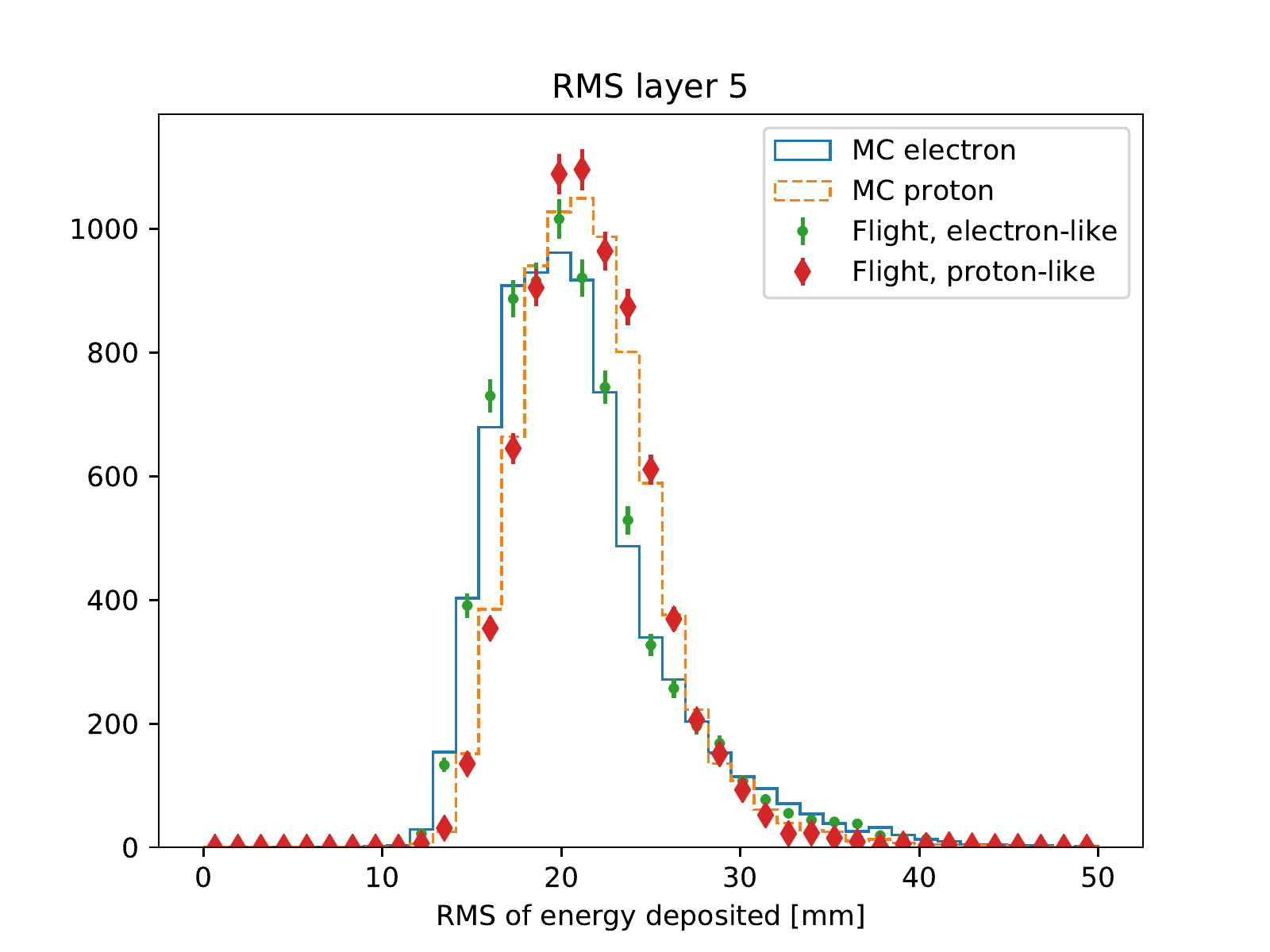}
    \includegraphics[width=.32\linewidth]{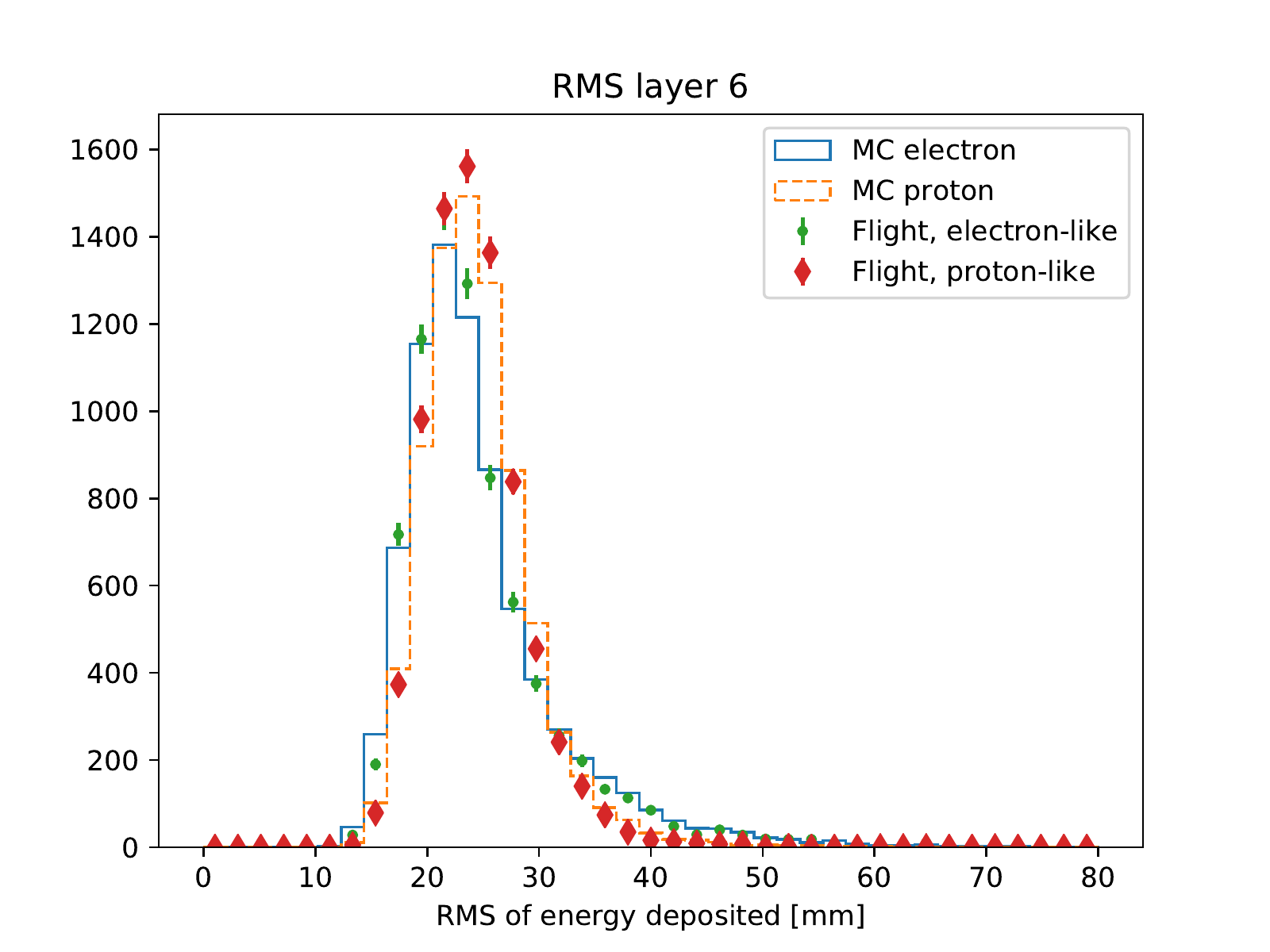}
    \includegraphics[width=.32\linewidth]{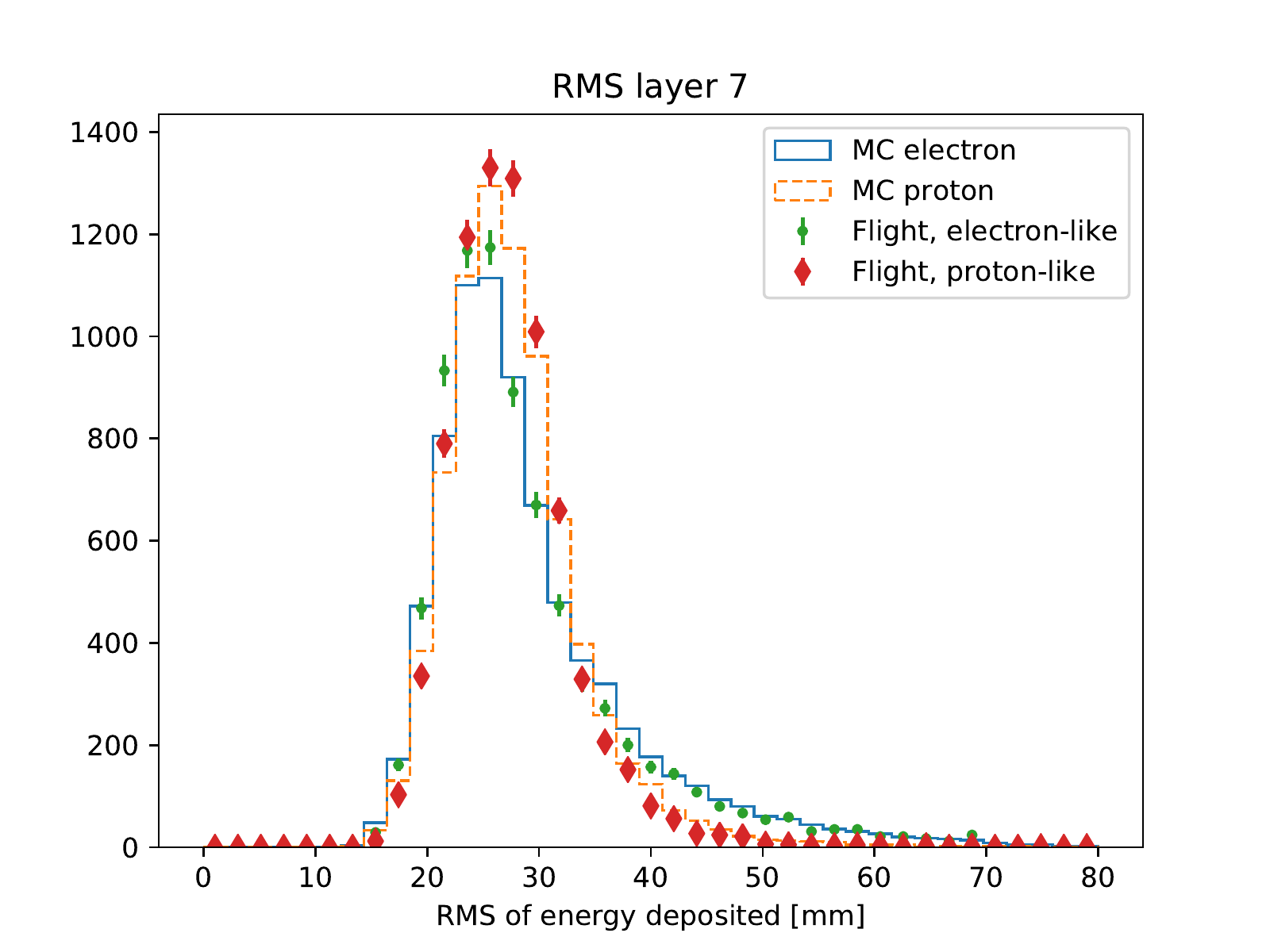}
    \includegraphics[width=.32\linewidth]{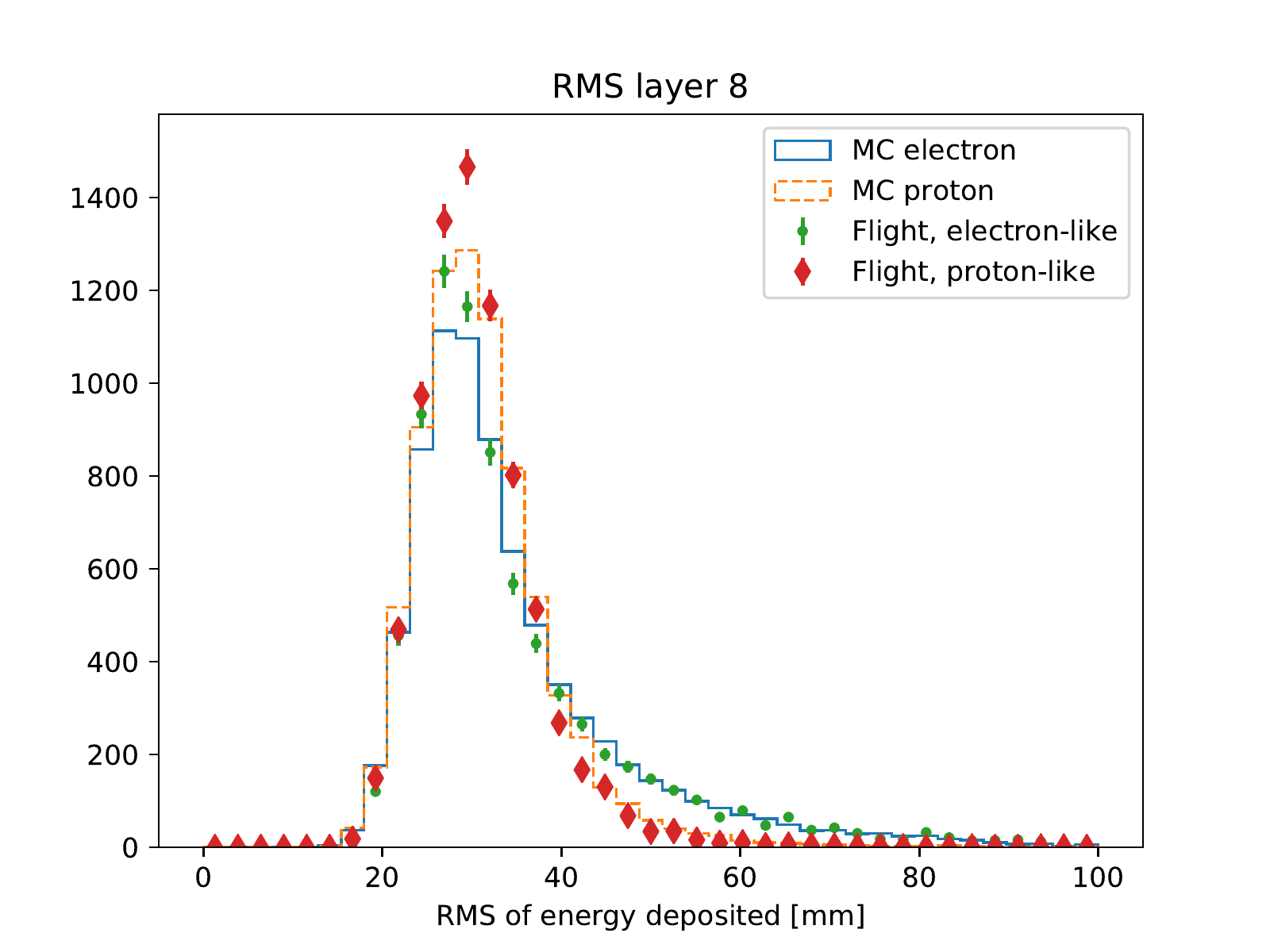}
    \includegraphics[width=.32\linewidth]{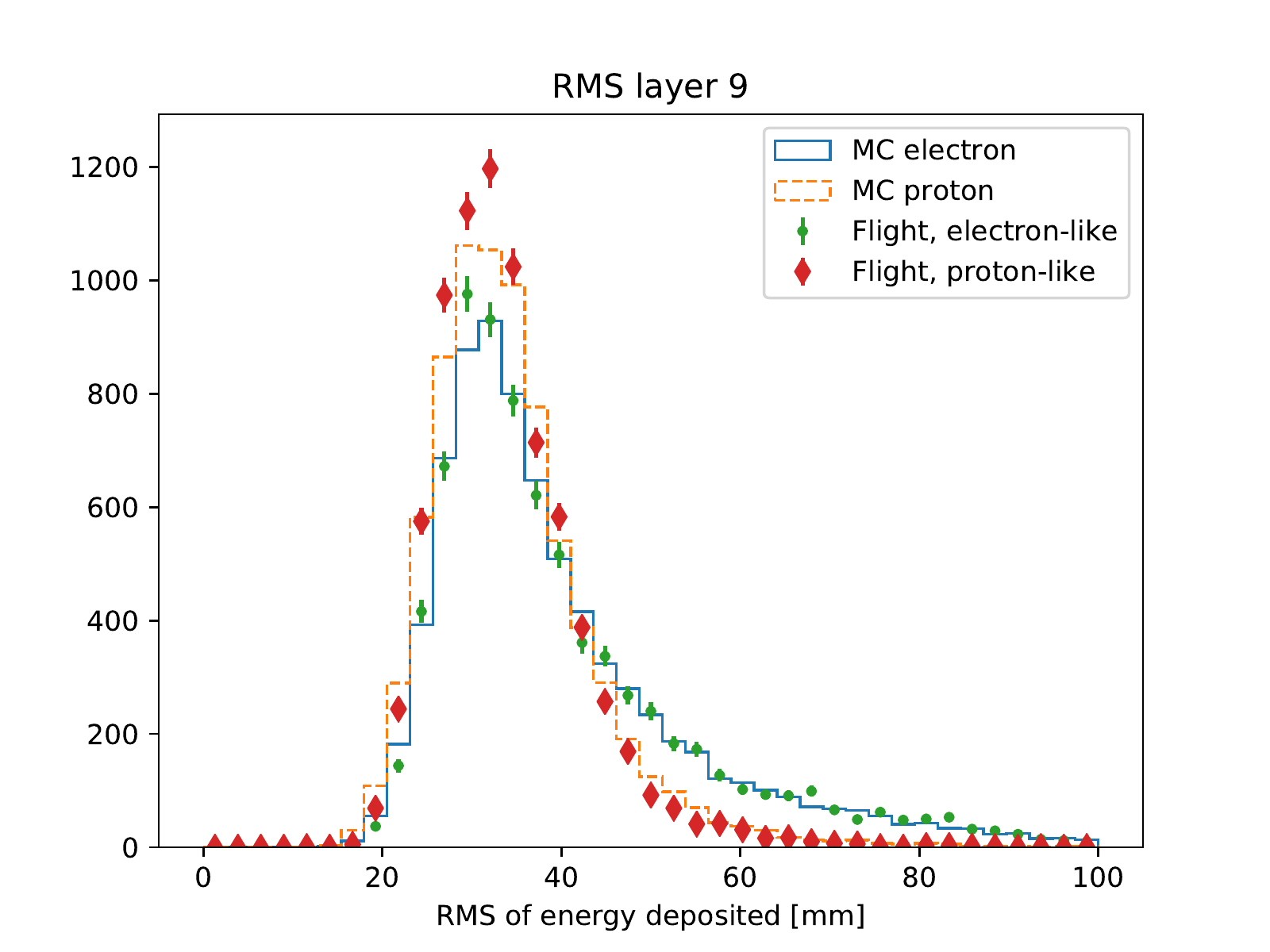}
    \includegraphics[width=.32\linewidth]{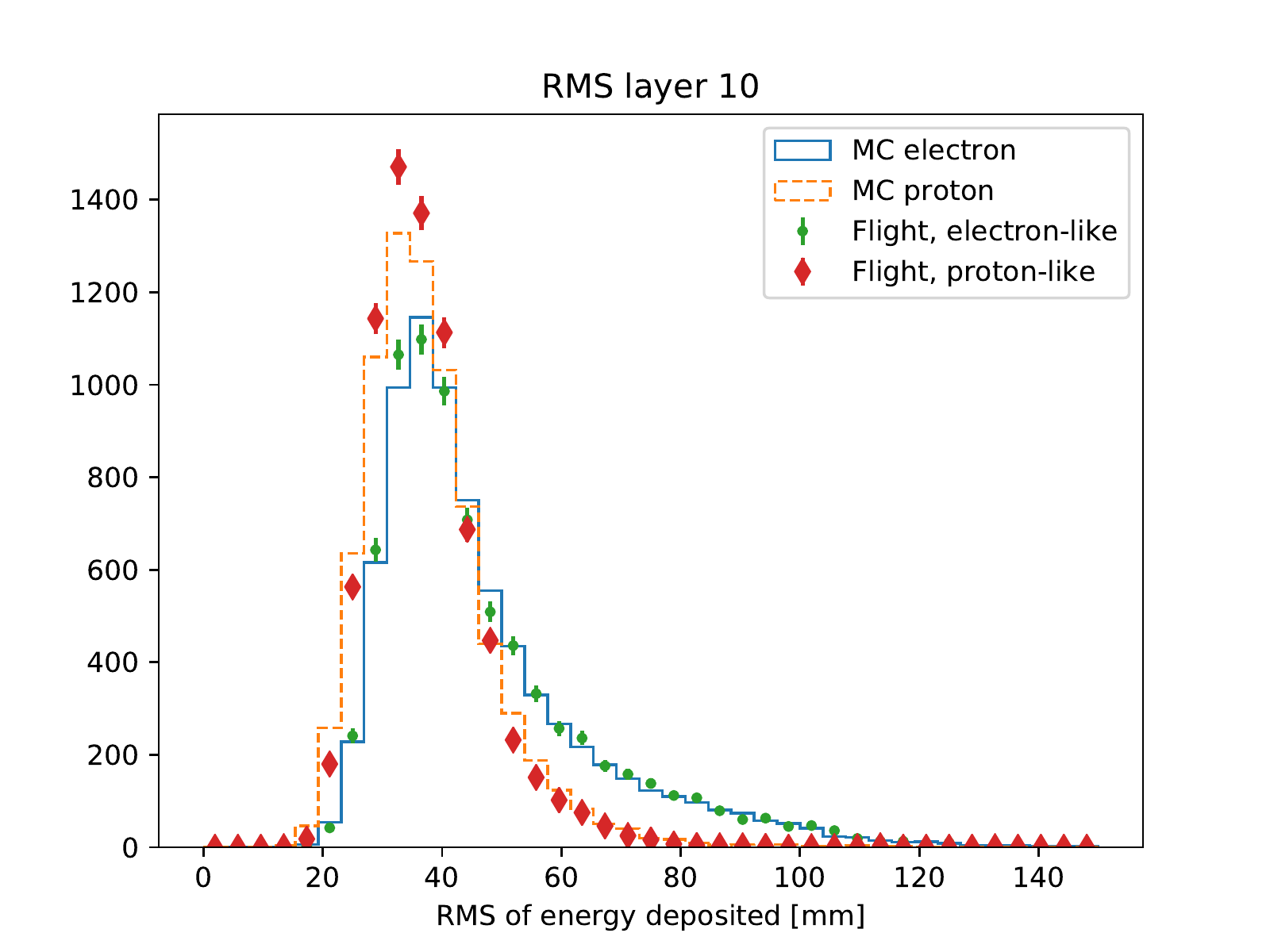}
    \includegraphics[width=.32\linewidth]{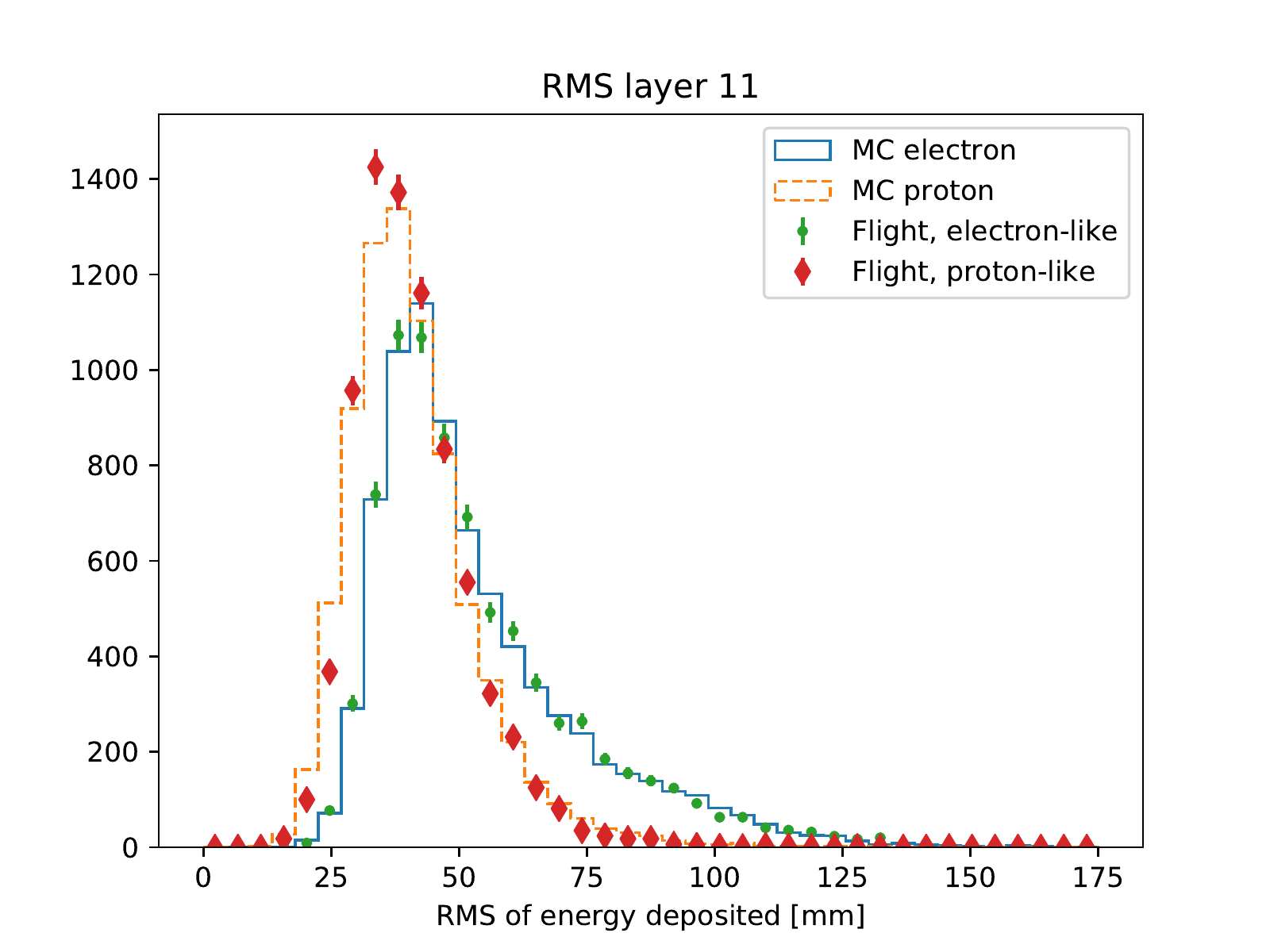}
    \includegraphics[width=.32\linewidth]{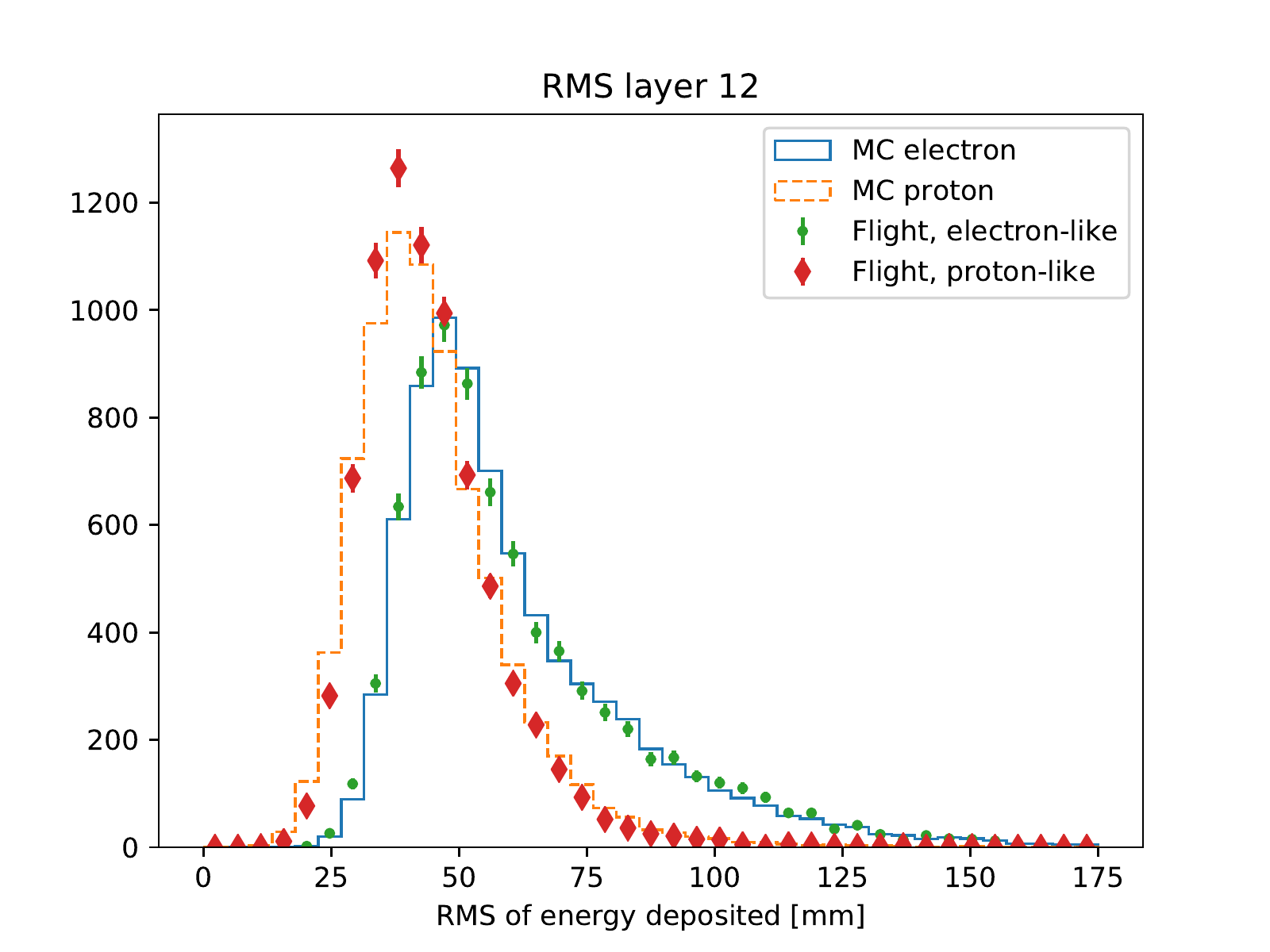}
    \includegraphics[width=.32\linewidth]{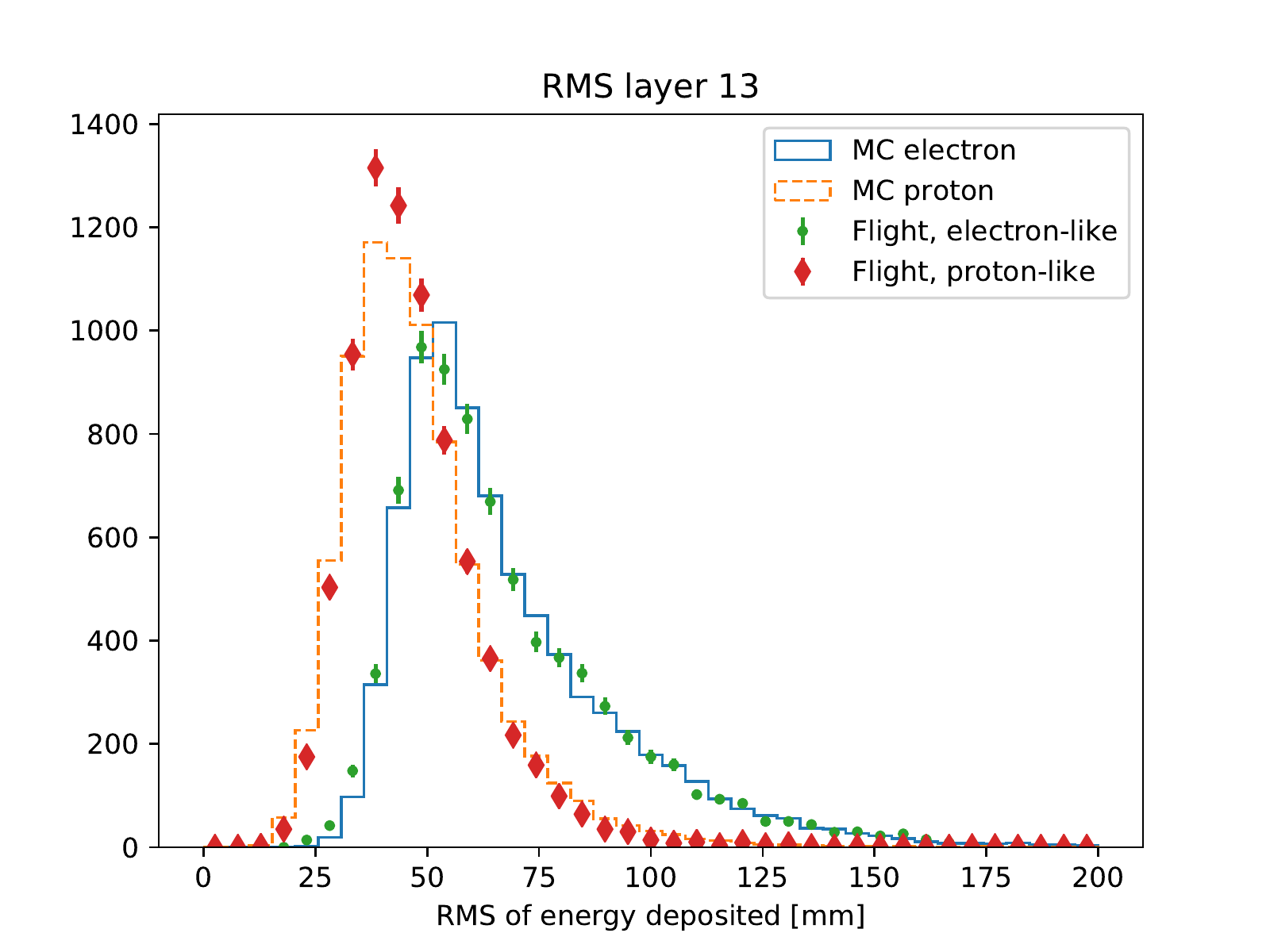}
    \includegraphics[width=.32\linewidth]{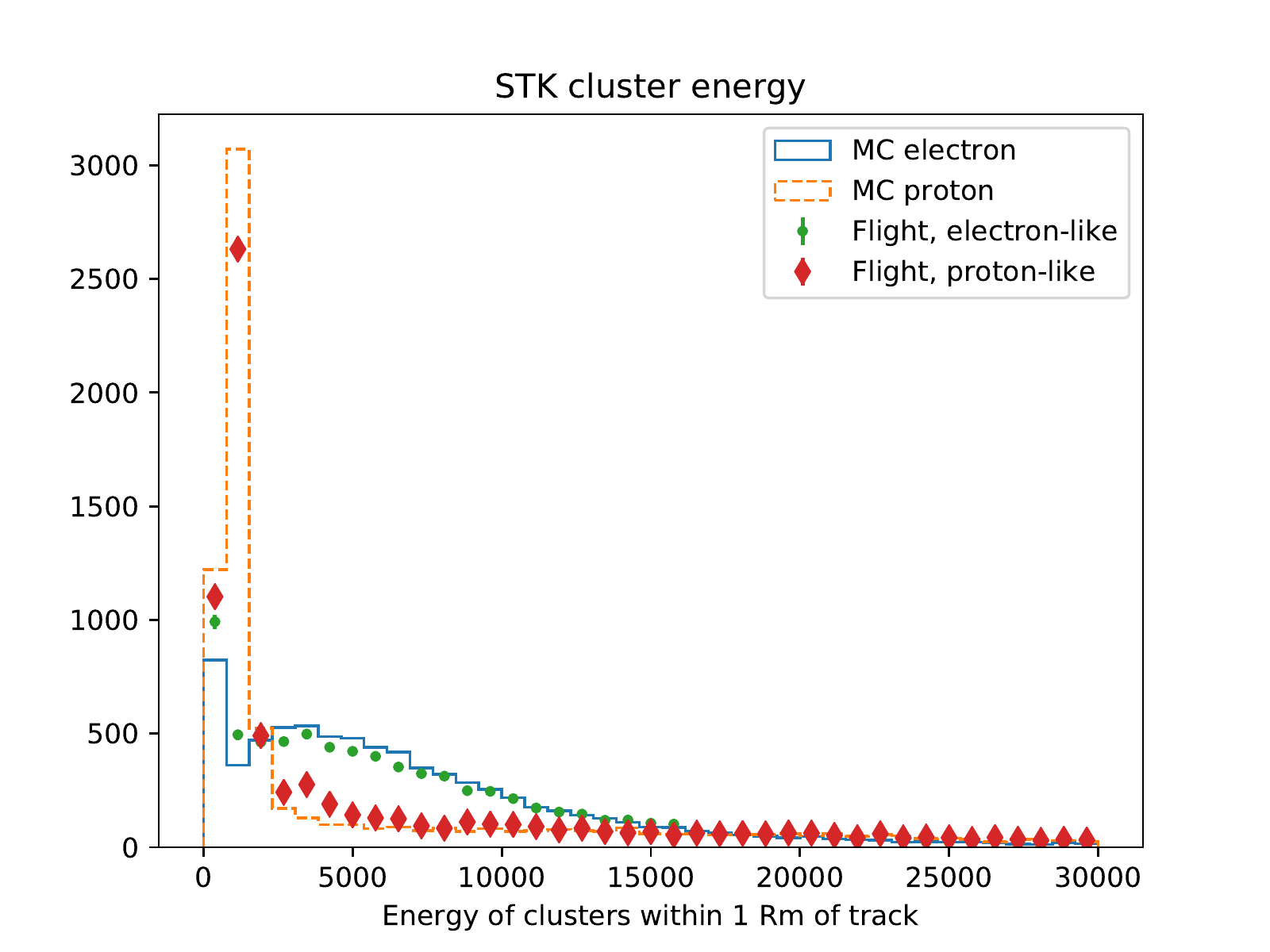}
    \caption{Distribution of selected DAMPE observables used as input of the neural network, comparing MC with real data (flight). Events are selected in the energy range 500 GeV to 1 TeV. Electron-like flight events are selected with $\zeta < 12$ and proton-like with $\zeta > 20$, where $\zeta$ is the classical discriminator (section \ref{sec:electronidentification}). The same cuts are applied on both MC and data. Error bars are purely statistical.}
    \label{fig:inputVariables2}
\end{figure}

We selected the input variables based on several criteria. First, the features had to be descriptive enough of the shower topology to provide as much information as possible to the neural network. Following published results on deep learning for particle physics \cite{Baldi:2014kfa}, we opted mostly for low-level variables to maximise said information, but we added in high-level variables as well such as the $\zeta$ classifier which we know provides an already powerful proton rejection. The idea being to give the network our best guess as expert, and then the low-level ingredients to improve on it. Adding $\zeta$ significantly improved the sub-TeV performances.  Finally, we required the input variables to be well simulated. We observed that depending on the set of features, the network could produce slightly different results between MC simulations and real data. We therefore followed an extensive campaign of empirical testing to find and remove the quantities that resulted in such differences, and select a set of variables that produces a satisfactory agreement between MC and data (section \ref{sec:results:val}). As a result, the features selected include the energy deposited and its RMS distribution in 12 out of 14 layers of the BGO calorimeter (excluding the top two), the reconstructed energy, the angle of the trajectory, the energy deposited in one Moliere radius of a STK track (STK cluster energy), and the classical $\zeta$ classifier. The physical motivation behind them is that they provide direct information on the shower topology and the characteristics of the event (incidence and energy). The STK variable accounts for showers that start before the calorimeter. The first two layers of BGO were excluded based on an observed offset of resulting MC proton distribution with respect to real data (section \ref{sec:results:val}). Figures \ref{fig:inputVariables1} and \ref{fig:inputVariables2} show the distributions and MC/data comparison of these variables for both particle species, excepting the reconstructed energy (power law), angle (BGO acceptance) and $\zeta$ (figure \ref{fig:XTRL}). These events were fully cleaned up by the aforementioned chain of cuts and filters. For the sake of visual comparison only, they have been separated into electron-like events with $\zeta < 12$ and proton-like events with $\zeta > 20$ (see figure \ref{fig:XTRL}). 

\begin{figure}
    \centering
    \includegraphics[width=.99\linewidth]{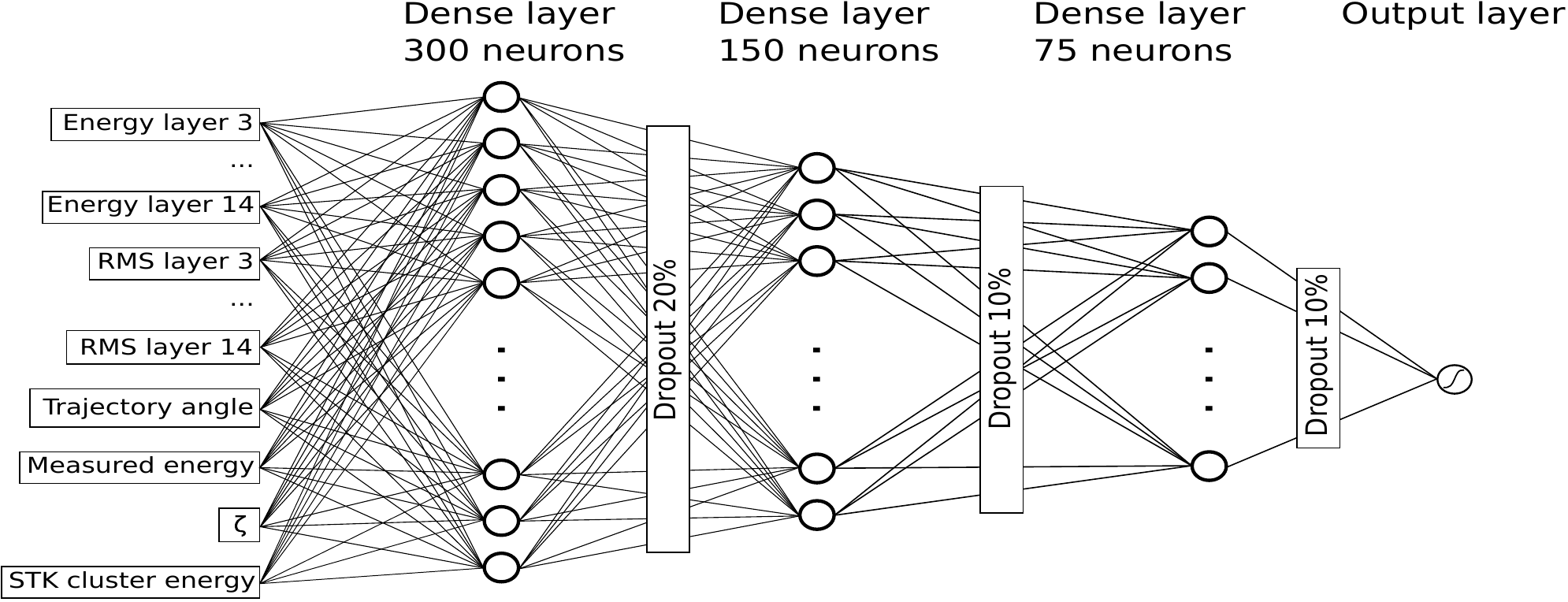}
    \caption{Schematic of the neural network model used in this work. The hidden layers use the ReLU activation function. The logistic sigmoid in the output layer is removed after training.}
    \label{fig:NNschema}
\end{figure}

Along with optimising the set of input variables we were also researching and optimising the architecture of the neural network itself (the model). Building a model indeed requires several parameter choices: the number of neurons and layers, the activation function, the optimiser and regularisers, etc. The optimisation of these hyper-parameters is somewhat of an art, and a common practice is to conduct a random gridsearch \cite{bergstra2012random}. We decided to follow this philosophy and tested hundreds of models against each other. The criterion used to select the best model was the area under the energy-dependent false positive curve (figure \ref{fig:MCperformances}). We found the optimal configuration to be a model consisting of 4 layers with 300, 150, 75 and 1 neuron, respectively, regularised with a 10-20\% dropout (technique consisting of randomly turning off neurons during the training) \cite{srivastava2014dropout}. We noted that a higher dropout resulted in a slightly higher false positive rate, of the order of 10\% more for 50\% dropout. The hidden layers use the Rectified Linear Unit (ReLU) \cite{nair2010rectified} activation function, and the output layer uses the logistic sigmoid function to map the network output to the $\left[0;1\right]$ range as is common to binary classification problems. Finally the model is optimised using the Adam gradient descent algorithm \cite{kingma2014adam} against the cross-entropy metric. The architecture is represented in figure \ref{fig:NNschema} and \mbox{table \ref{table:modelSummary}.} The network is trained for 50 epochs, longer training did not result in any perceptible change in performances and data matching (see further sections).

\begin{table}
\begin{center}
\begin{tabular}{lll}
\hline 
Layer         & Neurons & Parameters \\
\hhline{|=|=|=|}
Dense         & 300           & 9000             \\
\hline 
Dropout, 20\% &               & 0                \\
\hline 
Dense         & 150           & 45150            \\
\hline 
Dropout, 10\% &               & 0                \\
\hline 
Dense         & 75            & 11325            \\
\hline 
Dropout, 10\% &               & 0                \\
\hline 
Dense         & 1             & 76              \\
\hline  \\
\hline 
Total parameters: & 65,551 & \\
Trainable parameters: & 65,551  & \\
Non-trainable parameters: & 0 & \\
\hline 
\end{tabular}
\caption{ Layer-per-layer summary of the neural network used.}
\label{table:modelSummary}
\end{center}
\end{table}

The extensive training campaigns were conducted on the Baobab computer cluster of the University of Geneva, using Nvidia Titan X GPUs. On the software side, we used Nvidia cuDNN \cite{chetlur2014cudnn}, Keras \cite{chollet2015keras} with Theano \cite{theano} as a backend, and Scikit-Learn \cite{scikit-learn}. Google's Tensorflow \cite{tensorflow2015-whitepaper} was considered as well but internal benchmarks with our models and data showed no gain in performances, for a longer computing time.



\begin{figure}
    \centering
    \includegraphics[width=.49\linewidth]{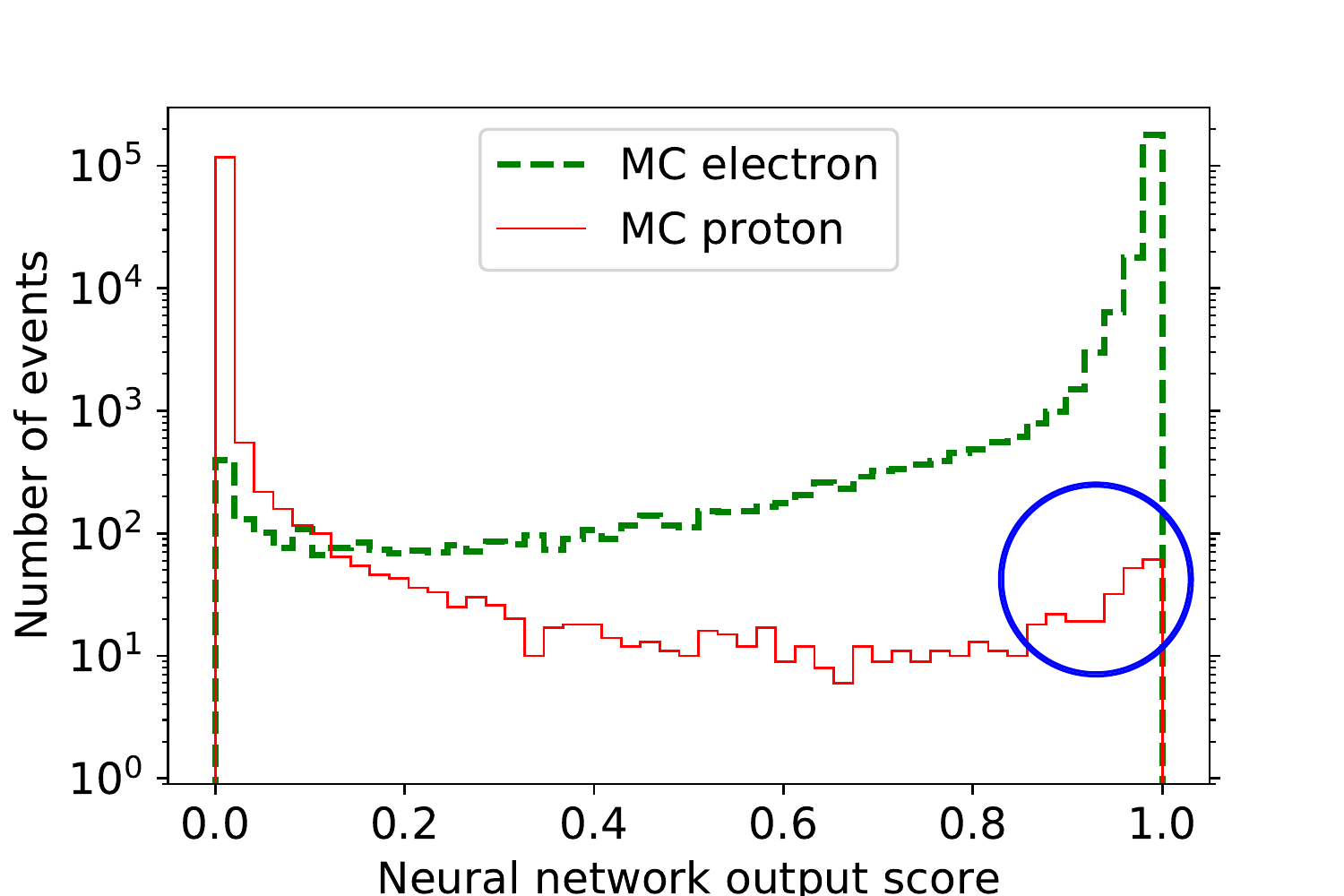}%
    \hfill
    \includegraphics[width=.49\linewidth]{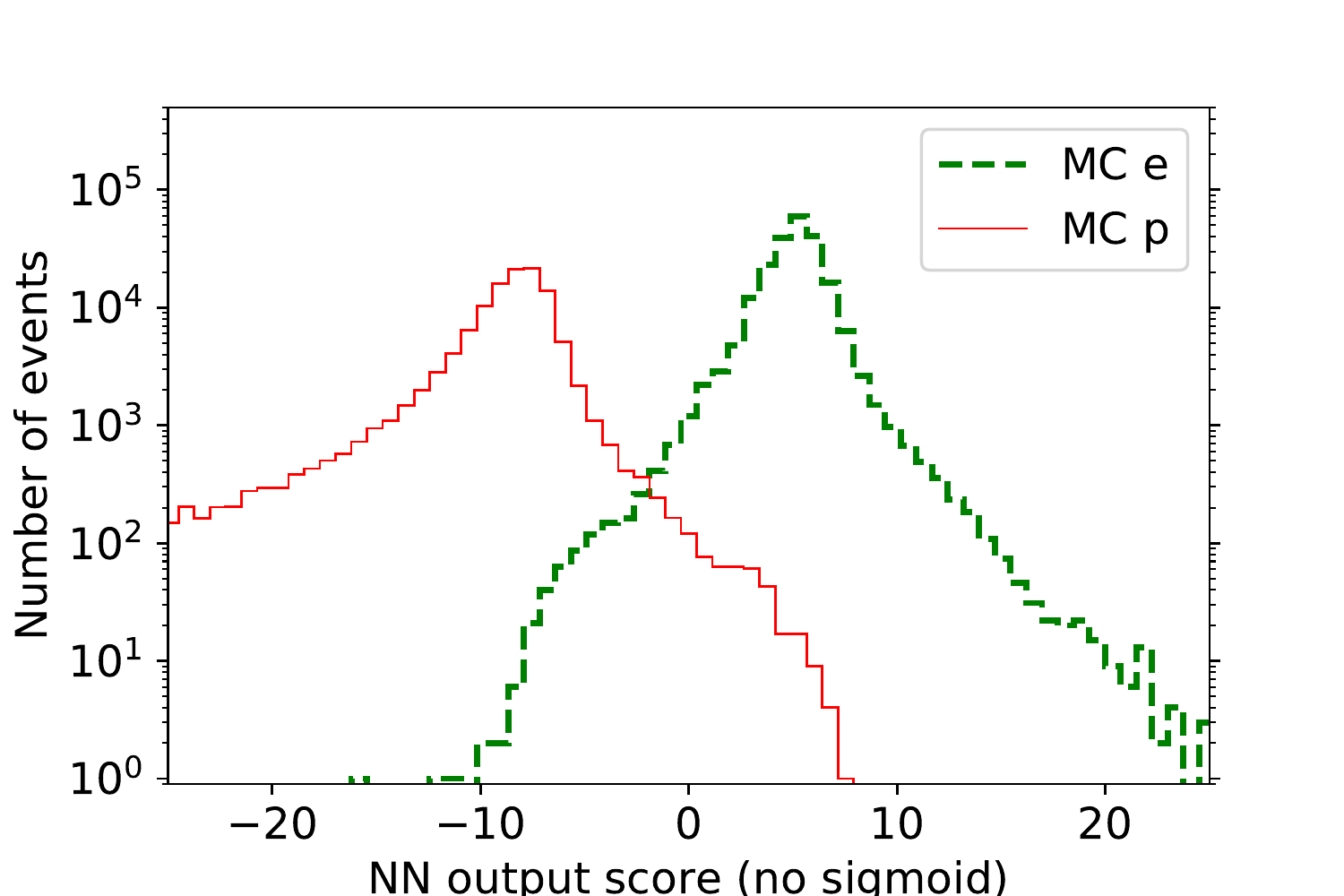}
    \caption{Histogram of the neural network output on electron and proton Monte Carlo: (left) standard model with a sigmoid activation function at the output of the last layer, and with a peak of false positive outcomes in the proton Monte Carlo highlighted with a circle ; (right) the same model after removing (post-training) the sigmoid at the last layer.}
    \label{fig:bkgpeak}
\end{figure}

A feature we noticed during the early stages of our optimisation procedure is that the neural network output values are either very close (or exactly equal to) 0.0 or 1.0, with only very few events classified in-between. This holds true for false positives and false negatives as well: figure \ref{fig:bkgpeak} (left) shows that the histogram of MC protons (background) exhibits two peaks: one at 0.0 (true negatives) and the second, much smaller, at 1.0 (false positives)\footnote{These events are likely protons that transfer most of their energy to one or several $\pi_0$, starting electromagnetic showers while the remaining energy yields a very small hadronic contribution. They are therefore the most difficult background to distinguish from electron-induced showers.}. The overall distribution is therefore non-monotonic, which introduces complications for the estimation of background on a cosmic ray electron measurement. For example, this behaviour prevents any sort of baseline background extrapolation.

The cause is a feature of neural networks for classification: the very last operation is a logistic sigmoid function that maps the output to the $[0;1]$ range:
\begin{equation}
f(x) = \frac{1}{1 + e^{-x}}
\end{equation}
Values $x \gg 0$ are mapped to $f(x) \simeq 1.0$, effectively compressing the output into a limited, finite space. Computer floating point accuracy also has an influence: for a 16-bits float, $f(18) = f(20) = 1.0$ exactly. Therefore the function is not bijective anymore.

\begin{figure}
    \centering
    \includegraphics[width=.65\linewidth]{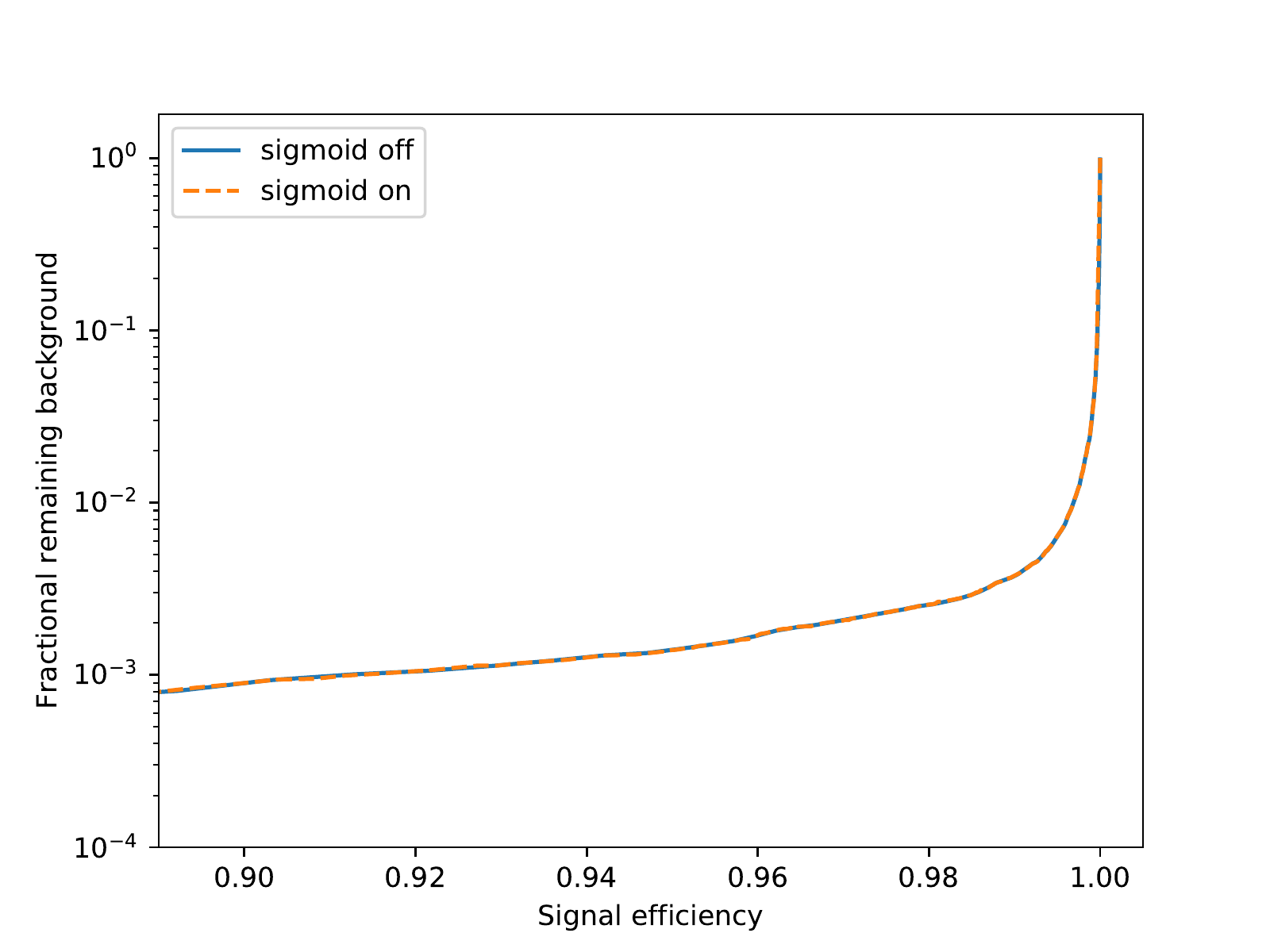}
    \caption{Sample Receiver Operating Characteristic (ROC) curve of a model with a sigmoid activation function at its output versus a model without. The curves compare the signal efficiency (true positive rate) versus the remaining background (false positive rate) at varying discrimination threshold. The perfectly superimposed curves prove that the performances are exactly the same.}
    \label{fig:bkgpeakROC}
\end{figure}

Our workaround is to remove the logistic sigmoid from the output layer. This must be done after training since the metric being optimised (binary crossentropy) assumes an output between 0 and 1. The result is shown on figure \ref{fig:bkgpeak} (right): the compression and non-monotonic features are gone, resulting in a distribution much easier to use for baseline extrapolation methods, without altering the classification performances (figure \ref{fig:bkgpeakROC}). By removing post-training the last activation function, we get a more monotonic behaviour for both classes. Another important gain is getting rid of the seemingly irreducible background: without the sigmoid we can select a small sample with 0\% background, by e.g. cutting away events with a score $\lesssim 10$. Whereas without this trick there are proton events mapped to the whole space $[0;1]$, meaning that even the tightest possible cut at 1.0 will still have a non-zero false positive rate. 

\section{Performances}
\label{sec:results:perf}

\begin{figure}
    \centering
    \includegraphics[width=.45\linewidth]{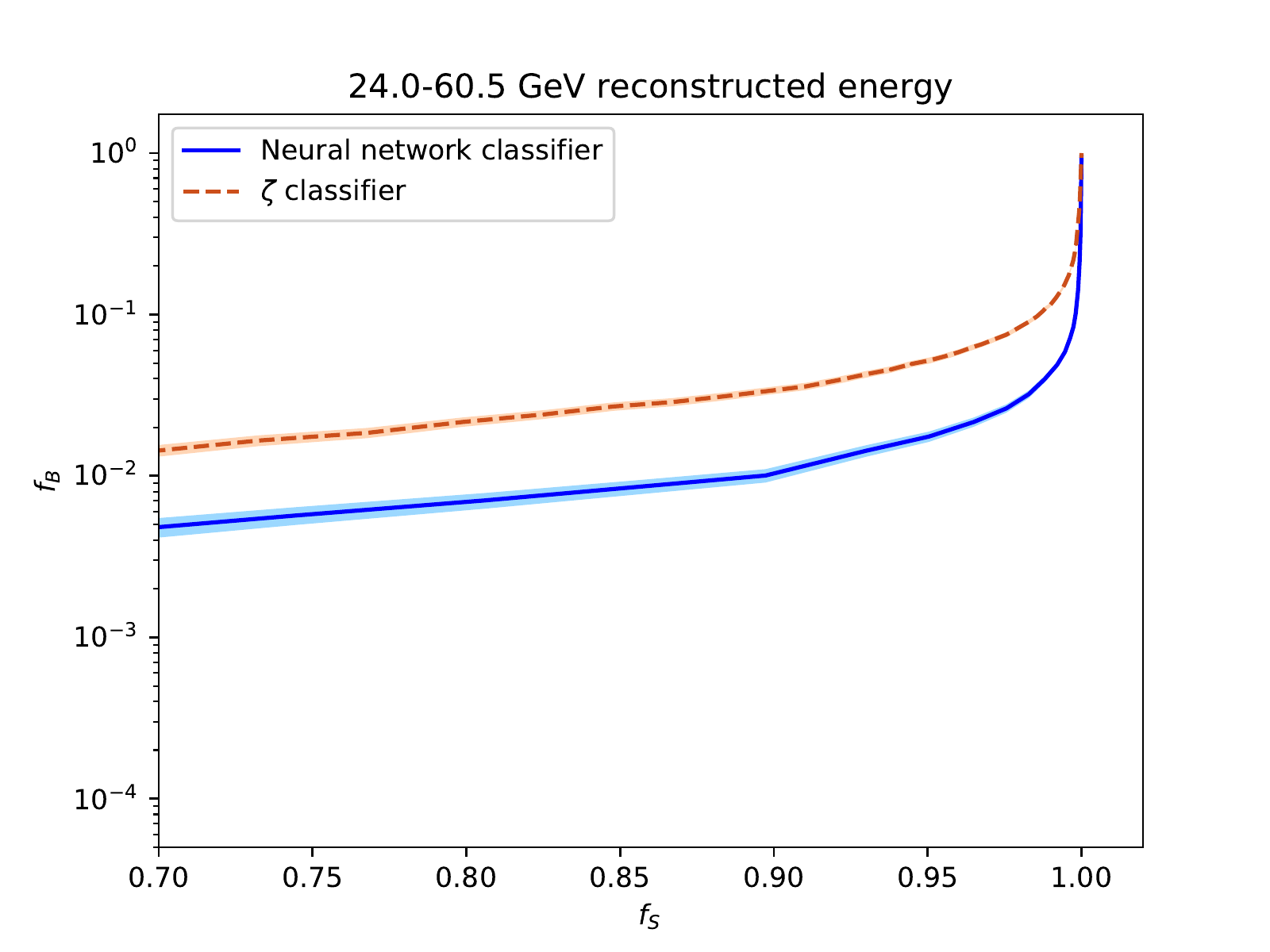}
    \includegraphics[width=.45\linewidth]{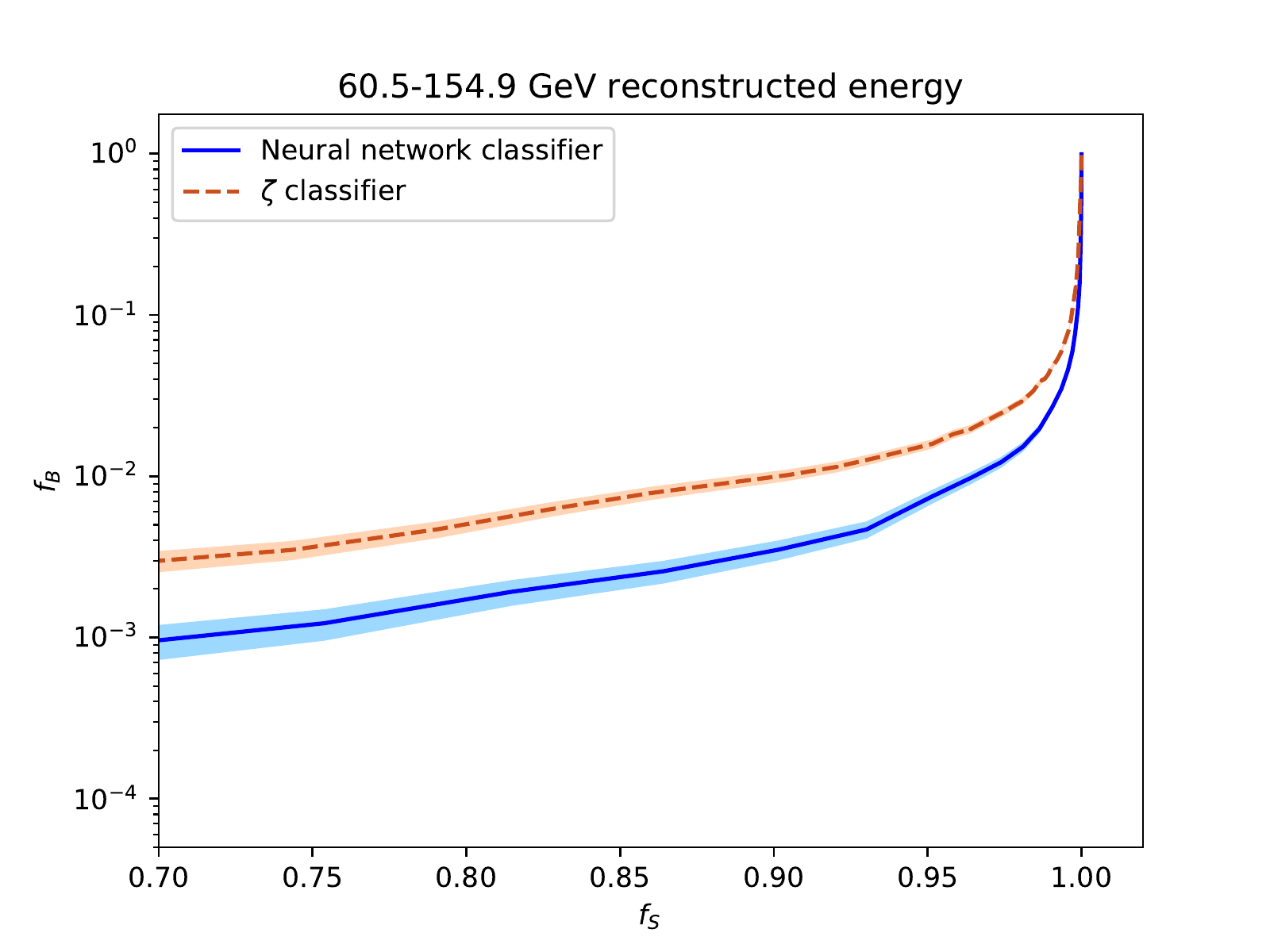}
    \includegraphics[width=.45\linewidth]{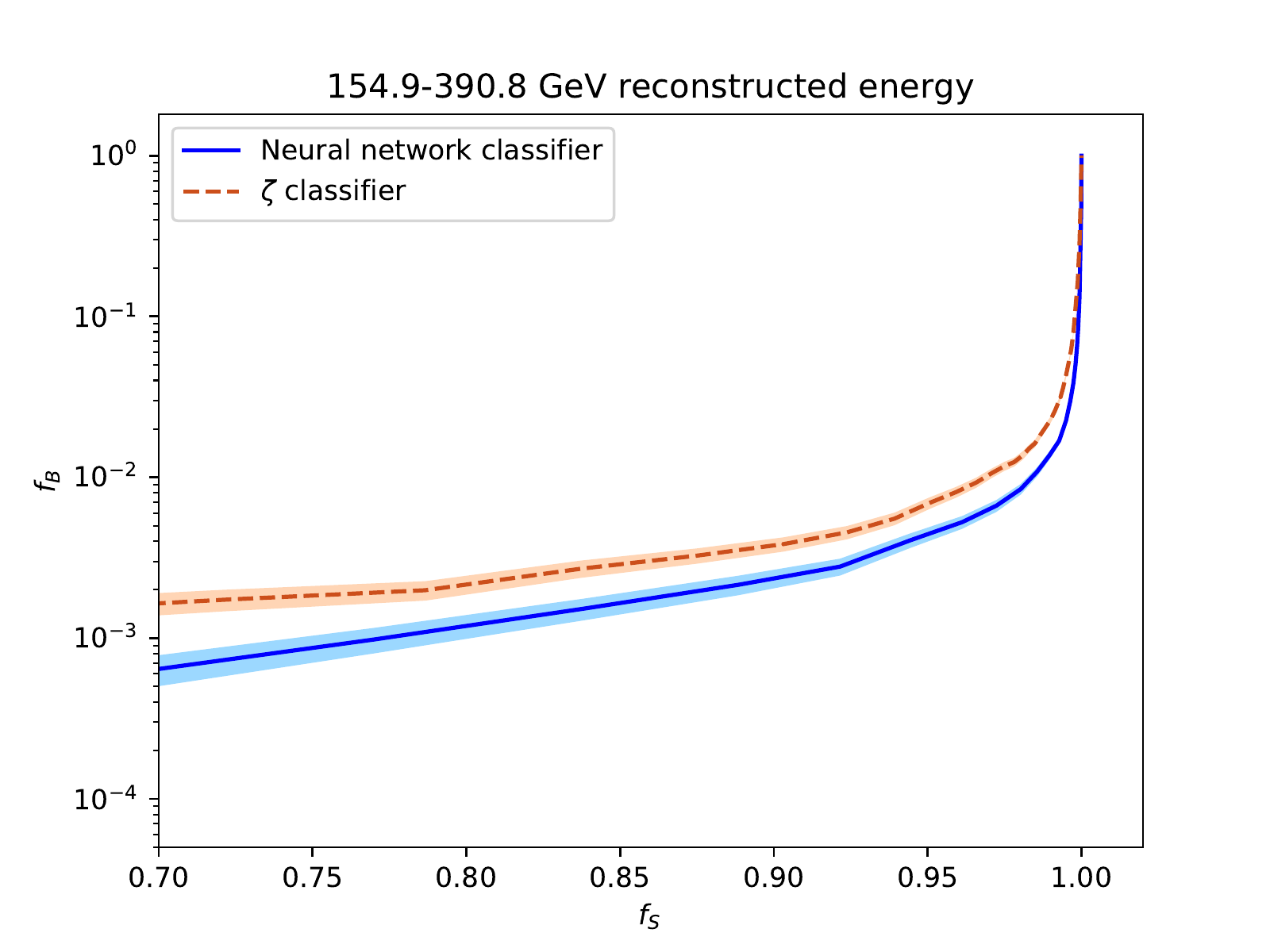}
    \includegraphics[width=.45\linewidth]{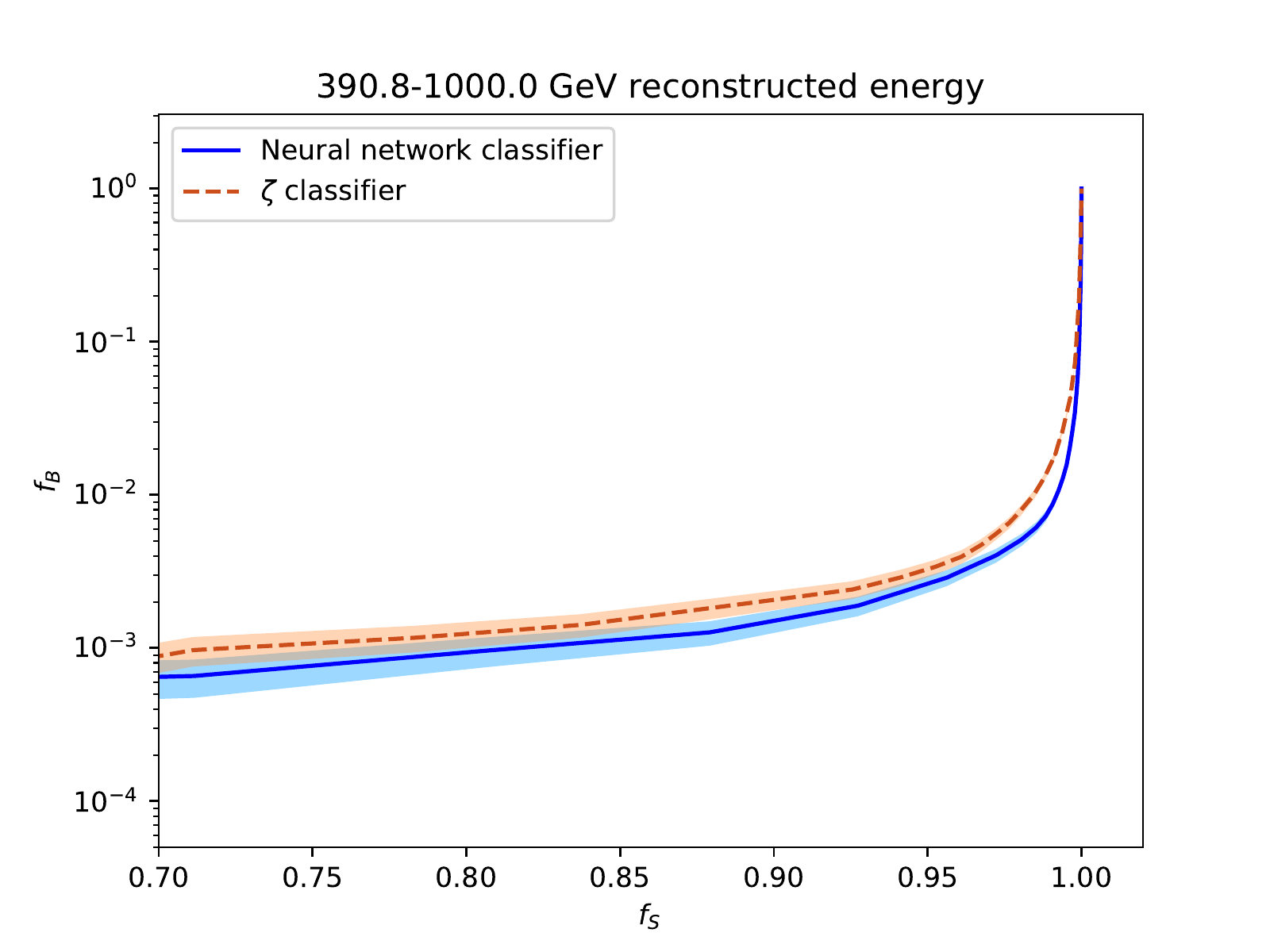}
    \includegraphics[width=.45\linewidth]{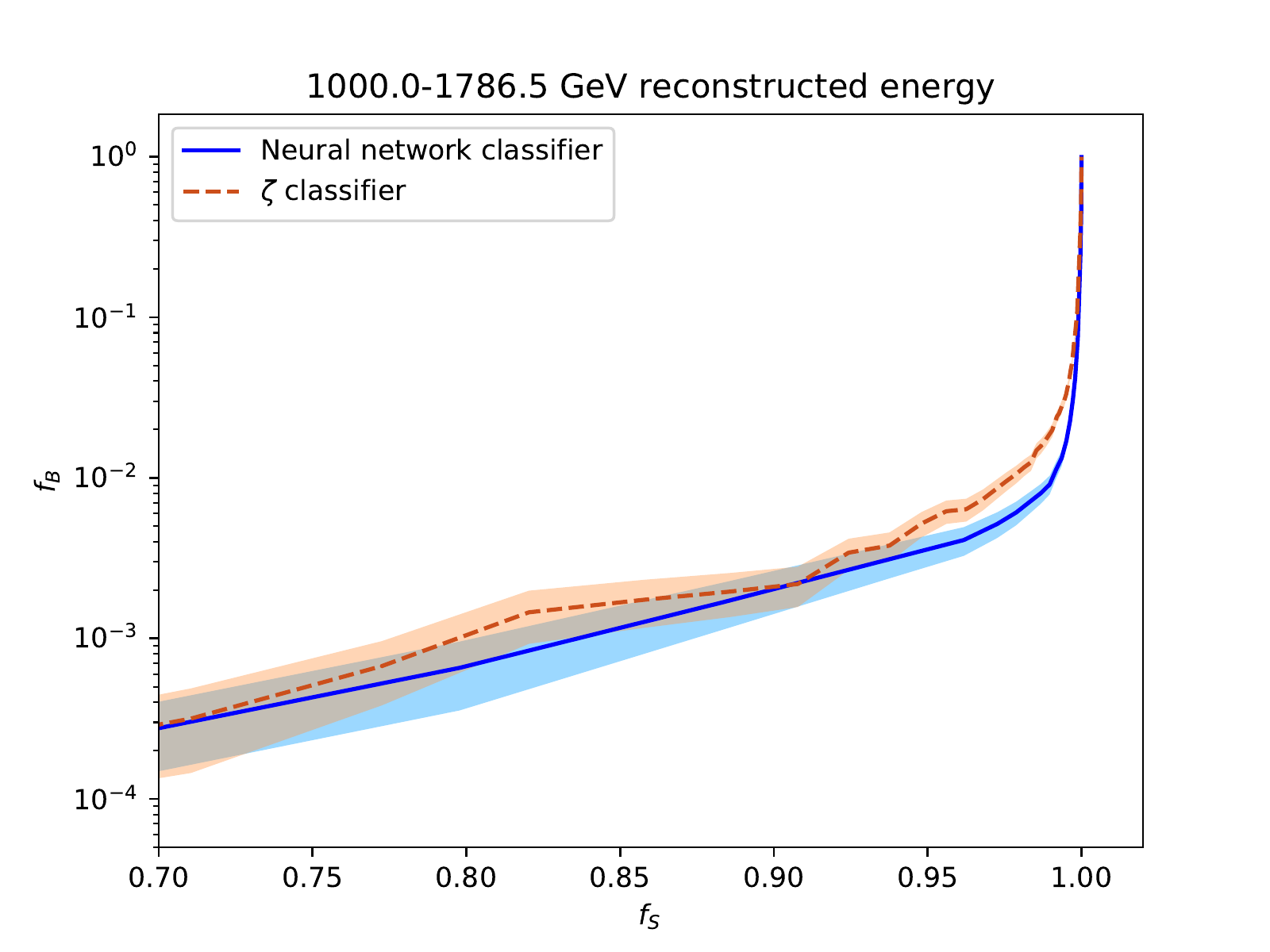}
    \includegraphics[width=.45\linewidth]{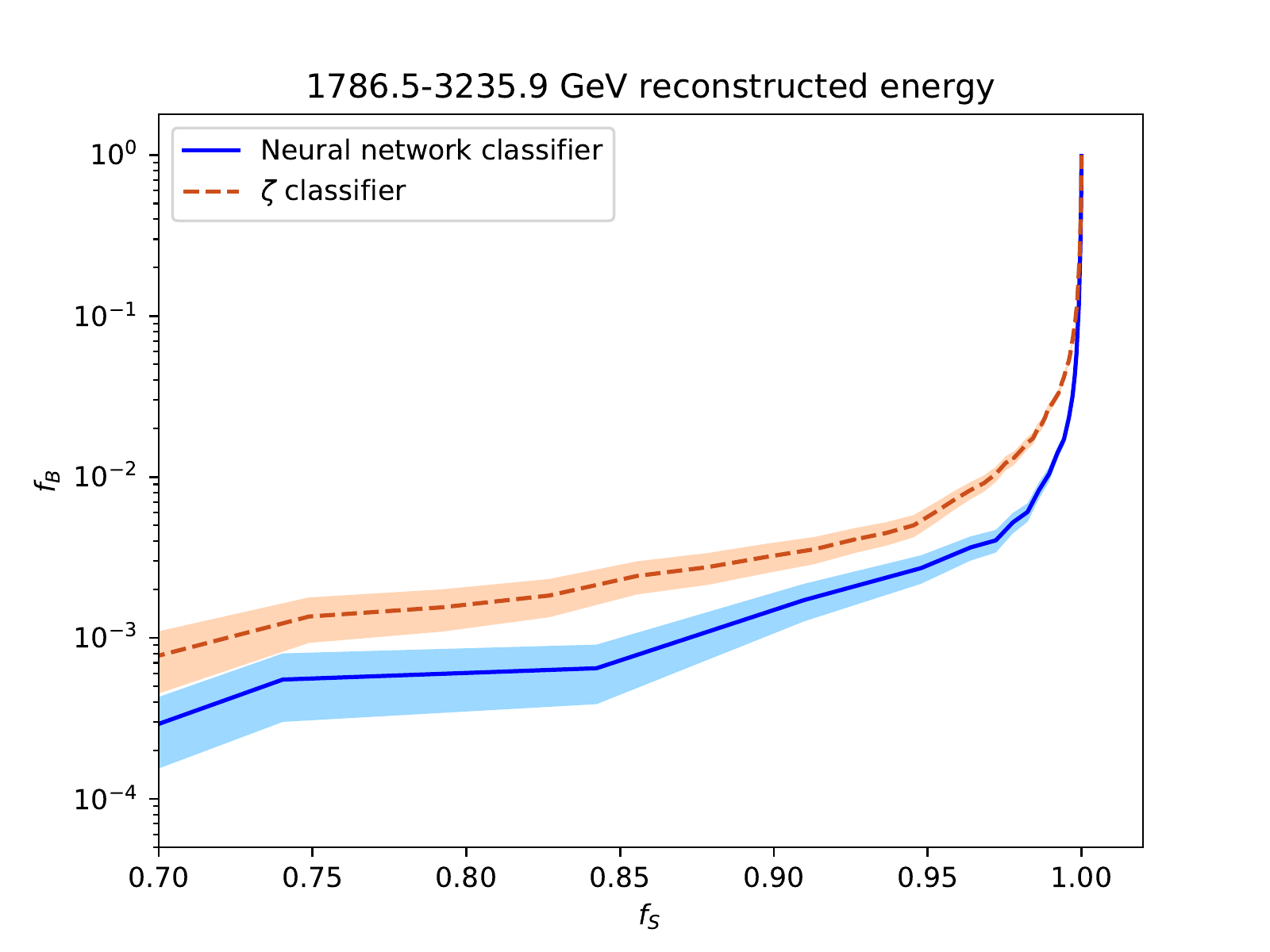}
    \includegraphics[width=.45\linewidth]{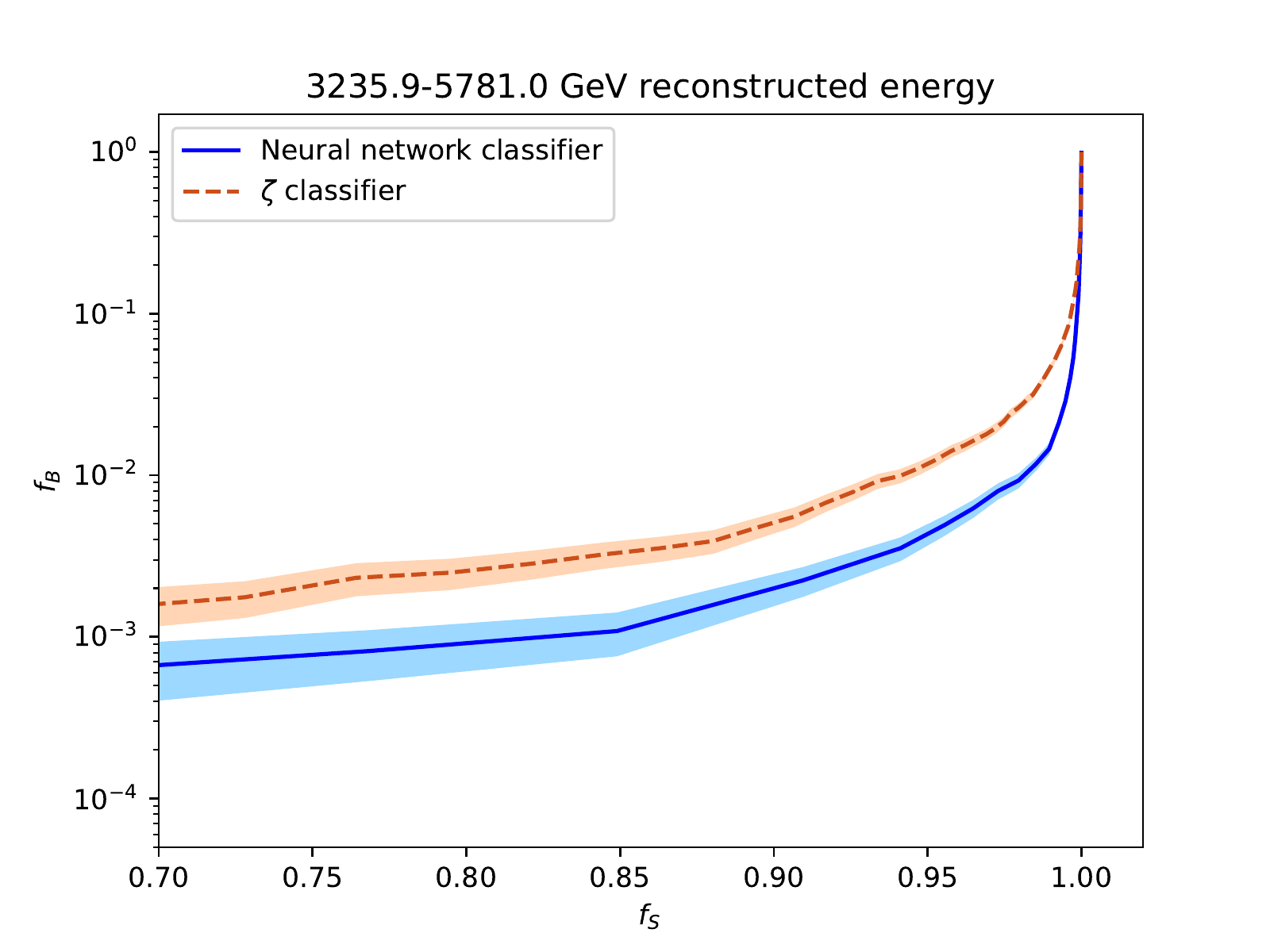}
    \includegraphics[width=.45\linewidth]{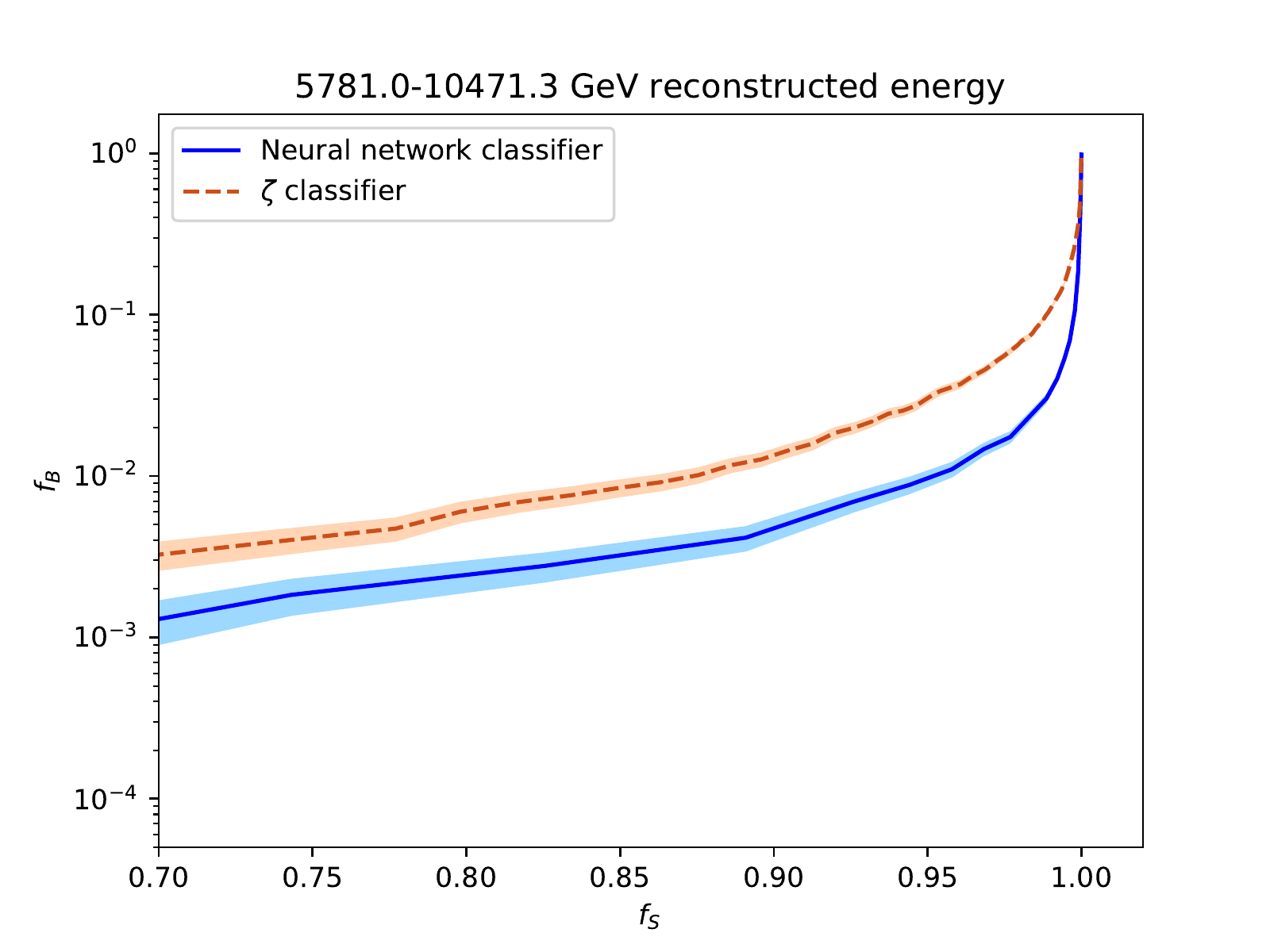}
    \caption{Sample ROC curves on Monte-Carlo for the neural network classifier and the classical $\zeta$ classifier, in 8 reconstructed energy bins from 24 GeV to 10.4 TeV, comparing the fractional surviving background $f_B$ with the signal efficiency $f_S$. The lowest curves have the lowest background for a fixed efficiency. The coloured band shows statistical uncertainty from Monte Carlo sampling. Logarithmic y-scale.}
    \label{fig:MCroc}
\end{figure}

\begin{figure}
    \centering
    \includegraphics[width=.49\linewidth]{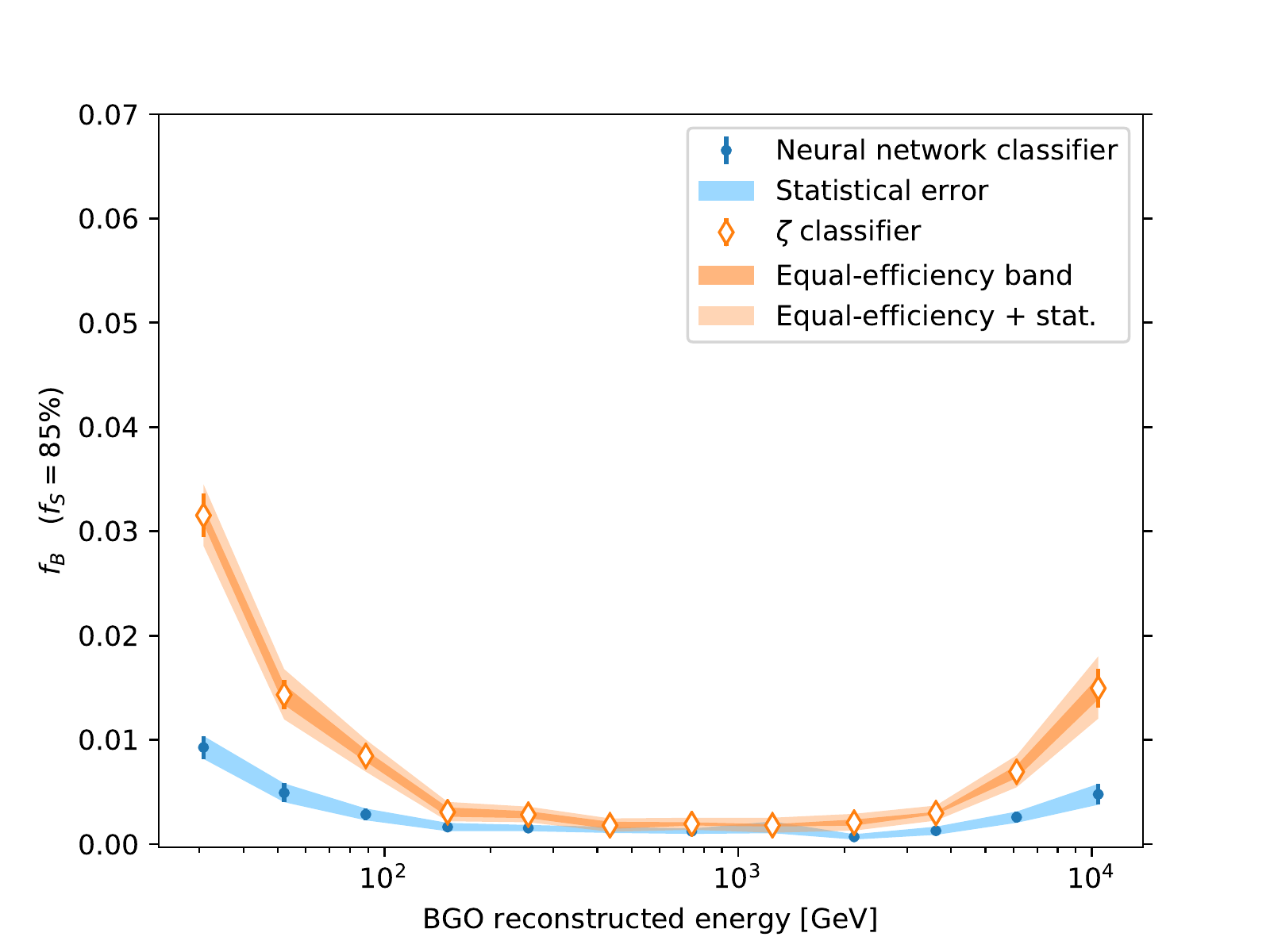}%
    \hfill
    \includegraphics[width=.49\linewidth]{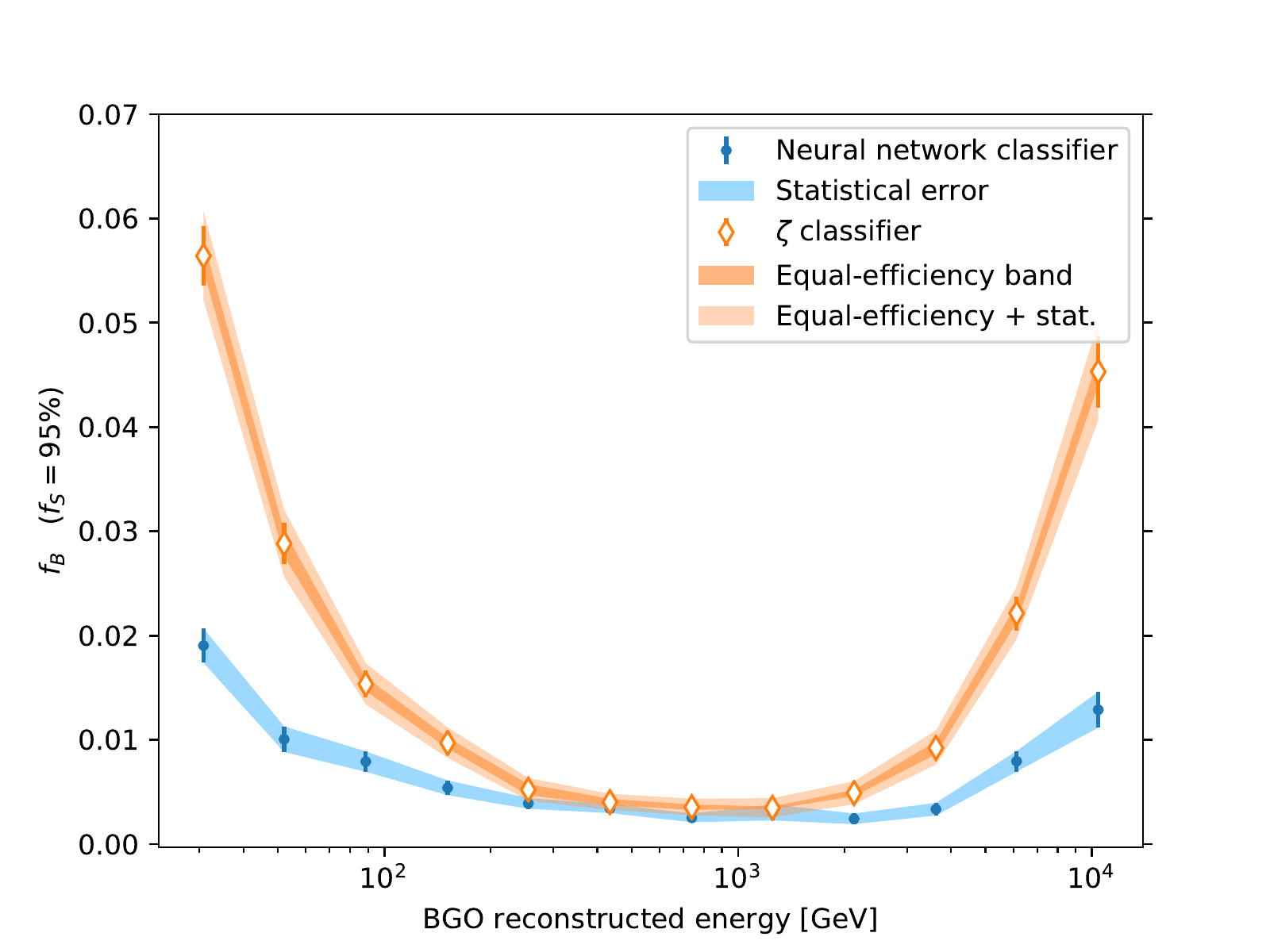}
    \caption{Energy dependency of the surviving background fraction $f_B$ for a fixed signal efficiency $f_S$ of 85\% (left) and 95\% (right), for the neural networks and the classical $\zeta$ method.}
    \label{fig:MCperformances}
\end{figure}

We report on figure \ref{fig:MCroc} several Receiver Operating Characteristics (ROC) curves for our neural networks, in comparison with the classical $\zeta$ method, covering the energy range from 24 GeV to 10.4 TeV. ROC curves are obtained by computing classification metrics at various discrimination thresholds, on the test sample. In our case we choose to plot the signal efficiency $f_S$ against the fractional remaining background $f_B$. Both quantities are defined as the ratio of passing events over total events for electrons and protons respectively:
\begin{align}
f_S &= \frac{N_{e^-,pass}}{N_{e^-}}  \\
f_B &= \frac{N_{p,pass}}{N_{p}}
\end{align}
These two metrics have the advantage of being independent from the relative abundance of electrons with respect to protons. A good classifier is one that maximises the first metric and/or minimises the second. This translates into a lower curve on figure \ref{fig:MCroc}: classifiers with the lowest curves have the smallest background for a set efficiency. The image shows that the neural network significantly outperforms the classical method in the lowest and highest energy ranges, while the performances appear roughly comparable at intermediate energies. Note that for both classifiers, $N_{e^-}$ and $N_{p}$ are taken after the $\zeta < 100$ cut for a fair comparison.

The performances are thus energy-dependent. To see this dependence and to better quantify the performances of both methods, we report on figure \ref{fig:MCperformances} the $f_B$ value when we set the discrimination threshold such as to have a 85\% or 95\% signal efficiency, as a function of the energy reconstructed from the BGO calorimeter. The comparison involves an uncertainty due to the efficiency of both classifiers not being perfectly equal. On the figure, the error bars associated to $\zeta$ show the statistical uncertainty from Monte Carlo sampling, the darker band shows the uncertainty associated to the choice of threshold to have compatible efficiency, and the lighter band is the combination of both.  The blue band associated to the neural networks is purely statistical. Figure \ref{fig:MCperformances} confirms the previous observation that the gains of neural networks are significant on both ends of the energy range, in the high efficiency regime. From a few hundred GeV to 2 TeV, the performances are within uncertainty of each other. Above 5 TeV, the proton rejection is improved by a factor at least 2.

\section{Model validation}
\label{sec:results:val}

\begin{figure}
    \centering
    \includegraphics[width=.48\linewidth]{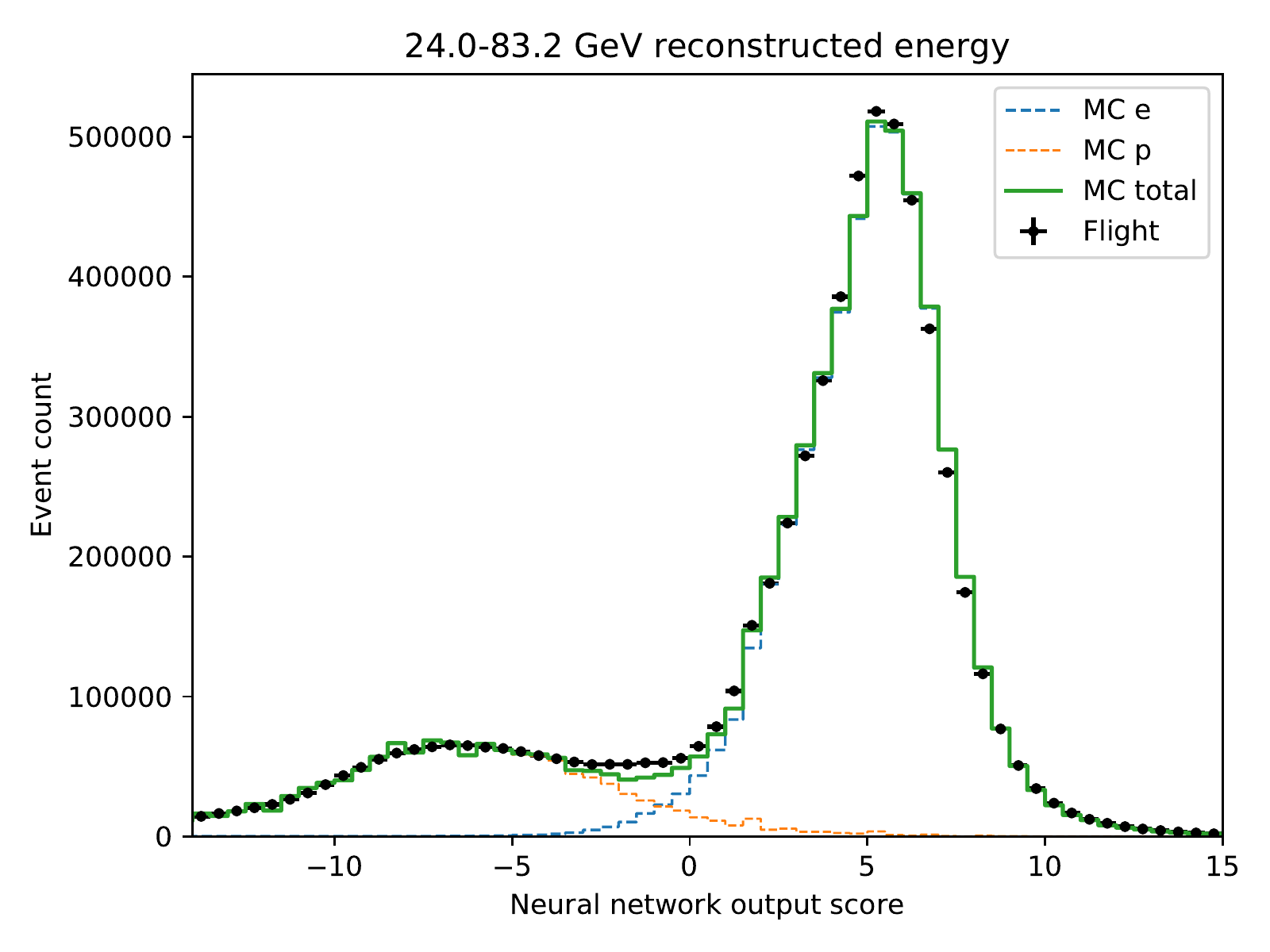}
    \includegraphics[width=.48\linewidth]{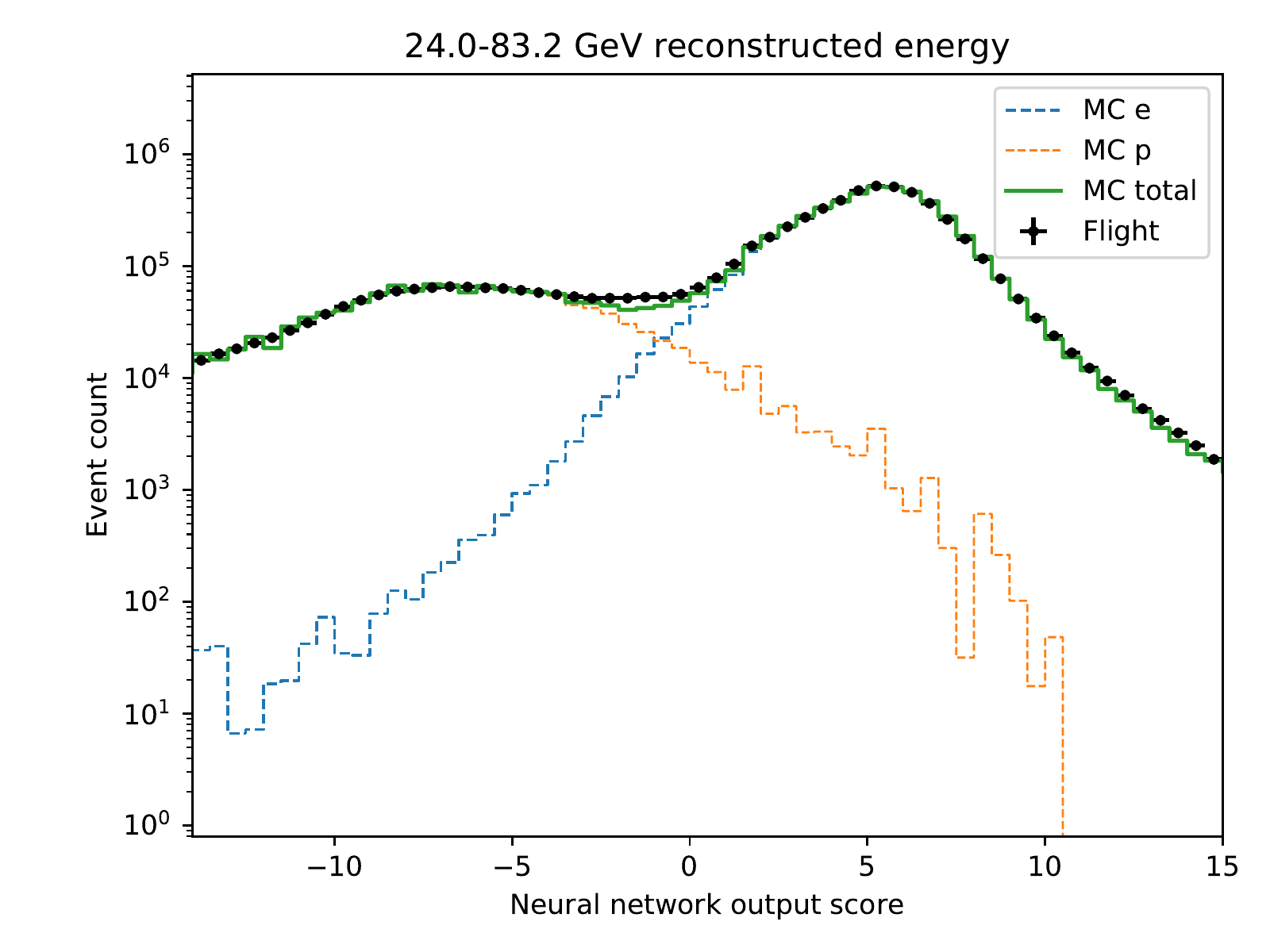}
    
    \includegraphics[width=.48\linewidth]{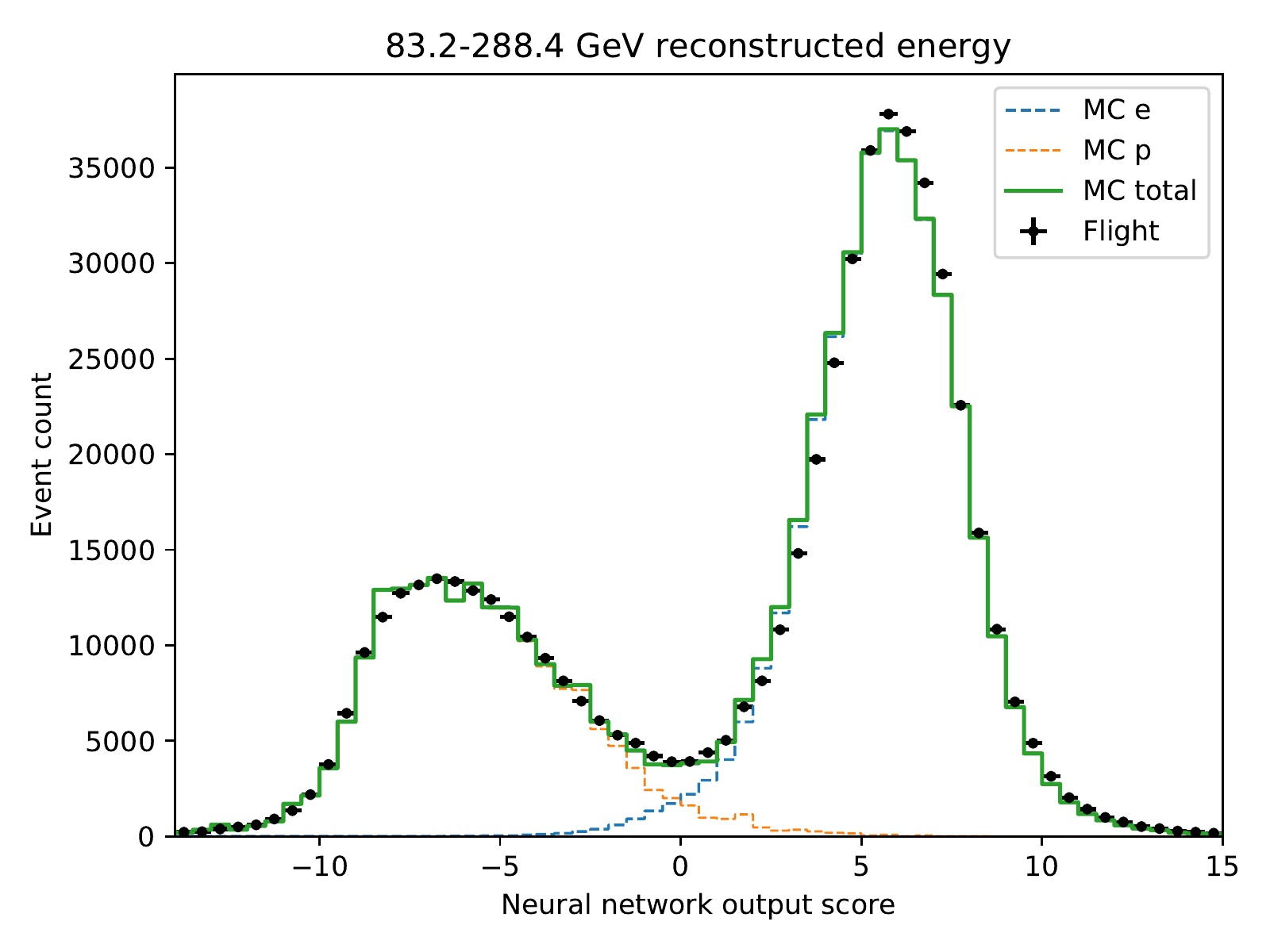}
    \includegraphics[width=.48\linewidth]{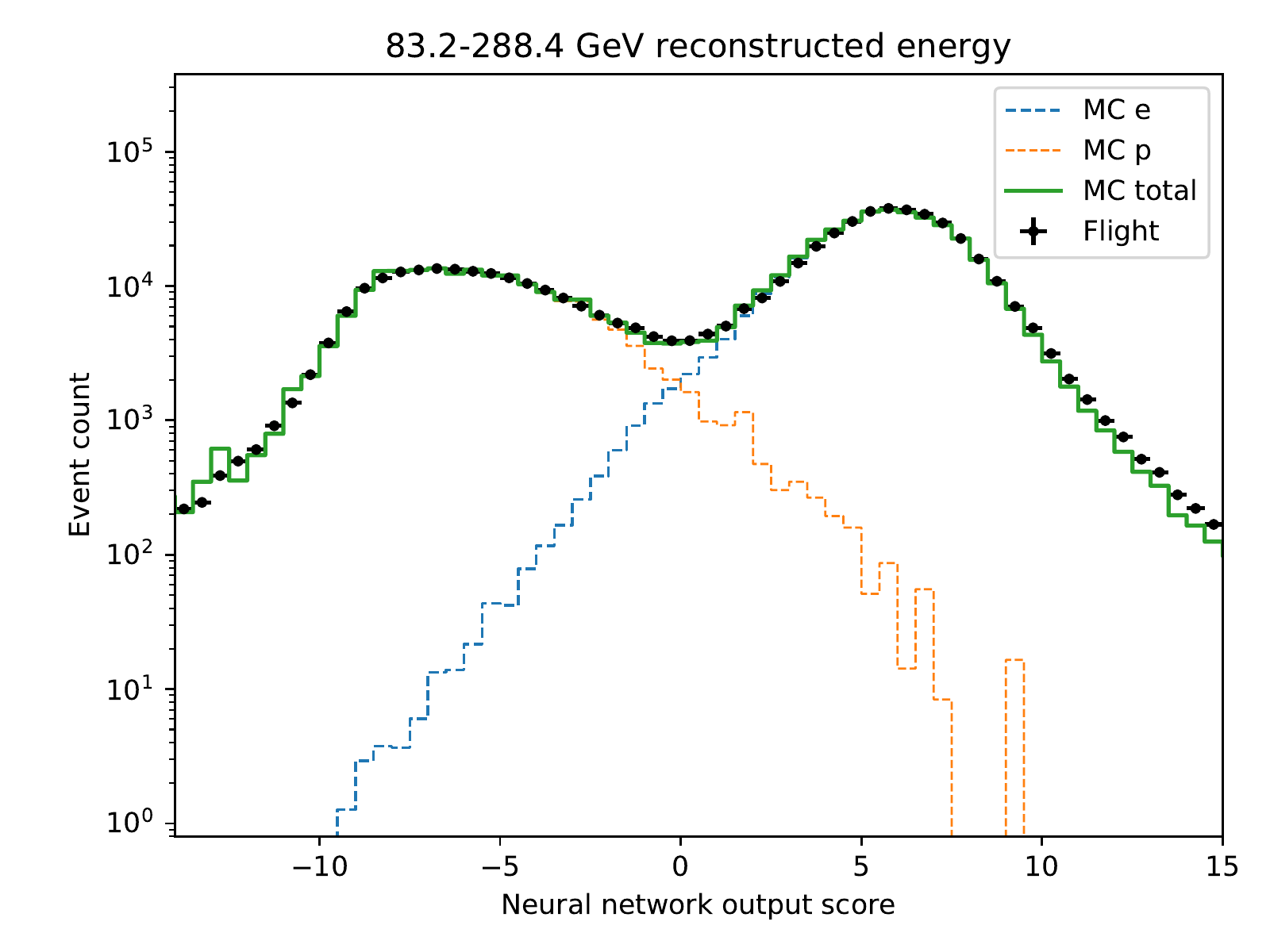}
    
    \includegraphics[width=.48\linewidth]{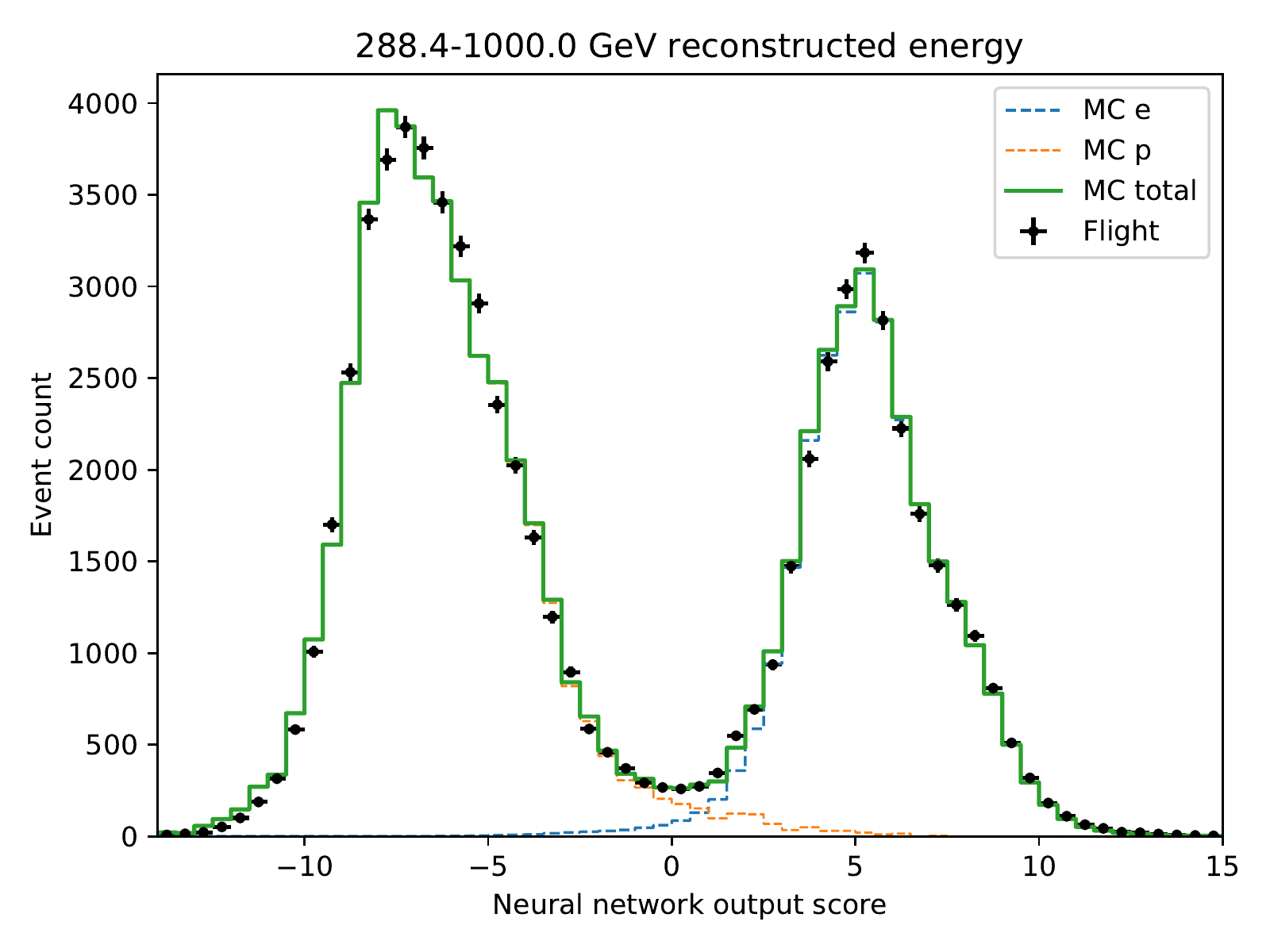}
    \includegraphics[width=.48\linewidth]{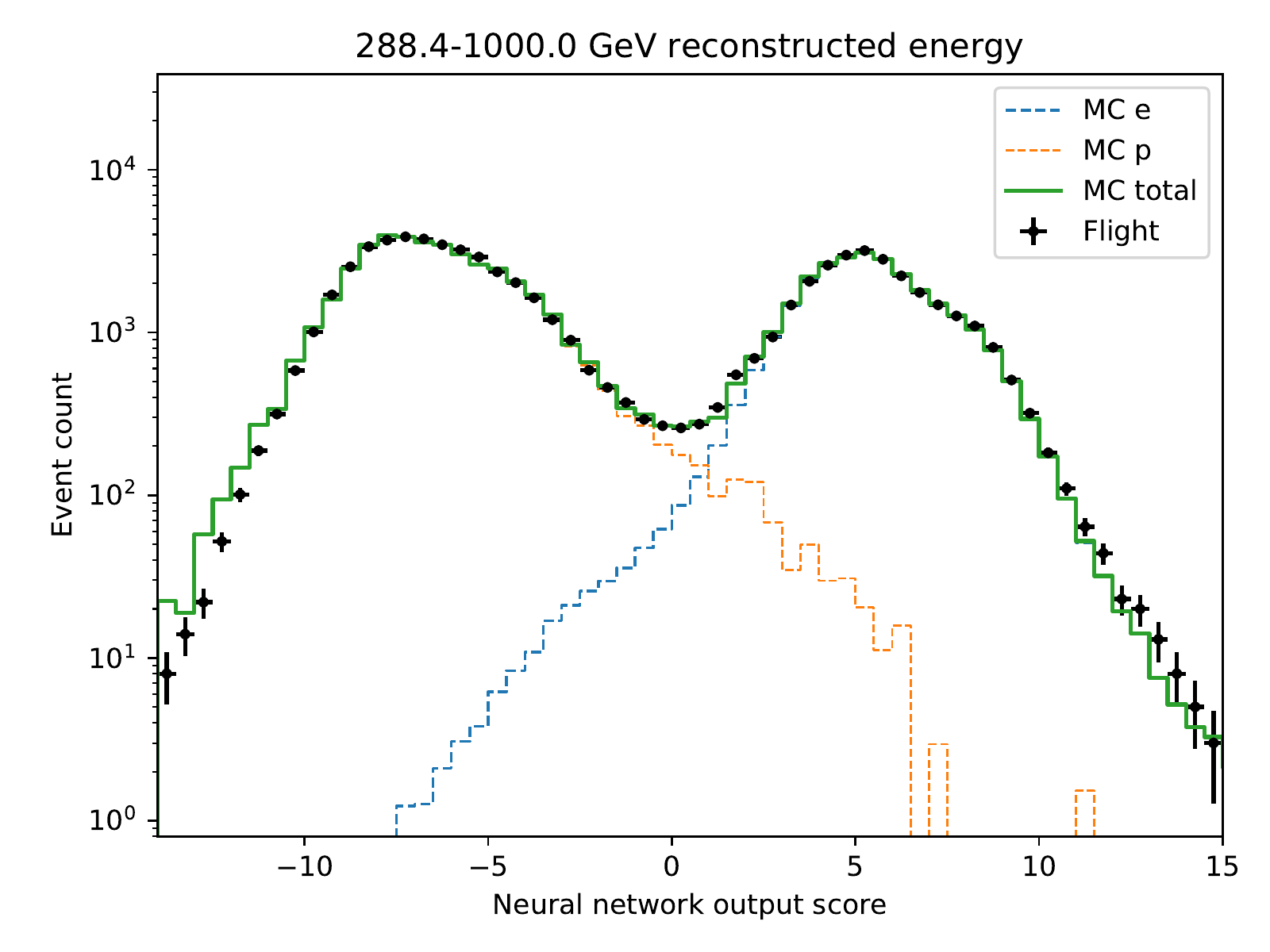}
    \caption{Comparison of the neural network output distribution between simulated Monte Carlo and real data, on three energy bins from 24 GeV to 1 TeV. Left panels are linear scale, right panels are log scale.}
    \label{fig:matching}
\end{figure}

\begin{figure}
    \centering
    \includegraphics[width=.48\linewidth]{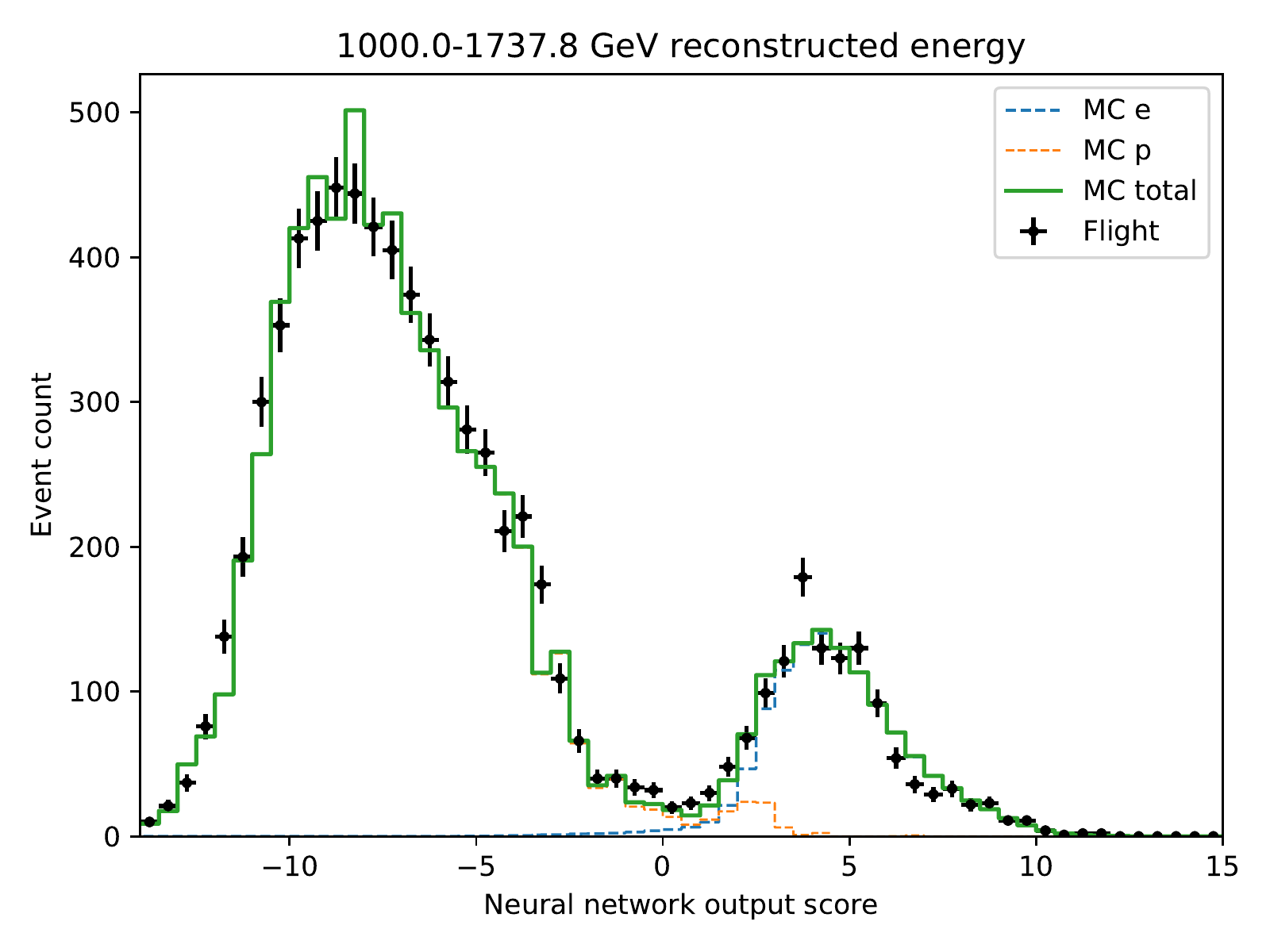}
    \includegraphics[width=.48\linewidth]{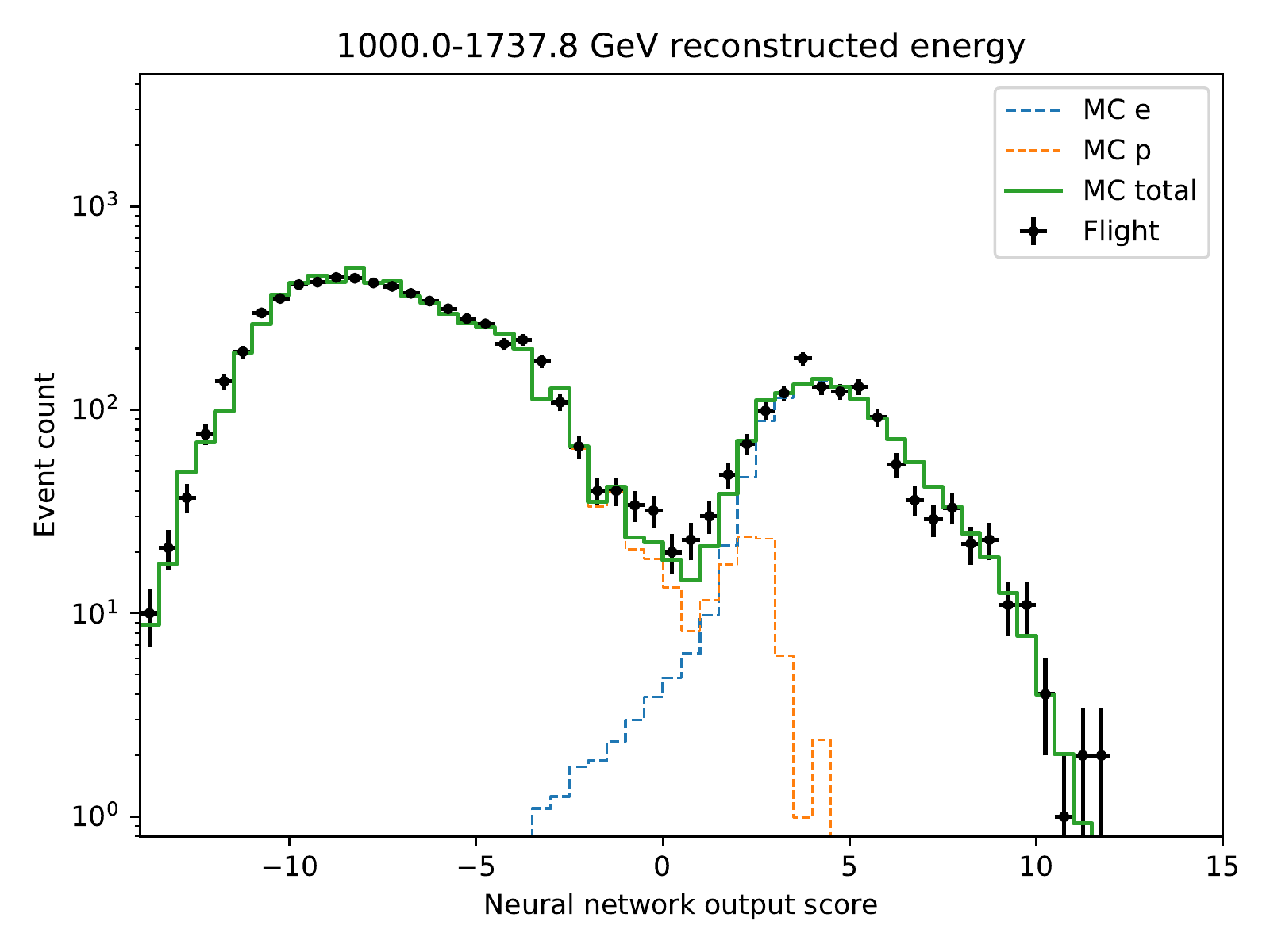}
    
    \includegraphics[width=.48\linewidth]{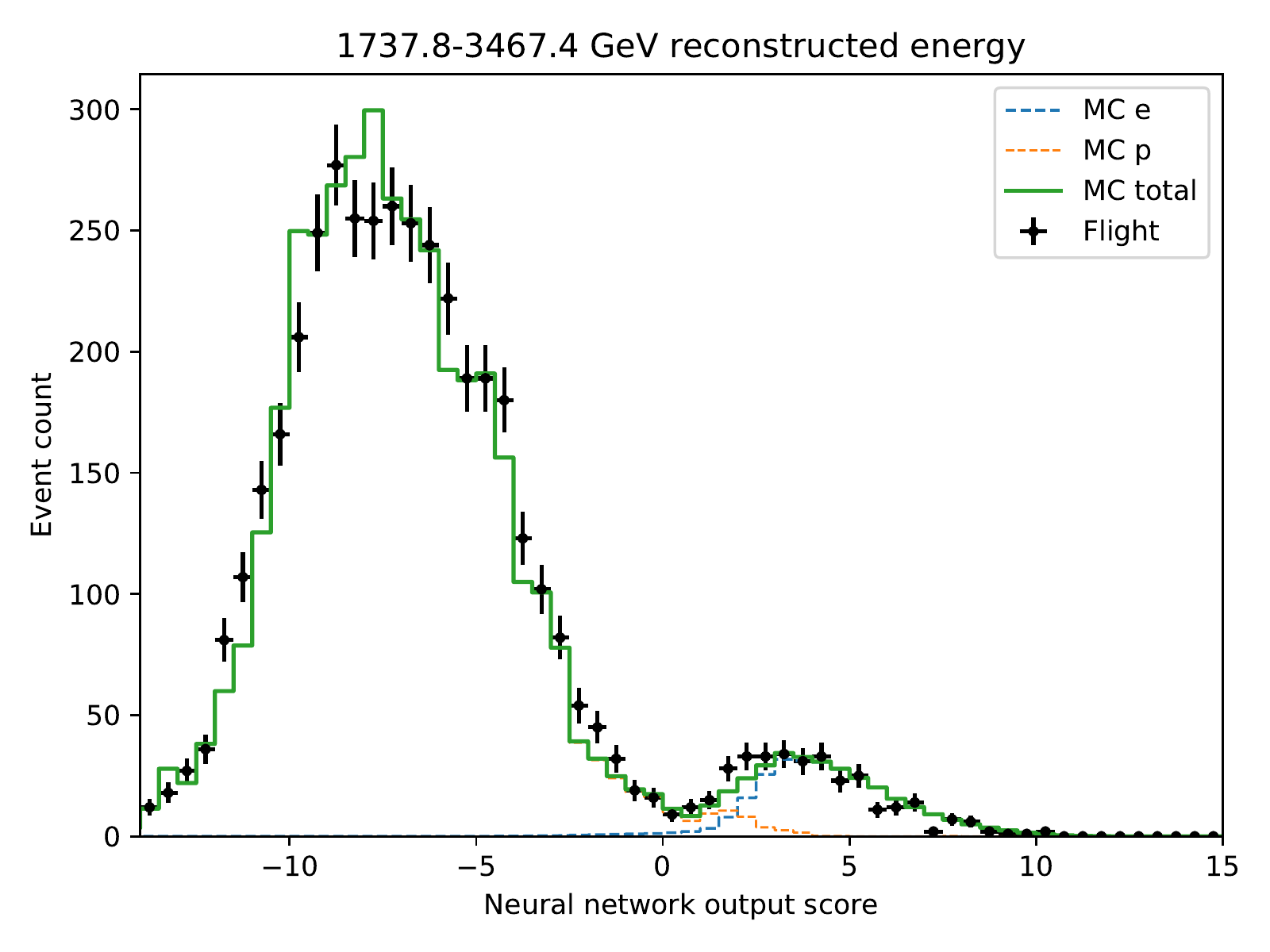}
    \includegraphics[width=.48\linewidth]{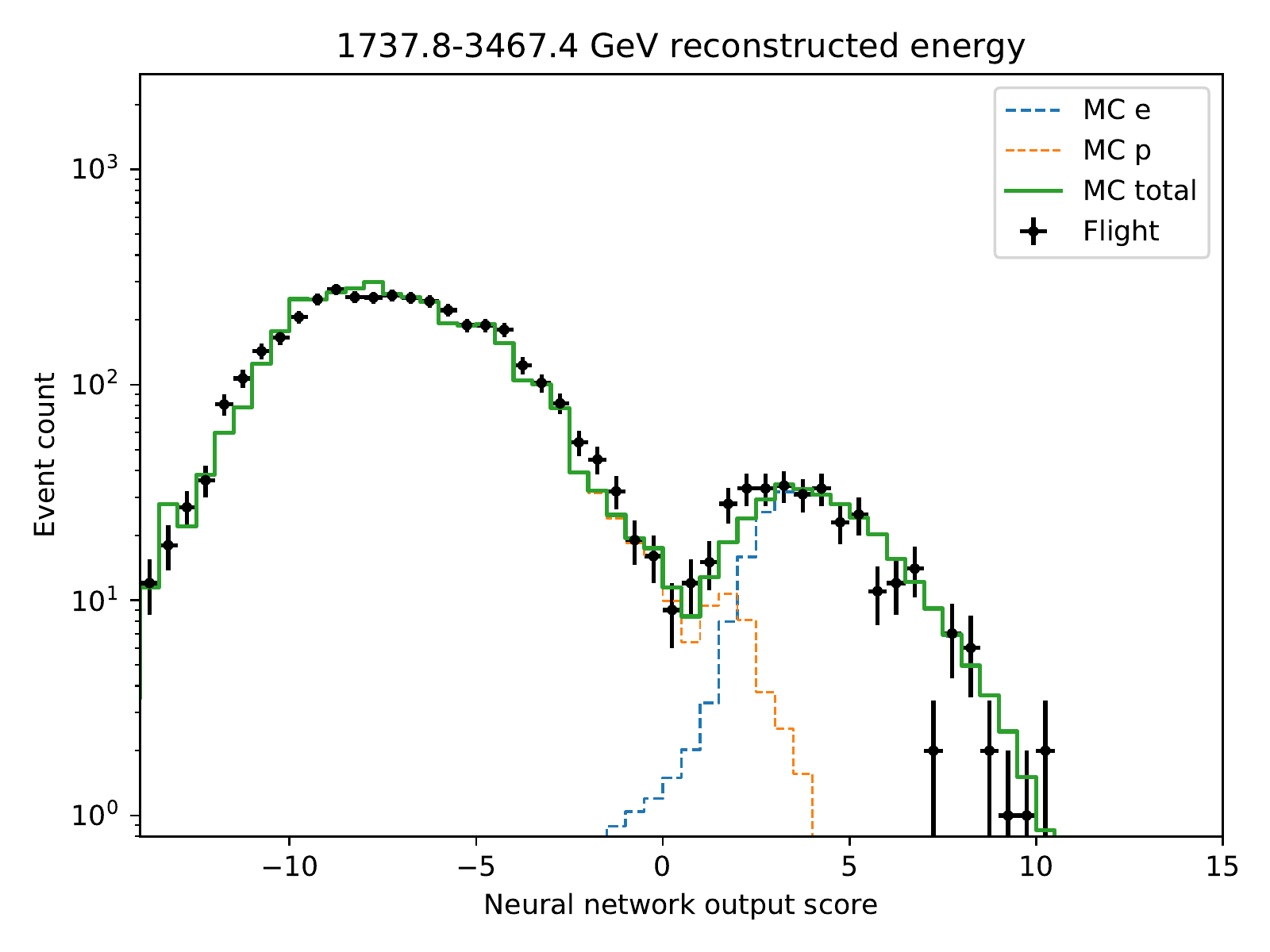}
    
    \includegraphics[width=.48\linewidth]{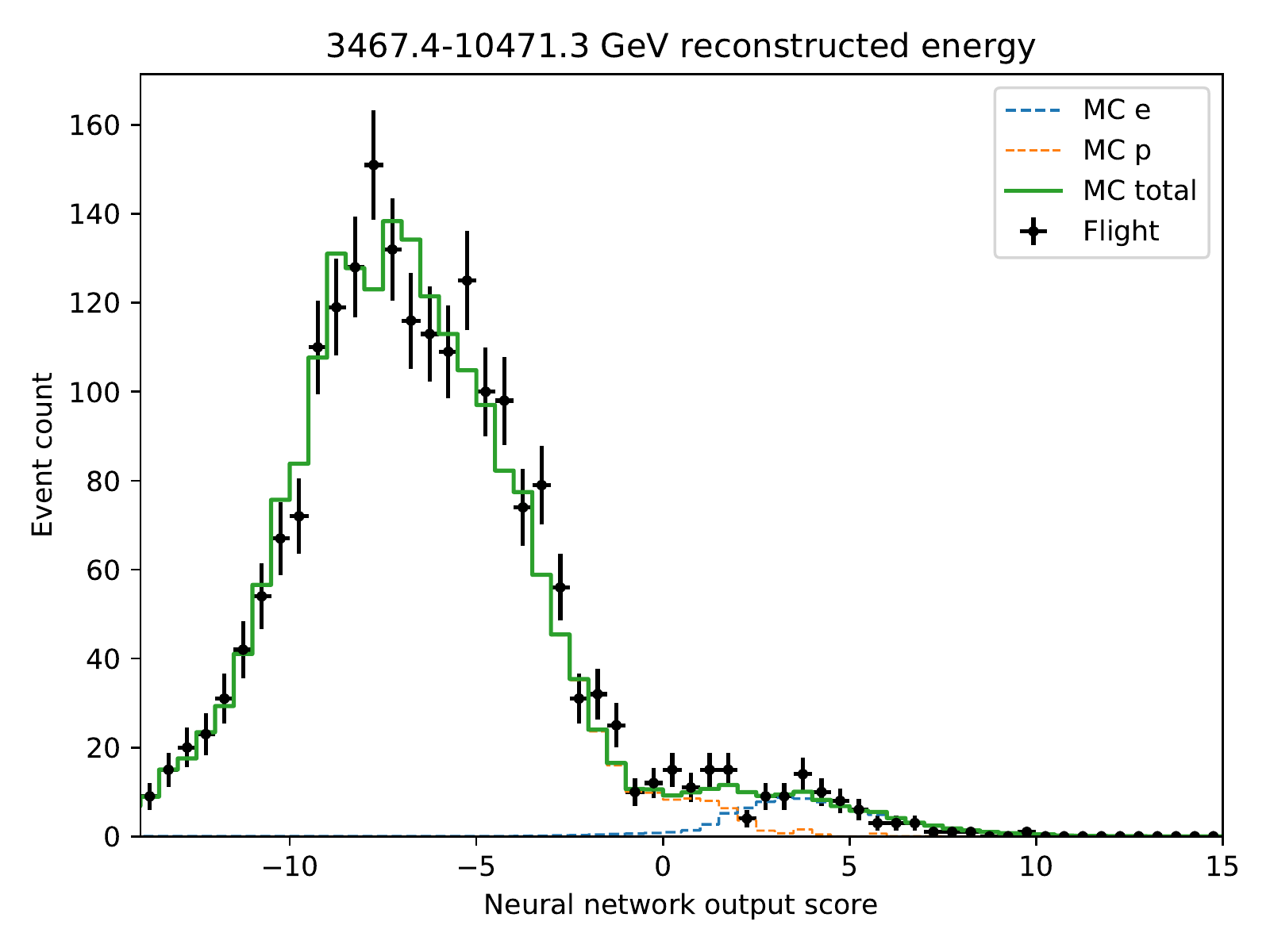}
    \includegraphics[width=.48\linewidth]{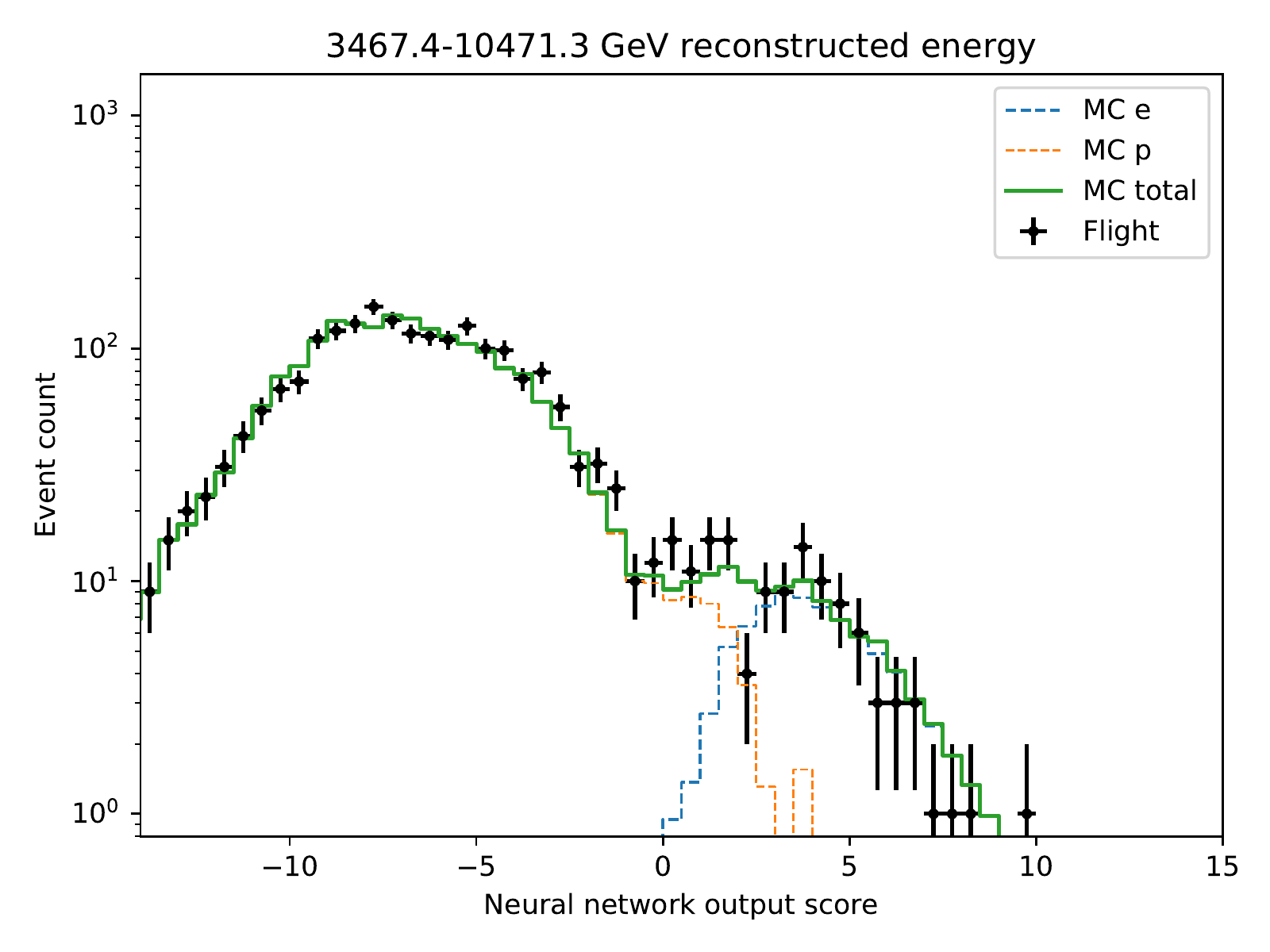}
    \caption{Comparison of the neural network output distribution between simulated Monte Carlo and real data, on three energy bins from 1 to 10.4 TeV. Left panels are linear scale, right panels are log scale.}
    \label{fig:matchingLog}
\end{figure}

Validating the modelling of the classifier on real data is an important step in order for its use in an analysis measurement. Namely, it allows us to validate the performances observed in figure \ref{fig:MCperformances}, to exclude a potential bias of the network with respect to MC, and to use the shape for e.g. baseline interpolation methods.

Both distributions of simulations and data are shown on figures \ref{fig:matching} and \ref{fig:matchingLog} in six logarithmatically-spaced energy bins. 
Both samples have been cleaned such that contributions from other cosmic species, notably Helium nuclei, are negligible (section \ref{sec:electronidentification}). A sum of proton and electron Monte-Carlo templates were fitted to the data in the following way: the proton sample was scaled to the data in the region $[-12;-4]$, while electron MC was scaled to the data in the region $>1$ after subtracting the proton template from the data. The scale factor is taken as the data-to-MC ratio of the integral over the relevant control region. This technique allows to scale each distribution independently, on regions where the contribution from the other species is negligible. The Monte Carlo distributions are weighted according to their expected flux, with a spectral index of -3 for electrons and -2.7 for protons. Changing these indices according to the observed DAMPE values~\cite{Ambrosi:2017wek,an2019measurement} and variating them within the corresponding uncertainties yielded similar figures and conclusions.
Due to some residual inaccuracies of modeling the input variables in the neural network, both Monte Carlo distributions are shifted by 0.2 to the right (less than half a bin size) in order to improve matching with the data. The final shift value is the result of a chi-square minimisation, iterating by steps of 0.1. Error bars show purely statistical uncertainties. Figures \ref{fig:matching} and \ref{fig:matchingLog} show a good agreement between simulations and DAMPE data, especially around the boundaries of signal and control regions. A small disagreement below 100 GeV is however noticeable, although thanks to the very low proton background at such energies it results in negligible uncertainties if applied to a CRE flux measurement. The neural network classifier is therefore reliable, in particular thanks to the transformation described in section \ref{sec:modeloptimisation} allowing the scaling of Monte Carlo shown in the figures. The remaining small differences between data and Monte-Carlo, in particular due to potential inherent limitations on accuracy of the hadronic models used in the simulations, can be further quantified and assigned a corresponding systematic uncertainty, reflected in the subsequent measurement of cosmic ray electrons. This is however beyond the scope of this work.

\section{Conclusions}
\label{sec:conclusions}

To tackle the problem of high energy cosmic electron identification and measurement with DAMPE, we developed a four-layers deep feed-forward neural network classifier, the output of which we transformed to suit the needs of particle physics analysis. On simulated data, the new classifier shows a strong background rejection power, outperforming the more traditional cut-based electron identification technique in the energy ranges where the latter shows its limits, thanks to a better exploitation of the information contained within DAMPE calorimeter and tracker. In particular,  the proton rejection improves by a factor $>$2 in the multi-TeV range at a signal efficiency of 95\%, and up to a factor 3 to 4 at 10 TeV, where the gain in accuracy is the most demanded due to the low relative event rate of cosmic electrons at such energies. A comparison between simulated and real data settles the reliability of the new method and paves the way towards high precision measurement of cosmic ray electron plus positron spectrum at the highest energies accessible by DAMPE.

\acknowledgments

The DAMPE mission was funded by the strategic priority science and technology projects in space science of the Chinese Academy of Sciences. In China, the data analysis was supported in part by the National Key Research and Development Program of China (no. 2016YFA0400200), the National Natural Science Foundation of China (nos. 11525313, 11622327, 11722328, U1738205, U1738207, and U1738208), the strategic priority science and technology projects of the Chinese Academy of Sciences (no. XDA15051100), the 100 Talents Program of Chinese Academy of Sciences, and the Young Elite Scientists Sponsorship Program. In Europe, the activities and the data analysis were supported by the Swiss National Science Foundation (SNSF), Switzerland, National Institute for Nuclear Physics (INFN), Italy and European Research Council (ERC) under the European Union’s Horizon 2020 research and innovation programme (grant agreement No 851103).

The computations presented in this document were performed at University of Geneva on the Baobab cluster, with significant help from computer engineer Y. Meunier and from the HPC team. Simulations were performed on INFN CNAF and ReCaS clusters, Italy, and on Swiss National Supercomputing Centre (CSCS) Piz Daint (project s979).

The corresponding author would like to acknowledge SARS-CoV-2 for making this work significantly harder than it should have been.




\bibliography{bib.bib} 
\bibliographystyle{unsrt}


\end{document}